\documentclass[sort&compress,a4paper,10pt]{edpsci} 

\usepackage{amsmath}
\usepackage{graphicx}
\usepackage{epstopdf}

\usepackage[numbers]{natbib}
\usepackage{hyperref}

\hypersetup{colorlinks=true,citecolor=blue,urlcolor=blue,linkcolor=blue}

\usepackage{physics}
\usepackage{algorithm}
\usepackage[noend]{algpseudocode}

\journalname{EPJ AP}
\articletype{Research article}

\begin{document}

\title{
Benchmarking analytical electron ptychography methods for the low-dose imaging of beam-sensitive materials
}

\author{
Hoelen L. Lalandec Robert\inst{1,2}$^{,}$\correspondingauthor{\email{\\hoelen.lalandecrobert@uantwerpen.be}} 
\and
Max Leo Leidl\inst{3}
\and
Knut Müller-Caspary\inst{3}
\and
Jo Verbeeck\inst{1,2}
}

\institute{
Electron Microscopy for Materials Science (EMAT), University of Antwerp, Groenenborgerlaan 171, 2020 Antwerp, Belgium
\and
NANOlight Center of Excellence, University of Antwerp, Groenenborgerlaan 171, 2020 Antwerp, Belgium
\and
Department of Chemistry and center for Nanoscience, Ludwig-Maximilians-Universität München, Butenandtstrasse 11, 81377 Munich, Germany
}

\abstract{
This publication presents an investigation of the performance of different analytical electron ptychography methods for low-dose imaging. In particular, benchmarking is performed for two model-objects, monolayer MoS$_2$ and apoferritin, by means of multislice simulations. Specific attention is given to cases where the individual diffraction patterns remain sparse. After a first rigorous introduction to the theoretical foundations of the methods, an implementation based on the scan-frequency partitioning of calculation steps is described, permitting a significant reduction of memory needs and high sampling flexibility. By analyzing the role of contrast transfer and illumination conditions, this work provides insights into the trade-off between resolution, signal-to-noise ratio and probe focus, as is necessary for the optimization of practical experiments. Furthermore, important differences between the different methods are demonstrated. Overall, the results obtained for the two model-objects demonstrate that analytical ptychography is an attractive option for the low-dose imaging of beam-sensitive materials.
}

\keywords{
Electron Ptychography, Beam Damage, Event-driven Detection, Low-dose Imaging
}

\maketitle

\section*{Introduction}
    
    Within recent years, scanning transmission electron microscopy (STEM) has evolved into an attractive tool for the investigation of beam-sensitive objects such as viruses \cite{Zhou2020,Lazic2022}, 2D materials \cite{Muller-Caspary2018,Wen2019,Chen2024c}, zeolites \cite{Liu2020b,Sha2023a,Zhang2023,Dong2023,Mitsuishi2023}, Li-rich oxides \cite{Lozano2018,Song2022,Song2024c}, polymers \cite{Hao2023}, perovskites \cite{Ma2023,Scheid2023,Schrenker2024} or metal-organic frameworks (MOF) \cite{Li2019b,Li2025}. When imaging such fragile materials, the damage following the transfer of energy from interacting electrons \cite{Egerton2019}, such as e.g. knock-on displacement of atoms \cite{King1987,Hobbs1994}, heating \cite{Bornes1944} or radiolysis \cite{Grubb1974}, imposes a critical electron dose \cite{Egerton2021a} beyond which the specimen structure is lost. This critical dose then constitutes the main experimental limitation, thus in practice determining the best achievable resolution \cite{Egerton2007,Rez2021}.
    
    More generally, the prevalence of beam damage requires both a re-evaluation of the maximum electron dose to be invested and an improvement in detector quantum efficiency (DQE). The latter was fulfilled by the introduction of direct electron detectors (DED) \cite{McMullan2007,Llopart2007,Ballabriga2011,Plackett2013,Poikela2014,Ryll2016,Tate2016,Philipp2022,Llopart2022,Zambon2023,Ercius2024} surpassing the capacities of conventional scintillator cameras \cite{Ruskin2013}, including in terms of their modulation transfer function (MTF) \cite{Milazzo2010,Mir2017,Paton2021}. The gain in recording speed, allowed by faster electronics, furthermore enabled the acquisition of a convergent beam electron diffraction (CBED) pattern at each scan position \cite{Muller2012a}, a technique often referred to as momentum-resolved STEM (MR-STEM) \cite{Muller-Caspary2018a} or 4D-STEM \cite{Yang2015a}. More recently, event-driven detection \cite{Fan1998}, based on the Timepix \cite{Llopart2007,Poikela2014,Frojdh2015,Llopart2022} technology, permitted the extension of this technique to sub-microsecond single-pattern acquisition times \cite{Jannis2022,Auad2023a,Kuttruff2024}, thus reaching the same speed as conventional STEM.
    
    The subsequent knowledge on the far-field intensity distribution furthermore enables the use of a class of computational imaging methods known as ptychography \cite{Hoppe1969,Hoppe1969a,Hoppe1969b} for the measurement of the projected electrostatic potential of the illuminated object, in the form of a phase shift map. Those methods consist in the correlative use of a series of coherent scattering experiments, in which a redundancy of imprinted specimen information permits the retrieval of a common illuminated object. They can be thus be considered as an extension of the well-established coherent diffractive imaging (CDI) technique \cite{Drenth1975,Fienup1978,Miao1998,Miao1999,Weierstall2002,Williams2006} to the situation where multiple independent recordings are employed and where, at least in the basic case, no prior information is available.
    
    Among ptychographic methods based on the focused-probe geometry, iterative phase retrieval \cite{Fienup1982} approaches have recently met some success e.g. with biological specimens \cite{Pelz2017,Zhou2020,Pei2023,Kucukoglu2024,Mao2024}. Those approaches consist in the probe position-dependent simulation of the elastic scattering of the incident electrons, thus leading to a repeated update of the multiplicative transmission function $T\left(\vec{r}\right)$ used to represent the specimen, given the error made against the experimental recordings and while cycling through the corresponding scan positions. This update is performed a number of times for each complete cycle, depending on the chosen batch size, and usually follows one of the several variants \cite{Elser2003,Faulkner2004,Rodenburg2004,Thibault2008,Thibault2009,Maiden2009,Maiden2017,Maiden2024} of the Gerchberg-Saxton algorithm \cite{Gerchberg1972,Gerchberg1974} for sequential projections or is given by the gradient of a specific loss function \cite{Guizar-Sicairos2008,Godard2012,Bian2016,Odstrcil2018}, i.e. the maximum likelihood approach. The process may also include a regularization term \cite{Thibault2012,Maiden2017,Pham2019a,Schloz2020,Lee2024,Herdegen2024} or be based on a parameterization strategy \cite{Diederichs2024,Yang2024b,Herdegen2024,Yang2025a}.
    
    Due to the wide range of parameters available, encompassing e.g. the coupled loss and regularization functions, the update strength, the batch size, the initial guess on the reconstructed object or the possible use of a supplementary momentum term \cite{Maiden2017}, iterative methods possess a high degree of flexibility. On the other hand, while a specific choice of parameter set may permit a degree of adaptation to particular cases, for instance with regards to the noise model \cite{Odstrcil2018,Chang2019b,Leidl2024,Seifert2024}, this also implies the need for a complex tuning step to achieve numerical convergence \cite{Melnyk2023}. Different results may otherwise be obtained through separate reconstruction processes or algorithms, hence creating reproducibility issues. Achieving convergence may furthermore prove more challenging in the low-dose case \cite{Godard2012,Katvotnik2013}, where the exploited far-field patterns are underdetermined, independently of the dose-efficiency demonstrated by the converged result in itself. Finally, iterative ptychography remains numerically intensive and often requires advanced computation capacities to avoid exceedingly long processing times \cite{Marchesini2016,Yu2022,Wang2022c,Mukherjee2022,Welborn2024}.
    
    For those reasons, there is still a high interest in pursuing work on analytical solutions \cite{Bates1989,Rodenburg1992,Rodenburg1993} which, since they lead to method-unique results through direct and well-understood imaging processes, arguably constitute more reproducible measurement approaches. In particular, as they are also fast, their application in a wide range of conditions or for large collections of specimens can be streamlined, hence making those methods especially useful for challenging experimental cases. In-line treatment options permitting live imaging \cite{Strauch2021,Bangun2023} have been reported as well, while this remains challenging in the framework of an iterative process \cite{Weber2024}. Analytical ptychography has moreover demonstrated a high dose-efficiency \cite{Pennycook2015,Yang2015b,OLeary2020,OLeary2021}, including with a sensitivity to light elements \cite{Yang2016,Yang2017,Wang2017a,Leidl2023}, and was successfully applied to a variety of beam-sensitive objects \cite{Lozano2018,Song2022,Dong2023,Mitsuishi2023,Hao2023,Scheid2023,Song2024c} in recent years.
    
    In this publication, the fundamental capacities of analytical ptychography methods to image a beam-sensitive specimen are explored in conditions of very low electron dose. In particular, interest is taken in the resolution achieved for different numerical apertures, in the dose-dependent precision of the measurements and in the fundamental frequency transfer capacity of different approaches. The methods investigated are the Wigner distribution deconvolution (WDD) \cite{Rodenburg1992,McCallum1992,Nellist1994,Li2014}, integrated center of mass (iCoM) \cite{Muller2014,Lazic2016} imaging as well as the sideband integration (SBI) \cite{Rodenburg1993,Pennycook2015,Yang2015b,Yang2016a} method, sometimes referred to as single sideband (SSB) \cite{OLeary2021}. Benchmarking is continued with an overfocused probe \cite{Hue2010,Song2019} and an adapted process permitting the direct correction of known aberrations. This specific recording geometry has recently attracted interests for the imaging of beam-sensitive objects and bears similarity with the original idea of ref. \cite{Vine2009a}. After an initial review of the theory in the fully coherent case, practical implementation is demonstrated through the newly introduced scan-frequency partitioning algorithm (SFPA), permitting a straightforward parallelization and offering high flexibility in the size and pixel resolution of the reconstruction window. All demonstrations made here are based on MR-STEM simulations, hence allowing direct control over the illumination parameters and the dose, while ensuring sparsity in the electron counts. Two model objects are employed: monolayer MoS$_2$ and ice-embedded apoferritin \cite{Leidl2023}.

\section{Theory of analytical ptychography and new highly parallelizable implementation}
    \label{sec:theory}
    
    \subsection{\textbf{Coherent and elastic interaction under the phase object approximation}}
        \label{subsec:POA}
        
        \subsubsection{Transmission function}
            
            In its conventional form \cite{Bates1989,Rodenburg1992}, analytical ptychography makes use of the phase object approximation (POA) \cite{Cowley1972}. In this context, the imaged material is considered thin enough so that no variation of wave amplitude occurs within it, thus making the scattering-induced phase shift additive along the propagation axis. For thicker objects, the applicability of the POA is limited due to the role of near-field propagation, leading to a finite depth of focus \cite{VanBenthem2005,VanBenthem2006} and dynamical diffraction effects such as channeling \cite{VanDyck1996}.
            
            Formally, the elastic interaction of the electron probe $P\left(\vec{r}_0\right)$ with the specimen is then modeled by a multiplication with a transmission function $T\left(\vec{r}_0\right)$, defined for each real-space position $\vec{r}_0$ in the specimen plane by
            \begin{equation}
                T\left(\vec{r}_0\right) \, = \, e^{ i \, \sigma \mu\left(\vec{r}_0\right) } \quad,
            \end{equation}
            where $\mu\left(\vec{r}_0\right)$ is the projected electrostatic potential of the specimen, i.e. the integral of the three-dimensional potential along the propagation axis. The interaction parameter $\sigma$, expressed in $\text{V}^{-1}\cdot\text{m}^{-1}$, is given by
            \begin{equation}
                \sigma \, = \, \frac{2\pi \, e}{h \, c} \frac{ \, mc^2 \, + \, eU \, }{\sqrt{ \, eU \, \left( \, 2mc^2 \, + \, eU \, \right) \, }} \quad,
                \label{eq:sigma}
            \end{equation}
            with $e$ is the elementary charge, $m$ the electron rest mass, $h$ the Planck constant and $c$ the speed of light. The product of $\sigma$ with $\mu\left(\vec{r}_0\right)$ thus represents the local phase shift imposed to the electron wavefunction by the specimen and is typically given in radians.
            
            Importantly, due to the dependence of $\sigma$ on $U$, the acceleration voltage affects this phase shift in a non-linear manner, independently of the specimen itself. An empirical absorption term may also be added to $\mu\left(\vec{r}_0\right)$, as an imaginary number, to improve the agreement with experimental results \cite{Humphreys1968,Cowley1972}, e.g. by accounting for amplitude variations in a computationally retrieved transmission function. Typically, this term is related to inelastic scattering \cite{Yoshioka1957,Wang1998d}, which otherwise leads to a diffuse component in the far-field \cite{Mkhoyan2008,Beyer2020,Barthel2020,Robert2022}, and specimen vibrations \cite{Hall1965,Hall1965a,VanDyck2009}. In this work, it is left out for simplicity, i.e. the interaction is assumed to not affect the coherence of the electron beam.
        
        \subsubsection{Convergent illumination}
            
            Continuing, given a fully coherent illumination, the electron probe $P\left(\vec{r}_0\right)$ is found equal to
            \begin{equation}
                P\left(\vec{r}_0\right) \, = \, \mathcal{F}^{-1}\left[ \, A\left(\vec{q}_0\right) \, e^{-i\chi\left(\vec{q}_0\right)} \, \right]\left(\vec{r}_0\right) \quad,
            \end{equation}
            with $\chi\left(\vec{q}_0\right)$ the geometrical aberration function and $A\left(\vec{q}_0\right)$ representing the aperture in the focal plane of the probe-forming lens, being equal to 1 for $\parallel\vec{q}_0\parallel\,<\,q_A$ and 0 otherwise. The quantity $q_A\,=\,\sin\left(\alpha\right)\,/\,\lambda$ introduced here is the reciprocal space cut-off imposed by the aperture, with $\alpha$ the semi-convergence angle and $\lambda$ the relativistically corrected wavelength \cite{Fujiwara1961}. Noteworthily, the aperture function actually used in the numerical implementation is further normalized, as being representative of the wavefunction in the focal plane.
            
            The notations $\mathcal{F}$ and $\mathcal{F}^{-1}$ respectively refer to a Fourier transform and an inverse Fourier transform, given by
            \begin{equation}
            \begin{split}
                \tilde{\varphi}\left(\vec{v}\right) \, = \, \mathcal{F}\left[\,\varphi\left(\vec{u}\right)\,\right]\left(\vec{v}\right) \, & = \, \sum\limits_{\vec{u}} \, e^{-i 2\pi \vec{v}\cdot\vec{u}} \, \varphi\left(\vec{u}\right) \\
                \varphi\left(\vec{u}\right) \, = \, \mathcal{F}^{-1}\left[\,\tilde{\varphi}\left(\vec{v}\right)\,\right]\left(\vec{u}\right) \, & = \, \sum\limits_{\vec{v}} \, e^{i 2\pi \vec{v}\cdot\vec{u}} \, \tilde{\varphi}\left(\vec{v}\right) \quad,
            \end{split}
            \end{equation}
            with Fourier normalization left implicit.
        
        \subsubsection{Scattered and measurable intensity}
            
            At a given scan position $\vec{r}_s$, a localized exit wave $\Psi_{\vec{r}_s}\left(\vec{r}_0\right)$ is formed which is given by
            \begin{equation}
                \Psi_{\vec{r}_s}\left(\vec{r}_0\right) \, = \, P\left( \vec{r}_0 - \vec{r}_s \right) \, T\left(\vec{r}_0\right) \quad,
            \end{equation}
            hence the diffracted intensity $I_{\vec{r}_s}\left(\vec{q}\right)$ accessible in the far-field, with $\vec{q}$ a scattering vector, is
            \begin{equation}
                I_{\vec{r}_s}\left(\vec{q}\right) \, = \, \mid \, \mathcal{F}\left[ \, \Psi_{\vec{r}_s}\left(\vec{r}_0\right) \, \right]\left(\vec{q}\right) \, \mid^2 \quad.
            \end{equation}
            As such, this quantity is interpretable as a probability distribution among the locations, at the concerned optical plane $\vec{q}$, for the collapse of the electron wave.
            
            Finally, the intensity $I^{det}_{\vec{r}_s}\left(\vec{q}_d\right)$ measured by the camera, across detector space $\vec{q}_d$, includes a possible point spread effect represented by the MTF $M\left(\vec{r}_d\right)$, $\vec{r}_d$ being the reciprocal dimension of $\vec{q}_d$. This leads to
            \begin{equation}
            \begin{split}
                I^{det}_{\vec{r}_s}\left(\vec{q}_d\right) \, & = \, \mathcal{F}\left[ \, M\left(\vec{r}_d\right) \,\mathcal{F}^{-1}\left[ \, I_{\vec{r}_s}\left(\vec{q}\right) \, \right]\left(\vec{r}_d\right) \, \right]\left(\vec{q}_d\right) \\
                & = \, \tilde{M}\left(\vec{q}_d\right) \, \otimes_{\vec{q}_d} \, I_{\vec{r}_s}\left(\vec{q}_d\right) \quad.
            \end{split}
            \end{equation}
            Given that $\tilde{M}\left(\vec{q}_d\right)\,=\,\mathcal{F}\left[\,M\left(\vec{r}_d\right)\,\right]\left(\vec{q}_d\right)$ is a real quantity, if the MTF-induced information spread effect is isotropic, then $\tilde{M}\left(\vec{q}_d\right)$ and $M\left(\vec{r}_d\right)$ are both real and point-symmetric. This assumption is implicit in the rest of this work.
    
    \subsection{\textbf{Wigner distribution formalism}}
        
        \subsubsection{Scattering data reformulation}
            
            Given the prior recording of $I^{det}_{\vec{r}_s}\left(\vec{q}_d\right)$ through an MR-STEM experiment, a Fourier transform with respect to the scan coordinates $\vec{r}_s$ towards an arbitrarily sampled spatial frequency space $\vec{Q}$ leads to a complex distribution $\tilde{J}_{\vec{Q}}\left(\vec{q}_d\right)$. As long as real-space is sampled finely enough by the scan points $\vec{r}_s$, i.e. under the condition of sufficient overlap ratio $\beta_{\delta \vec{r}_s}$ between successively illuminated areas \cite{Bunk2008}, $\tilde{J}_{\vec{Q}}\left(\vec{q}_d\right)$ can then be interpreted as a map of the specimen-dependent $\vec{Q}$-responses attributed to the camera pixels $\vec{q}_d$. In particular, each scattering vector $\vec{q}$ in the far-field is assimilated to a single conventional TEM image by arguments of reciprocity \cite{Cowley1969a,Krause2017}. More details on the redundancy condition and area overlap are provided in appendix A, including with a mathematical criterion.
            
            The distribution $\tilde{J}_{\vec{Q}}\left(\vec{q}_d\right)$ is found equal to
            \begin{equation}
            \begin{split}
                \tilde{J}_{\vec{Q}}\left(\vec{q}_d\right) \, & \, = \, \tilde{M}\left(\vec{q}_d\right) \, \otimes_{\vec{q}_d} \, \left( \tilde{P}\left(\vec{q}_d\right) \tilde{P}^*\left(\vec{q}_d+\vec{Q}\right) \right) \\
                & \,\,\,\,\,\,\,\,\,\,\,\,\,\,\,\,\,\,\,\,\,\, \otimes_{\vec{q}_d} \, \left( \tilde{T}\left(\vec{q}_d\right) \tilde{T}^*\left(\vec{q}_d-\vec{Q}\right) \right) \\
                & = \, \mathcal{F}\left[ \, M\left(\vec{r}_d\right) \, \Gamma\left(\vec{Q}\,;\,\vec{r}_d\right) \, \Upsilon\left(\vec{Q}\,;\,\vec{r}_d\right) \, \right]\left(\vec{q_d}\right) \quad,
                \label{eq:FTalongscan}
            \end{split}
            \end{equation}
            where $\Gamma\left(\vec{Q}\,;\,\vec{r}_d\right)$ and $\Upsilon\left(\vec{Q}\,;\,\vec{r}_d\right)$ are Wigner distributions \cite{Wigner1932}, i.e. autocorrelations of the probe and of the transmission function.
            
            Formally, they are given by
            \begin{equation}
            \begin{split}
                \Upsilon\left(\vec{Q}\,;\,\vec{r}_d\right) \, & = \, \mathcal{F}\left[ \, T\left(\vec{r}_0+\vec{r}_d\right) \, T^*\left(\vec{r}_0\right) \, \right]\left(\vec{Q}\right) \\
                & = \, \mathcal{F}^{-1}\left[ \, \tilde{T}\left(\vec{q}_0\right) \, \tilde{T}^*\left(\vec{q}_0-\vec{Q}\right) \, \right]\left(\vec{r}_d\right) \quad,
            \end{split}
            \end{equation}
            and
            \begin{equation}
                \Gamma\left(\vec{Q}\,;\,\vec{r}_d\right) \, = \, \mathcal{F}^{-1}\left[ \, A\left(\vec{q}_0\right) \, A\left(\vec{q}_0+\vec{Q}\right) \, \theta\left(\vec{q}_0\,;\,\vec{q}_0+\vec{Q}\right) \, \right]\left(\vec{r}_d\right) \quad.
                \label{eq:wignerdistrib}
            \end{equation}
            The function $\theta\left({\vec{q}_0}\,;\,\vec{q}_0+\vec{Q}\right)$ encompasses the imperfections in the illumination and is equal to 
            \begin{equation}
                \theta\left({\vec{q}_0}\,'\,;\,\vec{q}_0\right) = e^{-i\left( \, \chi\left({\vec{q}_0}\,'\right) \, - \, \chi\left(\vec{q}_0\right) \, \right)} \quad.
                \label{eq:thetatheta}
            \end{equation}
            As such, the insertion of this term in a ptychographic processing allows correcting for geometrical aberrations.
            
            A subsequent inverse Fourier transform from the camera dimensions $\vec{q}_d$ to an arbitrary set of reciprocal real-space coordinates $\vec{R}$ leads to $J_{\vec{Q}}\left(\vec{R}\right)$, a new complex four-dimensional distribution, equal to the product of the Wigner distributions with the MTF, i.e.
            \begin{equation}
                J_{\vec{Q}}\left(\vec{R}\right) \, = \, M\left(\vec{R}\right) \, \Gamma\left(\vec{Q}\,;\,\vec{R}\right) \, \Upsilon\left(\vec{Q}\,;\,\vec{R}\right) \quad.
                \label{eq:truthofWigner}
            \end{equation}
        
        \subsubsection{Wigner distribution deconvolution and direct extraction of the transmission function}
            
            The WDD method for analytical ptychography \cite{Rodenburg1992,McCallum1992,Nellist1994,Li2014} thus first consists in the recovery of $\Upsilon\left(\vec{Q}\,;\,\vec{R}\right)$ through
            \begin{equation}
                \Upsilon\left(\vec{Q}\,;\,\vec{R}\right) \, \equiv \, \frac{ M\left(\vec{R}\right) \Gamma^*\left(\vec{Q}\,;\,\vec{R}\right) \, J_{\vec{Q}}\left(\vec{R}\right) }{ \epsilon \, + \, \mid M\left(\vec{R}\right) \Gamma\left(\vec{Q}\,;\,\vec{R}\right) \mid^2 } \quad,
                \label{eq:oldstyleWDD}
            \end{equation}
            where $\epsilon$ is a small number introduced to avoid divisions by zero, i.e. the actual deconvolution is done via a Wiener filter process \cite{Bates1986}. Noteworthily, a careful choice of the Wiener parameter $\epsilon$ is also important to avoid an amplification of the noise propagated from detector space \cite{OLeary2021}.
            
            As a second step in the processing, the summation of $\Upsilon\left(\vec{Q}\,;\,\vec{R}\right)$ along $\vec{R}$ permits the recovery of the transmission function by
            \begin{equation}
            \begin{split}
                f\left(\vec{Q}\right) \, & = \, \sum\limits_{\vec{R}} \, \Upsilon\left(\vec{Q}\,;\,\vec{R}\right) \, = \, \tilde{T}\left(\vec{0}\right) \tilde{T}^*\left(-\vec{Q}\right) \\
                T^{WDD}\left(\vec{r}\right) \, & = \, \left( \, \mathcal{F}^{-1}\left[ \, \frac{f\left(\vec{Q}\right)}{\sqrt{ f\left(\vec{0}\right) }} \, \right]\left(\vec{r}\right) \, \right)^* \quad,
                \label{eq:transmfuncextract}
            \end{split}
            \end{equation}
            where $f\left(\vec{Q}\right)$ is introduced as an intermediary result. With the deconvolution done, $\sigma \mu^{WDD}\left(\vec{r}\right)$, i.e. the measurement of the phase shift map $\sigma \mu\left(\vec{r}\right)$ by WDD ptychography, can be obtained from $T^{WDD}\left(\vec{r}\right)$ by extracting its angle. In most practical cases, including those presented here, the values obtained remain small enough to avoid phase discontinuities, thus making an unwrapping process unnecessary.
        
        \subsubsection{Zero-frequency component}
            
            As illustrated by equations \ref{eq:transmfuncextract}, the measurement is performed such that $\text{arg}\left[f\left(\vec{0}\right)\right]\,=\,0$. Consequently, $f\left(\vec{0}\right)$ is a real number, which implies that the mean of the ptychographic phase in the reconstruction window, i.e. its DC component, remains inaccessible. This is consistent with the fact that phase, as a mathematical abstraction rather than a significant physical quantity, remains unmeasurable unless compared to a reference, e.g. by wave interference like in the case of off-axis electron holography \cite{Mollenstedt1955,Tonomura1978}.
            
            In particular, in the coherent and elastic interaction regime, only relative local phase shifts created within the illuminated patch, thus requiring gradients in the specimen-induced phase shift map, can lead to measurable changes in the momentum distribution \cite{Winkler2020}. As a side-note, the normalization of $T^{WDD}\left(\vec{r}\right)$ by the real constant $\sqrt{f\left(\vec{0}\right)}$ only modifies its amplitude, and not its angle. It thus does not affect the measurement of the projected potential itself.
        
        \subsubsection{Resolution limit}
            
            Continuing, since $\Gamma\left(\vec{Q}\,;\,\vec{R}\right)\,=\,0$ for $\parallel\vec{Q}\parallel \, \geq \, 2 q_A$, owing to formula \ref{eq:wignerdistrib}, it appears at first sight as though the best resolution achievable by the WDD approach is equal to half the conventional Abbe criterion $\delta r_{\text{Abbe}} = 0.5 / q_A$. In reality, super-resolution, i.e. the transfer of frequencies exceeding the $2 q_A$ diffraction limit \cite{Sayre1952,Gerchberg1974,Rodenburg1992,Nellist1995,Maiden2011,Humphry2012}, is still possible based on the so-called stepping out method \cite{Rodenburg1992,Li2014}.
            
            This is nevertheless done at a high cost in dose \cite{Jiang2018}, as it is specifically dependent on the availability of dark field electrons. Since this publication focuses particularly on the dose-efficiency of analytical ptychography, this aspect is left for future work. The WDD method can otherwise make use of the dark field electrons outside of the stepping out paradigm, as apparent in the equations. Moreover, and as long as the interaction can still be faithfully described using the POA, geometrical aberrations are corrected through the introduction of the term $\theta\left(\vec{q}_0\,;\,\vec{q}_0+\vec{Q}\right)$, defined in equation \ref{eq:thetatheta}.
    
    \subsection{\textbf{Sideband formalism for a weak scatterer}}
        \label{subsec:SBISBI}
        
        \begin{figure}
            \centering
            \includegraphics[width=1.0\columnwidth]{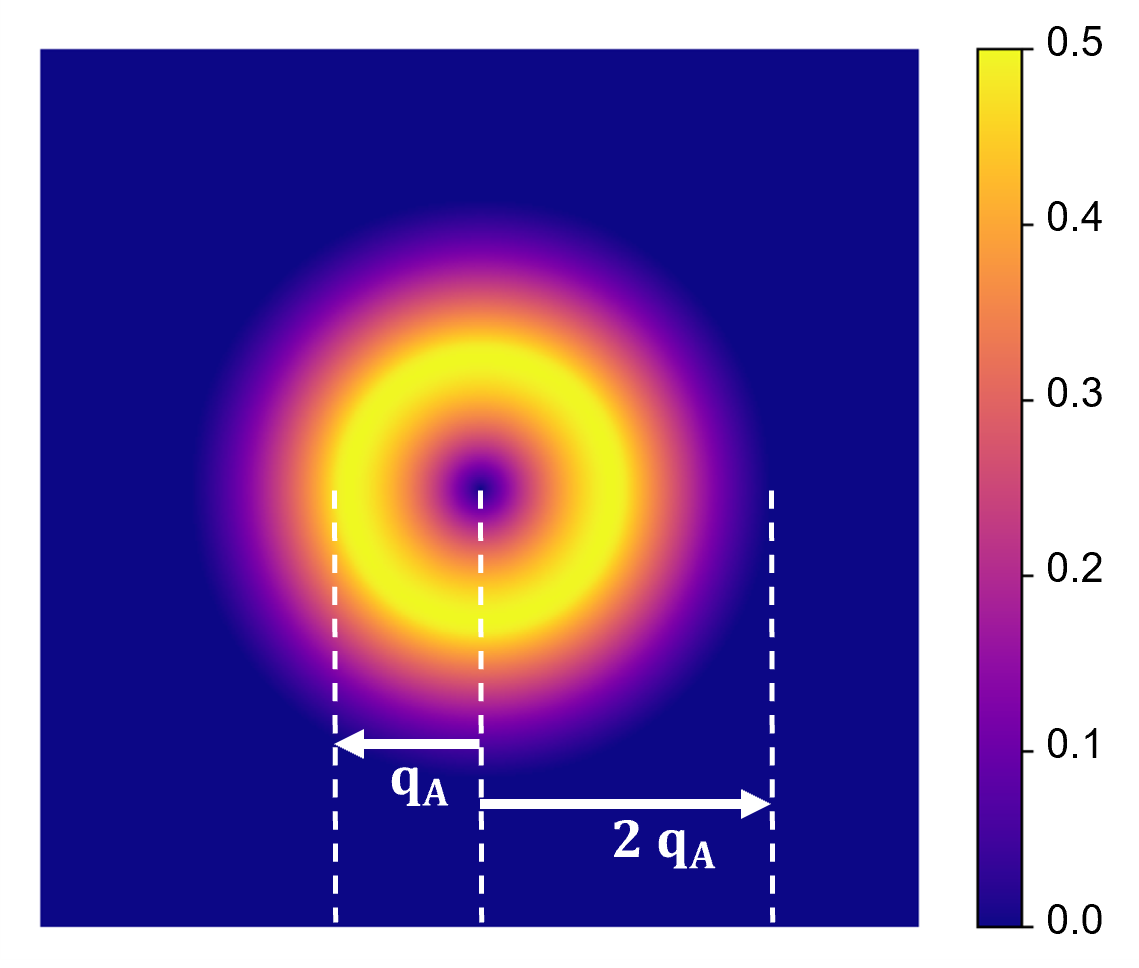}
            \caption{Depiction of the PCTF $\tilde{\zeta}\left(\vec{Q}\right)$.}
            \label{fig:CTF_SBI_S}
        \end{figure}
        
        \subsubsection{Simplification of Wigner distribution formalism via a first order Taylor expansion}
            
            The special case where the specimen can be considered weakly scattering, in addition to fulfilling the POA, occurs when the range of phase shift covered by the measurable $\sigma \mu\left(\vec{r}_0\right)$, i.e. accounting for the reduction due to resolution limit, is significantly smaller than 1. Equivalently to the well-known small-angle approximation, the transmission function may be then replaced by the following first order Taylor expansion
            \begin{equation}
                T\left(\vec{r}_0\right) \, \approx \, 1 \, + \, i \sigma \mu\left(\vec{r}_0\right) \quad.
            \end{equation}
            This condition is usually referred to as the weak phase object approximation (WPOA) \cite{Cowley1972}.
            
            It then follows that
            \begin{equation}
            \begin{split}
                T\left(\vec{r}_0+\vec{r}_d\right) T^*\left(\vec{r}_0\right) \, \approx \, 1 \, & + \, i \sigma \mu\left(\vec{r}_0+\vec{r}_d\right) \\
                & - \, i \sigma \mu\left(\vec{r}_0\right) \\
                & + \, \sigma^2 \mu\left(\vec{r}_0+\vec{r}_d\right)\mu\left(\vec{r}_0\right) \quad.
            \end{split}
            \end{equation}
            As the condition of a weakly scattering object also implies that $\sigma^2 \mu\left(\vec{r}_0+\vec{r}_d\right)\mu\left(\vec{r}_0\right) \, \ll \, 1$, equation \ref{eq:FTalongscan} leads to
            \begin{equation}
                \tilde{J}_{\vec{Q}}\left(\vec{q}_d\right) \, \approx \, B\left(\vec{q}_d\right) \, \delta\left(\vec{Q}\right) \, + \, i \, \sigma \, \omega\left(\vec{Q}\,;\,\vec{q}_d\right) \, \tilde{\mu}\left(\vec{Q}\right) \quad.
                \label{eq:FTalongscanWPOA}
            \end{equation}
            Notably, this is also justified by the realness of $\mu\left(\vec{r}_0\right)$, which implies that $\tilde{\mu}^*\left(-\vec{Q}\right)\,=\,\tilde{\mu}\left(\vec{Q}\right)$.
            
            The functions $B\left(\vec{q}_d\right)$ and $\omega\left(\vec{Q}\,;\,\vec{q}_d\right)$ are given by
            \begin{equation}
            \begin{split}
                B\left(\vec{q}_d\right) \, & = \, \mathcal{F}\left[ \, M\left(\vec{r}_d\right) \, \mathcal{F}^{-1}\left[ \, A^2\left(\vec{q}_0\right) \, \right]\left(\vec{r}_d\right) \, \right]\left(\vec{q}_d\right) \\
                \omega\left(\vec{Q}\,;\,\vec{q}_d\right) \, & = \, \mathcal{F}\left[ \, M\left(\vec{r}_d\right) \, \Gamma\left(\vec{Q}\,;\,\vec{r}_d\right) \, \left( e^{i2\pi\vec{Q}\cdot\vec{r}_d} - 1 \right) \, \right]\left(\vec{q}_d\right) \quad.
                \label{eq:omegainSBI}
            \end{split}
            \end{equation}
            $\tilde{J}_{\vec{Q}}\left(\vec{q}_d\right)$ thus consists of a zero-frequency term, associated to the unscattered portion of the electron beam, and a two-sideband term resulting from the Fourier transform of the distribution $\Gamma\left(\vec{Q}\,;\,\vec{r}_d\right)$.
        
        \subsubsection{Deconvolutive extraction of the potential}
            
            Equation \ref{eq:FTalongscanWPOA} serves as a basis for the sideband method of analytical ptychography \cite{Rodenburg1993,Pennycook2015,Yang2015b,Yang2016a}, in this work referred to as SBI. It can thus be understood as a special case of the Wigner distribution approach, applicable when the object fulfills the WPOA. Here, a deconvolutive form is used, similarly to e.g. ref. \cite{Yang2016a}. For clarity, it will be referred to as SBI-D in the rest of this text, while the conventional summative form \cite{Pennycook2015,Yang2015b} will be referred to as SBI-S.
            
            The SBI-D process thus consists in performing
            \begin{equation}
            \begin{split}
                g\left(\vec{Q}\right) \, & = \, \frac{1}{i \, \Omega} \, \sum\limits_{\vec{q}_d} \, \frac{ \omega^*\left(\vec{Q}\,;\,\vec{q}_d\right) \, \tilde{J}_{\vec{Q}}\left(\vec{q}_d\right) }{ \epsilon \, + \, \mid\omega\left(\vec{Q}\,;\,\vec{q}_d\right)\mid^2 } \\
                \mu^{SBI}\left(\vec{r}\right) \, & = \, \frac{1}{\sigma} \, \mathcal{F}^{-1}\left[ \, g\left(\vec{Q}\neq\vec{0}\right) \, \right]\left(\vec{r}\right) \quad.
                \label{eq:SBIdeconv}
            \end{split}
            \end{equation}
            $g\left(\vec{Q}\right)$ is an intermediary result and $\Omega \, = \, \sum\limits_{\vec{q}_d} \, A^2\left(\vec{q}_d\right)$ is introduced for normalization, i.e. the projected potential is obtained by calculating a mean among the scattering coordinates $\parallel\vec{q}_d\parallel\,<\,q_A$, post-division by $\omega\left(\vec{Q}\,;\,\vec{q}_d\right)$.
            
            Importantly, like for the WDD case, the DC component is not recoverable, since $\omega\left(\vec{0}\,;\,\vec{q}_d\right)\,=\,0$, as is shown by formula \ref{eq:omegainSBI}. The inclusion of $\theta\left(\vec{q}_0\,;\,\vec{q}_0+\vec{Q}\right)$ also permits the correction of aberrations, at least as long as the WPOA is fulfilled.
        
        \subsubsection{Summative extraction of the potential}
            
            If the influence of the MTF is neglected, i.e. $M\left(\vec{r}_d\right)\,=\,1$, then equation \ref{eq:FTalongscanWPOA} becomes
            \begin{equation}
            \begin{split}
                \tilde{J}_{\vec{Q}}\left(\vec{q}_d\right) \, \approx \, & A^2\left(\vec{q}_d\right) \, \delta\left(\vec{Q}\right) \\
                & + \, i \, \sigma \, A\left(\vec{q}_d-\vec{Q}\right) A\left(\vec{q}_d\right) \theta\left(\vec{q}_d-\vec{Q}\,;\,\vec{Q}\right) \, \tilde{\mu}\left(\vec{Q}\right) \\
                & - \, i \, \sigma \, A\left(\vec{q}_d\right) A\left(\vec{q}_d+\vec{Q}\right) \theta\left(\vec{q}_d\,;\,\vec{q}_d+\vec{Q}\right) \, \tilde{\mu}\left(\vec{Q}\right) \quad.
                \label{eq:sidebandsideband}
            \end{split}
            \end{equation}
            Hence, $\tilde{J}_{\vec{Q}}\left(\vec{q}_d\right)$ can be described as a superposition of two sidebands terms with a zero-frequency component.
            
            In practice, this sideband-like geometry means that, upon visualizing the values across the $\vec{q}_d$-dimensions, for a given specimen frequency $\vec{Q}$ and as long as $\chi\left(\vec{q}_0\right)\,=\,0$, the double overlap area will be homogeneously equal to $\pm\sigma\tilde{\mu}\left(\vec{Q}\right)$. This constitutes the basis of the conventional SSB workflow \cite{Pennycook2015,Yang2015b} and provides an opportunity for straightforward aberration correction \cite{Yang2016,Li2025a}.
            
            In this context, the SBI-S process consists in performing a summation within the double overlaps, while excluding triple overlaps where the terms cancel out. Formally, it consists in
            \begin{equation}
            \begin{split}
                & g'\left(\vec{Q}\right) \, = \, \frac{1}{i} \,\sum\limits_{\vec{q}_d} \, \left( \beta^{+}\left(\vec{Q}\,;\,\vec{q}_d\right) - \beta^{-}\left(\vec{Q}\,;\,\vec{q}_d\right) \right) \tilde{J}_{\vec{Q}}\left(\vec{q}_d\right) \\
                & \zeta\left(\vec{r}\right) \, \otimes_{\vec{r}} \, \mu^{SBI}\left(\vec{r}\right) \, = \, \frac{1}{\sigma} \, \mathcal{F}^{-1}\left[ \, g'\left(\vec{Q}\right) \, \right]\left(\vec{r}\right) \quad,
                \label{eq:SBIsum}
            \end{split}
            \end{equation}
            where the $\beta^{\pm}\left(\vec{Q}\,;\,\vec{q}_d\right)$ terms are given by
            \begin{equation}
            \begin{split}
                \beta^{+}\left(\vec{Q}\,;\,\vec{q}_d\right) \, & = \, \frac{A\left(\vec{q}_d\right) \, A\left(\vec{q}_d-\vec{Q}\right) \, \left( 1 - A\left(\vec{q}_d+\vec{Q}\right) \right)}{\theta\left(\vec{q}_d-\vec{Q}\,;\,\vec{Q}\right)} \\
                \beta^{-}\left(\vec{Q}\,;\,\vec{q}_d\right) \, & = \, \frac{A\left(\vec{q}_d\right) \, A\left(\vec{q}_d+\vec{Q}\right) \, \left( 1 - A\left(\vec{q}_d-\vec{Q}\right) \right)}{\theta\left(\vec{q}_d\,;\,\vec{q}_d+\vec{Q}\right)} \quad.
                \label{eq:betainSBI}
            \end{split}
            \end{equation}
            Each thus aims to access one of the double overlap areas within the $\tilde{J}_{\vec{Q}}\left(\vec{q}_d\right)$ distribution, with a phase shift term inserted to compensate aberrations.
        
        \subsubsection{Phase contrast transfer function}
            
            The contrast transfer function (CTF) $\tilde{\zeta}\left(\vec{Q}\right)$, depicted in figure \ref{fig:CTF_SBI_S}, is introduced due to the summation over $\beta^{\pm}\left(\vec{Q}\,;\,\vec{q}_d\right)$ and is given by
            \begin{equation}
                \tilde{\zeta}\left(\vec{Q}\right) \, = \, \sum\limits_{\vec{q}_0} \, A\left(\vec{q}_0\right) A\left(\vec{q}_0-\vec{Q}\right) \, \left( 1 - A\left(\vec{q}_0+\vec{Q}\right) \right) \quad.
            \end{equation}
            As such, it is equal to the surface of the double overlap region, across camera space $\vec{q}_d$, corresponding to each spatial frequency $\vec{Q}$. It is interesting to note that, while this CTF is peaked at intermediary frequencies, i.e. close to $q_A$, it decays for both higher and lower frequencies.
            
            Here, it should be furthermore highlighted that, while the CTF of SBI-S is explicit, SBI-D possesses the same fundamental characteristics with regards to frequency transfer, as shown by the dependence of $\omega\left(\vec{Q}\,;\,\vec{q}_d\right)$ on $\vec{Q}$. In essence, it is not the choice between the summative or the deconvolutive forms that leads to the frequency weighting, but rather the assumption of a weakly scattering object in itself.
            
            In particular, if equation \ref{eq:sidebandsideband} is fulfilled, and if no geometrical aberrations are present, the parts of the distribution $\tilde{J}_{\vec{Q}}\left(\vec{q}_d\right)$ found outside the double overlap areas are expected to carry no useful information on the specimen, and thus to contain only noise. $\tilde{\zeta}\left(\vec{Q}\right)$ then reflects the information content in the scattering data itself, being equal to the proportion of available scattering vectors $\vec{q}$ that are useful to recover a specific frequency component $\vec{Q}$ of the specimen, i.e. it constitutes the phase contrast transfer function (PCTF) of the experiment in the sense of e.g. ref. \cite{Yang2015b}. When the WPOA is fulfilled, it is thus expected to intrinsically apply to all STEM-based phase retrieval methods, irrespective of whether their formulation assumes a weakly scattering object in the first place.
            
            As such, the SBI method, which consists in a treatment based explicitly on equation \ref{eq:FTalongscanWPOA}, permits to exclude all pixels outside double overlaps, thus in principle minimizing the total noise in the real-space result. As explained in details in ref. \cite{Seki2018}, the SBI-based treatment of counts in the detector space $\vec{q}_d$, which follow Poisson statistics \cite{Luczka1991}, then leads to a predictable noise level added to the frequency spectrum of the recovered object, given by the square root of the PCTF.
            
            In this context, the option to deconvolve the reconstructed phase shift $\sigma\mu\left(\vec{r}\right)$ post-process with $\mathcal{F}^{-1}\left[\sqrt{\tilde{\zeta}\left(\vec{Q}\right)}\right]\left(\vec{r}\right)$ has been proposed as an effective noise normalization strategy \cite{Seki2018,OLeary2021}, i.e. rendering the noise level homogeneous across spatial frequencies. Deconvolving by the complete $\zeta\left(\vec{r}\right)$, which in the conventional SSB workflow \cite{Pennycook2015,Yang2015b} is equivalent to averaging $\tilde{J}_{\vec{Q}}\left(\vec{q}_d\right)$ within the double overlap areas instead of performing a summation, may otherwise permit to homogenize frequency transfer. This is nevertheless only practical when the dose is high enough, as the amplitude of the frequency components to be amplified may then be below the noise level.
        
        \subsubsection{Applicability of the weak scatterer approximation}
            
            It should be understood that the specific situation where $\sigma\mu\left(\vec{r}\right)$ possesses the low value range of a weak phase object only occurs in a handful of cases. This may not only be due to excessive atomic potentials or material thicknesses, but also because lower acceleration voltages $U$ imply non-linearly higher values for the interaction parameter $\sigma$, as shown by equation \ref{eq:sigma}. In the case where the illuminated object is not a weak scatterer, as noted e.g. in ref. \cite{OLeary2021,Clark2023}, the SBI process still imposes a frequency-wise attenuation following $\tilde{\zeta}\left(\vec{Q}\right)$, due to the forms of $\omega\left(\vec{Q}\,;\,\vec{q}_d\right)$ and $\beta^{\pm}\left(\vec{Q}\,;\,\vec{q}_d\right)$. The CTF is then method-induced rather than reflective of the PCTF of the experiment itself.
            
            In this situation, the underlying sideband-like geometry in the distribution $\tilde{J}_{\vec{Q}}\left(\vec{q}_d\right)$ also cannot be expected to occur strictly, i.e. the values taken by $\vec{q}_d$-coordinates within the $\vec{Q}$-wise double overlap areas may not be homogeneously equal to $\pm\sigma\tilde{\mu}\left(\vec{Q}\right)$ anymore and exploitable information may be present outside as well. On that second aspect, it is noteworthy that, under the more general POA, non-zero coordinates of $\tilde{J}_{\vec{Q}}\left(\vec{q}_d\right)$ include both the triple overlap areas and the dark field. In contrast, the scattering of electrons outside the primary beam is not possible in the framework of the WPOA, as directly noticeable in equations \ref{eq:FTalongscanWPOA} and \ref{eq:sidebandsideband}. Whereas this is not the case for the iCoM and WDD methods, the SBI-S and SBI-D processes are thus unable to exploit dark field electrons.
            
            As a side-note, the CTF $\tilde{\zeta}\left(\vec{Q}\right)$ leads to artificial image features, e.g. negative halos around atomic sites \cite{Hofer2023}. If the specimen is not weakly scattering, such artificial features are not expected to occur through methods based only on the POA, hence demonstrating the possible violation of the WPOA upon comparison of results.
        
        \subsubsection{Interests of the deconvolutive approach}
            
            Finally, when comparing the two forms of sideband ptychography, SBI-D presents a few advantages compared to the already established SBI-S approach. First, the MTF $M\left(\vec{r}_d\right)$ can be explicitly included. Second, SBI-D permits the use of an arbitrarily shaped aperture \cite{Yang2016a} where the selection of specific overlapping regions would be less obvious, and thus the straightforward application to specific phase plate designs \cite{Verbeeck2018,VegaIbanez2023,Yu2023}.
    
    \subsection{\textbf{Center of mass imaging}}
        \label{subsec:iCoMiCoM}
        
        \begin{figure}
            \centering
            \includegraphics[width=1.0\columnwidth]{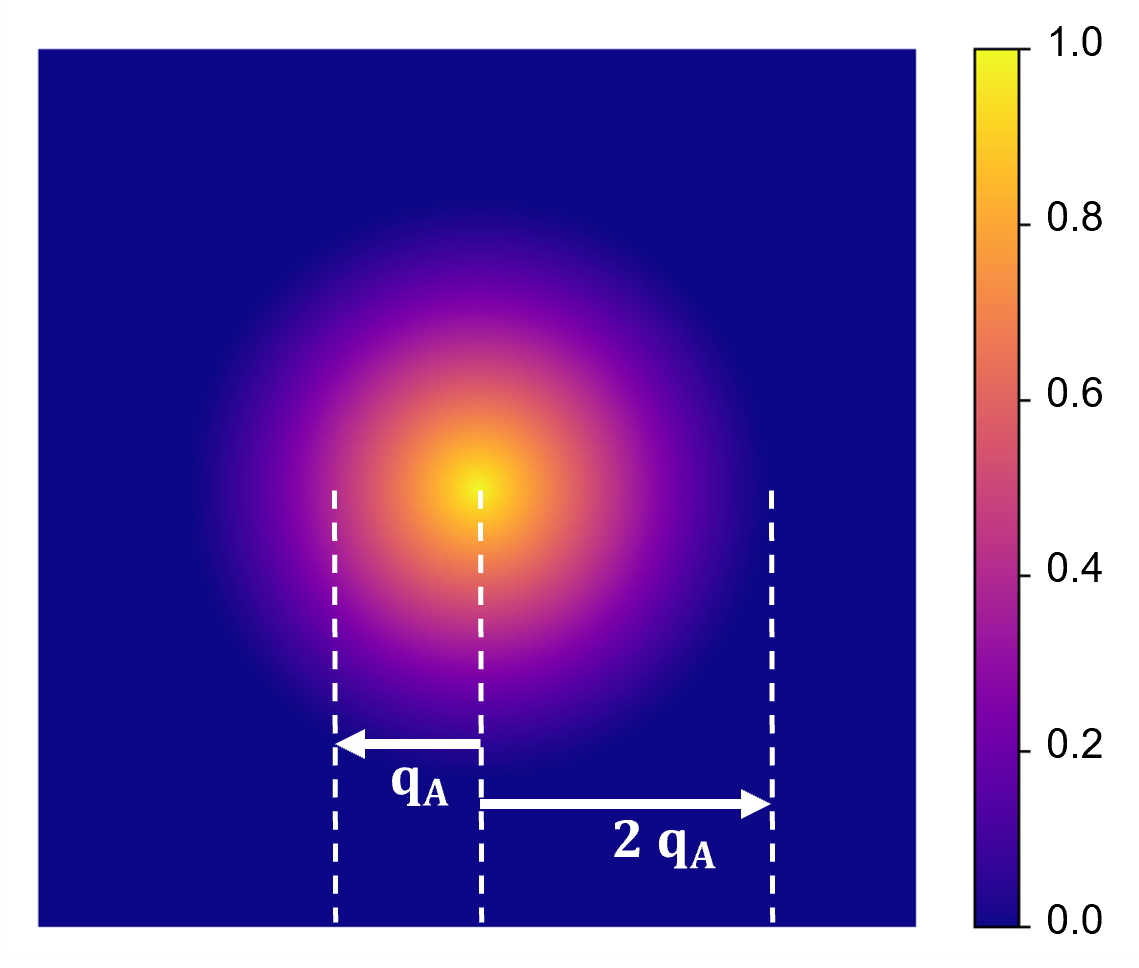}
            \caption{Depiction of the OTF $\tilde{\gamma}\left(\vec{Q}\right)$.}
            \label{fig:CTF_iCoM}
        \end{figure}
        
        \subsubsection{Scan position-wise average momentum transfer}
            
            The center of mass (CoM) $\langle\,\vec{q}\,\,\rangle_{\vec{r}_s}$ of the scan position-dependent CBED patterns constitutes a measurement of the average momentum transfer between the scattered electrons and the specimen. Under the POA, the CoM is linearly related to the local gradient of the projected potential \cite{Muller2014}. Formally, this means that
            \begin{equation}
            \begin{split}
                \langle\,\vec{q}\,\,\rangle_{\vec{r}_s} \, & = \, \sum\limits_{\vec{q}_d} \, \vec{q}_d \, I^{det}_{\vec{r}_s}\left(\vec{q}_d\right) \\
                & = \, \gamma\left(\vec{r}_s\right) \, \otimes_{\vec{r}_s} \, \left( \frac{\sigma}{2\pi} \, \widehat{\vec{\nabla}}_{\vec{r}_s} \, \mu\left(\vec{r}_s\right) \right) \quad,
                \label{eq:calculCoM}
            \end{split}
            \end{equation}
            with $\tilde{\gamma}\left(\vec{Q}\right)$ a CTF given by
            \begin{equation}
                \tilde{\gamma}\left(\vec{Q}\right) \, = \, \sum\limits_{\vec{q}_0} \, A\left(\vec{q}_0\right) A\left(\vec{q}_0-\vec{Q}\right) \quad.
            \end{equation}
            $\tilde{\gamma}\left(\vec{Q}\right)$ is depicted in fig. \ref{fig:CTF_iCoM}. This CTF is peaked at low frequencies and smoothly decays as a function of $\vec{Q}$, thus indicating difficulties in transferring higher frequencies.
            
            In the absence of a fast DED to perform an MR-STEM experiment, the average momentum transfer is conventionally approximated by using quadrants of a segmented detector \cite{Shibata2010,Lohr2012}, a technique usually referred to as differential phase contrast (DPC) in relation to historical references \cite{Dekkers1974,Rose1977}. In that context, CoM imaging can be understood as a more accurate approach to measuring the DPC signal \cite{Muller-Caspary2018a}, in particular considering that the use of segmented detectors leads to a non-isotropic CTF \cite{Seki2017}. As a side-note, another existing detector paradigm consists in a position-sensitive non-pixelated design \cite{Schwarzhuber2018}.
        
        \subsubsection{Fourier-based extraction of the potential}
            
            Following the measurement of $\langle\,\vec{q}\,\,\rangle_{\vec{r}_s}$, an extraction of the projected potential can be performed through a simple Fourier integration scheme, as conventionally used e.g. for the integrated DPC (iDPC) \cite{Lazic2016,Yucelen2018} counterpart to iCoM. This consists in
            \begin{equation}
            \begin{split}
                h\left(\vec{Q}\right) \, & = \, \frac{ \vec{Q} \, \cdot \, \mathcal{F}\left[ \, \langle\,\vec{q}\,\,\, \rangle\left(\vec{r}_s\right) \right]\left(\vec{Q}\right) }{ i \, \left( \, \epsilon \, + \, \parallel\vec{Q}\parallel^2 \, \right) } \\
                \gamma\left(\vec{r}\right) \, \otimes_{\vec{r}} \, \mu^{iCoM}\left(\vec{r}\right) \, & = \, \frac{1}{\sigma} \, \mathcal{F}^{-1}\left[ \, h\left(\vec{Q}\right) \, \right]\left(\vec{r}\right) \quad,
                \label{eq:integrCoM}
            \end{split}
            \end{equation}
            with $h\left(\vec{Q}\right)$ an intermediary result. Importantly, just like for the WDD and SBI methods, the DC component is inaccessible, as shown by the scalar product with $\vec{Q}$.

            Importantly, in contrast to analytical ptychography, aberration correction does not seem straightforward with CoM imaging. The dependence on focus and thickness has nevertheless been investigated in recent years \cite{Close2015,Addiego2020,Burger2020,Robert2021,Liang2023}, in particular with the objective of maintaining an interpretable contrast when the POA is not strictly fulfilled anymore.
        
        \subsubsection{Optical transfer function}
            
            Continuing, in contrast to the PCTF $\tilde{\zeta}\left(\vec{Q}\right)$, which is applicable in the situation where the WPOA is fulfilled and is then reflective of the information content of the scattering data itself, $\tilde{\gamma}\left(\vec{Q}\right)$ is fully process-induced and is derivable in the more general context of the POA. In particular, it is equal to the Fourier transform of the unaberrated probe intensity \cite{Muller2014} and, as such, can be termed as an optical transfer function (OTF) in the sense of ref. \cite{Black1957}. Beyond that, the immediate consequence of this OTF is the higher weighting of low-frequency features, compared to the rest of the object spectrum, being then attenuated.
            
            As a result, the iCoM imaging mode is prone to low-frequency artefacts \cite{Yucelen2018}, which may constitute a limit to the use in the low-dose case \cite{Gao2022}. This is nevertheless not problematic for many of the common applications of DPC and CoM consisting in the imaging of long-range features, such as e.g. charge density gradients \cite{Shibata2015}, magnetic domain structures \cite{Chapman1984}, large proteins \cite{Lazic2022}, interfaces between materials \cite{Mahr2022}, particle shapes \cite{Wu2017}, skyrmions \cite{McVitie2018} or stray electrostatic fields \cite{Dushimineza2023}.
            
            In principle, and like in the SBI case, it should furthermore be possible to compensate this effect by directly deconvolving the real-space result with $\gamma\left(\vec{r}\right)$, though the limitation is then whether the frequency components to be amplified have been brought below the noise level. Hence, such a solution is not practical at low doses. In the specific situation where iCoM imaging is employed on a weak phase object, both the PCTF $\tilde{\zeta}\left(\vec{Q}\right)$ and the OTF $\tilde{\gamma}\left(\vec{Q}\right)$ can be expected to apply.
    
    \subsection{\textbf{Scan-frequency partitioning algorithm}}
        \label{subsec:SFPA}
        
        \begin{algorithm}
        \begin{algorithmic}[1]
            
            \caption{SFPA}
            \label{alg:SFPA}
            
            \State Choose imaging method
            \State Partition $\vec{r}_s$-coordinates in packets $P_{\vec{r}_s}$
            \State Define $\vec{Q}$-grid
            \State Partition $\vec{Q}$-coordinates in domains $D_{\vec{Q}}$
            
            \If{imaging method is WDD}
                \State Initialize intermediary result as $f\left(\vec{Q}\right)\,=\,0$
            \ElsIf{imaging method is SBI-D}
                \State Initialize intermediary result as $g\left(\vec{Q}\right)\,=\,0$
            \ElsIf{imaging method is SBI-S}
                \State Initialize intermediary result as $g'\left(\vec{Q}\right)\,=\,0$
            \ElsIf{imaging method is iCoM}
                \State Initialize intermediary result as $h\left(\vec{Q}\right)\,=\,0$
            \EndIf
            
            \State Distribute $P_{\vec{r}_s} / D_{\vec{Q}}$ couples asynchronously
            
            \For{each packet $P_{\vec{r}_s}$}
                \For{each domain $D_{\vec{Q}}$}
                    
                    \If{imaging method is WDD}
                        
                        \State Calculate $J^{P_{\vec{r}_s}}_{\vec{Q} \in D_{\vec{Q}}}\left(\vec{R}\right)$
                        
                        \State Calculate $f_{P_{\vec{r}_s}}\left(\vec{Q} \in D_{\vec{Q}}\right)$
                        \State Add $f_{P_{\vec{r}_s}}\left(\vec{Q} \in D_{\vec{Q}}\right)$ to $f\left(\vec{Q} \in D_{\vec{Q}}\right)$
                    
                    \Else
                        
                        \State Calculate $\tilde{J}^{P_{\vec{r}_s}}_{\vec{Q} \in D_{\vec{Q}}}\left(\vec{q}_d\right)$
                        
                        \If{imaging method is SBI-D}
                            \State Calculate $g_{P_{\vec{r}_s}}\left(\vec{Q} \in D_{\vec{Q}}\right)$
                            \State Add $g_{P_{\vec{r}_s}}\left(\vec{Q} \in D_{\vec{Q}}\right)$ to $g\left(\vec{Q} \in D_{\vec{Q}}\right)$
                        \ElsIf{imaging method is SBI-S}
                            \State Calculate $g'_{P_{\vec{r}_s}}\left(\vec{Q} \in D_{\vec{Q}}\right)$
                            \State Add $g'_{P_{\vec{r}_s}}\left(\vec{Q} \in D_{\vec{Q}}\right)$ to $g'\left(\vec{Q} \in D_{\vec{Q}}\right)$
                        \ElsIf{imaging method is iCoM}
                            \State Calculate $h_{P_{\vec{r}_s}}\left(\vec{Q} \in D_{\vec{Q}}\right)$
                            \State Add $h_{P_{\vec{r}_s}}\left(\vec{Q} \in D_{\vec{Q}}\right)$ to $h\left(\vec{Q} \in D_{\vec{Q}}\right)$
                        \EndIf
                    
                    \EndIf
                
                \EndFor
            \EndFor
            
            \If{imaging method is WDD}
                \State Divide $f\left(\vec{Q}\right)$ by $\sqrt{f\left(\vec{0}\right)}$
                \State Inverse Fourier transform along $\vec{Q}$
                \State \textbf{Transmission function is measured}
                \State Extract angle of transmission function
                \State \textbf{Phase shift map is measured}
            \Else
                \State Inverse Fourier transform along $\vec{Q}$
                \State \textbf{Phase shift map is measured}
            \EndIf
        
        \end{algorithmic}
        \end{algorithm}
        
        \subsubsection{Motivations and numerical basis}
            
            One of the main limiting factor for the practical implementation of analytical ptychography is the necessity to first load the full dataset in computer memory, in order to perform a collective treatment consisting in a succession of fast Fourier transforms (FFT) and deconvolution/summation steps. Such a process requires a large available memory and makes e.g. GPU implementation difficult. This publication thus proposes a new scan-frequency partitioning algorithm, i.e. the SFPA solution mentioned in the introduction, which constitutes a straightforward, memory-limited and parallelizable implementation of the WDD, SBI and iCoM methods. As explained in more details below, this approach also relaxes sampling conditions that would normally be imposed by the scan grid.
            
            The basis for the algorithm is the replacement of the FFT leading from the scan dimension $\vec{r}_s$ to the spatial frequencies $\vec{Q}$ with an explicit, term-by-term, summation, e.g. following the Einstein notation. This is conventionally referred to as the einsum algorithm, as included e.g. in several Python packages. A similar explicit construction of the Fourier series was used for instance in ref. \cite{Strauch2021} for live processing. The formal procedure is described in the following paragraphs, and is otherwise provided in algorithm \ref{alg:SFPA}. Noteworthily, under the current implementation developed for this work, the PyTorch package \cite{Paszke2019} was chosen for its capacities in straightforward GPU-based programming.
        
        \subsubsection{Partitioning of calculation steps}
            
            The SFPA encompasses two distinct levels of partitioning among the calculation steps needed for the complete process. A first one is ensured by cutting the complete four-dimensional dataset $I^{det}_{\vec{r}_s}\left(\vec{q}_d\right)$ in packets of scan positions $P_{\vec{r}_s}$, each containing a user-defined number of arbitrarily chosen coordinates $\vec{r}_s$. The packets are treated individually, in particular the einsum-based Fourier transform, itself done for specific spatial frequency domains $D_{\vec{Q}}$. This then represents the second partitioning introduced in the algorithm.
            
            The complete set of calculations is thus divided in a number of single independent operations, each involving a specific $P_{\vec{r}_s} / D_{\vec{Q}}$ couple. Those operations yield partial Fourier transformed datasets given by
            \begin{equation}
            \begin{split}
                \tilde{J}^{P_{\vec{r}_s}}_{\vec{Q} \in D_{\vec{Q}}}\left(\vec{q}_d\right) \, & = \, \sum\limits_{\vec{r}_s \in P_{\vec{r}_s}} \, e^{-i 2\pi \vec{Q}\cdot\vec{r}_s} \, I^{det}_{\vec{r}_s}\left(\vec{q}_d\right) \\
                J^{P_{\vec{r}_s}}_{\vec{Q} \in D_{\vec{Q}}}\left(\vec{R}\right) \, & = \, \sum\limits_{\vec{r}_s \in P_{\vec{r}_s}} \, e^{-i 2\pi \vec{Q}\cdot\vec{r}_s} \, \mathcal{F}^{-1}\left[ I^{det}_{\vec{r}_s}\left(\vec{q}_d\right) \right]\left(\vec{R}\right) \quad,
                \label{eq:EinsteinEinstein}
            \end{split}
            \end{equation}
            depending on the type of reconstruction performed, i.e. $J_{\vec{Q}}\left(\vec{R}\right)$ is the input of a WDD process while $\tilde{J}_{\vec{Q}}\left(\vec{q}_d\right)$ is needed for SBI and iCoM. Note that in equation \ref{eq:EinsteinEinstein}, the term $\mathcal{F}^{-1}$ indicates an inverse Fourier transform done over the camera space, for the few CBED patterns in $P_{\vec{r}_s}$.
            
            Each partial Fourier transformed dataset is used for one of the following calculations
            \begin{equation}
            \begin{split}
                f\left(\vec{Q} \in D_{\vec{Q}}\right) \, & \leftarrow \, f\left(\vec{Q} \in D_{\vec{Q}}\right) \, + \, f_{P_{\vec{r}_s}}\left(\vec{Q} \in D_{\vec{Q}}\right) \\
                g\left(\vec{Q} \in D_{\vec{Q}}\right) \, & \leftarrow \, g\left(\vec{Q} \in D_{\vec{Q}}\right) \, + \, g_{P_{\vec{r}_s}}\left(\vec{Q} \in D_{\vec{Q}}\right) \\
                g'\left(\vec{Q} \in D_{\vec{Q}}\right) \, & \leftarrow \, g'\left(\vec{Q} \in D_{\vec{Q}}\right) \, + \, {g'}_{P_{\vec{r}_s}}\left(\vec{Q} \in D_{\vec{Q}}\right) \\
                h\left(\vec{Q} \in D_{\vec{Q}}\right) \, & \leftarrow \, h\left(\vec{Q} \in D_{\vec{Q}}\right) \, + \, h_{P_{\vec{r}_s}}\left(\vec{Q} \in D_{\vec{Q}}\right) \quad,
            \end{split}
            \end{equation}
            where the packet-specific intermediary result $f_{D_{\vec{r}_s}}\left(\vec{Q} \in D_{\vec{Q}}\right)$ is obtained through equations \ref{eq:oldstyleWDD} and \ref{eq:transmfuncextract}, $g_{D_{\vec{r}_s}}\left(\vec{Q} \in D_{\vec{Q}}\right)$ through equation \ref{eq:SBIdeconv}, ${g'}_{D_{\vec{r}_s}}\left(\vec{Q} \in D_{\vec{Q}}\right)$ through equation \ref{eq:SBIsum} and $h_{D_{\vec{r}_s}}\left(\vec{Q} \in D_{\vec{Q}}\right)$ through equations \ref{eq:calculCoM} and \ref{eq:integrCoM}. Performing the same process for the entirety of the $\vec{r}_s$-to-$\vec{Q}$ components of the Fourier series finally yields the full reconstruction result.
            
            Noteworthily, in the case of the WDD method, the complete, four-dimensional, Wigner distribution $\Upsilon\left(\vec{Q}\,;\,\vec{R}\right)$ is not explicitly retrieved. Instead, in the implementation described by algorithm \ref{alg:SFPA}, each $P_{\vec{r}_s} / D_{\vec{Q}}$ couple leads to an increment of $f\left(\vec{Q} \in D_{\vec{Q}}\right)$ directly. The same einsum-based Fourier transform strategy could nevertheless be used for this purpose, i.e. without an immediate summation step across $\vec{R}$, straightforwardly as well.
        
        \subsubsection{Opportunity for parallelization}
            
            A first interest of the scan-frequency partitioning algorithm is its low need in active memory, since the size of the packets $P_{\vec{r}_s}$ and domains $D_{\vec{Q}}$ are chosen by the user directly. This in turn permits to adapt the process to the computer memory available, including as part of a straightforward implementation on a GPU, e.g. involving specialized Python-based procedures \cite{Paszke2019}. Furthermore, since the treatment of each individual $P_{\vec{r}_s} / D_{\vec{Q}}$ couple is independent of all others, parallelization is possible along both the $\vec{r}_s$ and $\vec{Q}$ dimensions. In comparison, the implementation reported in ref. \cite{Strauch2021} only allowed it along $\vec{r}_s$, though it was already enough for live processing using a computer with sufficient performance.
            
            Given the low memory requirement of a single $P_{\vec{r}_s} / D_{\vec{Q}}$ calculation, such a parallelization strategy is in principle implementable on a wider range of devices, including low-end. Though extensive numerical benchmarking was left for future work, it should be noted that avoiding the two-dimensional $\vec{r}_s$-to-$\vec{Q}$ FFT can be expected to lead to an increment in the numerical complexity of the complete process. Specifically, it then goes from the typical $O\left( N_{s\,;\,x} \cdot N_{s\,;\,y} \cdot \log\left( N_{s\,;\,x} \cdot N_{s\,;\,y} \right) \right)$ to $O\left( N_{s\,;\,x} \cdot N_{s\,;\,y} \cdot N_{\vec{Q}} \right)$, with $N_{s\,;\,x} / N_{s\,;\,y}$ the number of positions along the two scan axes and $N_{\vec{Q}}$ the total number of frequencies $\vec{Q}$ used. This number is equal to
            \begin{equation}
                N_{\vec{Q}} \, = \, 4 \pi S_{rec} {q_A}^{\,\,\,2} \quad,
                \label{eq:numbfreq}
            \end{equation}
            where $\pi \left(2 q_A\right)^2$ is the reconstructed frequency surface and $S_{rec}$ is the reconstructed real-space surface. Specifically, the discretized frequencies $\vec{Q}$ are distributed within a disk of radius $2 q_A$, with a pixel density determined by the real-space extent of the reconstruction window.
            
            Note that, in order to avoid aliasing and periodicity artefacts, $S_{rec}$ has to be made sufficiently larger than the scanned surface $S$, with a portion of it maintained at a value of zero in the real-space reconstruction window. The same is true in the frequency space window \cite{Self1983}. Overall, the preparation of the reconstructed object in both $\vec{Q}$- and $\vec{r}$-space follows the conventional approach used, for instance, in typical multislice simulations, as is described e.g. in ref. \cite{Kirkland2020}.
        
        \subsubsection{Decorrelation of scan and reconstruction grids}
            
            Perhaps most importantly, and as implied by equation \ref{eq:numbfreq}, the employment of the einsum algorithm permits the explicit decorrelation of the scan and frequency dimensions. As such, the real-space reconstruction grid, and thus the actual choice of $\vec{Q}$-coordinates for which the result is calculated, is prepared independently of the scan grid, and the formal contribution of each given $\vec{r}_s$-coordinate to a single arbitrary frequency $\vec{Q}$ may be determined separately from all others. In contrast, in the conventional full FFT solution \cite{Li2014,Pennycook2015}, as well as in ref. \cite{Strauch2021}, each scan point equates one pixel in the result. The SFPA however permits a calculation at frequencies exceeding e.g. the maximum that would then be allowed by the finite scan interval. In that context, it becomes possible, for instance, to perform a reconstruction given a strongly defocused probe and less scan positions \cite{Song2019}, while conserving an appropriately resolved reconstruction window in real-space.
            
            At first sight, this development may be understood as breaking the Nyquist-Shannon sampling theorem. The possibility of retrieving information beyond the Nyquist frequency of the scan grid should however be seen as resulting from the usage of the information available along the $\vec{q}_d / \vec{R}$ dimensions. In particular, in defocused conditions, the amount of details contained in the far-field intensity is greater as it then consists in a shadow image of the specimen \cite{Cowley1979a}. As such, including the appropriate aberration function in $\theta\left(\vec{q}_0\,;\,\vec{q}_0+\vec{Q}\right)$ permits to correctly translate this information back into the result and extend the area, in the real-space reconstruction window, that may be informed by a single scan position. Beyond this, since such a measurement geometry is already commonly used in combination with iterative approaches \cite{Hue2010,Song2019}, where a similar decorrelation of the scan and reconstruction pixels is implicit and where the same scattering data is used, this fundamental ability of analytical ptychography is expected.
            
            Another interest of using an arbitrary set of reconstruction frequencies is the facilitated implementation of high-pass and low-pass filtering, as the concerned $\vec{Q}$-coordinates can be omitted from the calculation, hence reducing $N_{\vec{Q}}$ as well. A limitation to this practice is however that, in order to perform an extraction of the phase shift map from the WDD-retrieved transmission function, it should have a defined zero-frequency component, i.e. the mean of the complex numbers $T^{WDD}\left(\vec{r}\right)$ should not be equal to zero. For this reason, only the inherent high-pass filtering, and not both the high- and low-pass ones, may be used for the SFPA-based WDD calculation.
            
            The employment of an orthonormal Fourier transform may moreover be useful in ensuring appropriate numerical normalization, as the number of pixels in the reconstruction grid and the scan grid are likely to differ. This proper choice of convention is important to conserve consistent results among slightly different scan grids among recordings.
            
            Finally, the operations involving the dimensions $\vec{q}_d$ and $\vec{R}$, as in equations \ref{eq:transmfuncextract}, \ref{eq:SBIdeconv}, \ref{eq:SBIsum} and \ref{eq:calculCoM}, are performed in the camera space. As such, $\vec{R}$ represents a spatially limited kernel, similarly to e.g. ref. \cite{Maiden2009}, in which numerical artefacts are prevented by a simple interpolation or zero-padding step. In that manner, the process can also precisely account for elliptical distortions observed in the far-field pattern \cite{Capitani2006} and prevent them from affecting the result, based on an initial calibration of the $\vec{q}_d$ dimension \cite{Robert2021}.

\section{Atomic-resolution imaging of MoS$_2$}
    \label{sec:MoS2}
    
    \subsection{\textbf{Conventional focused-probe conditions}}
        \label{subsec:MoS2FP}
        
        \begin{figure}
            \centering
            \includegraphics[width=1.0\columnwidth]{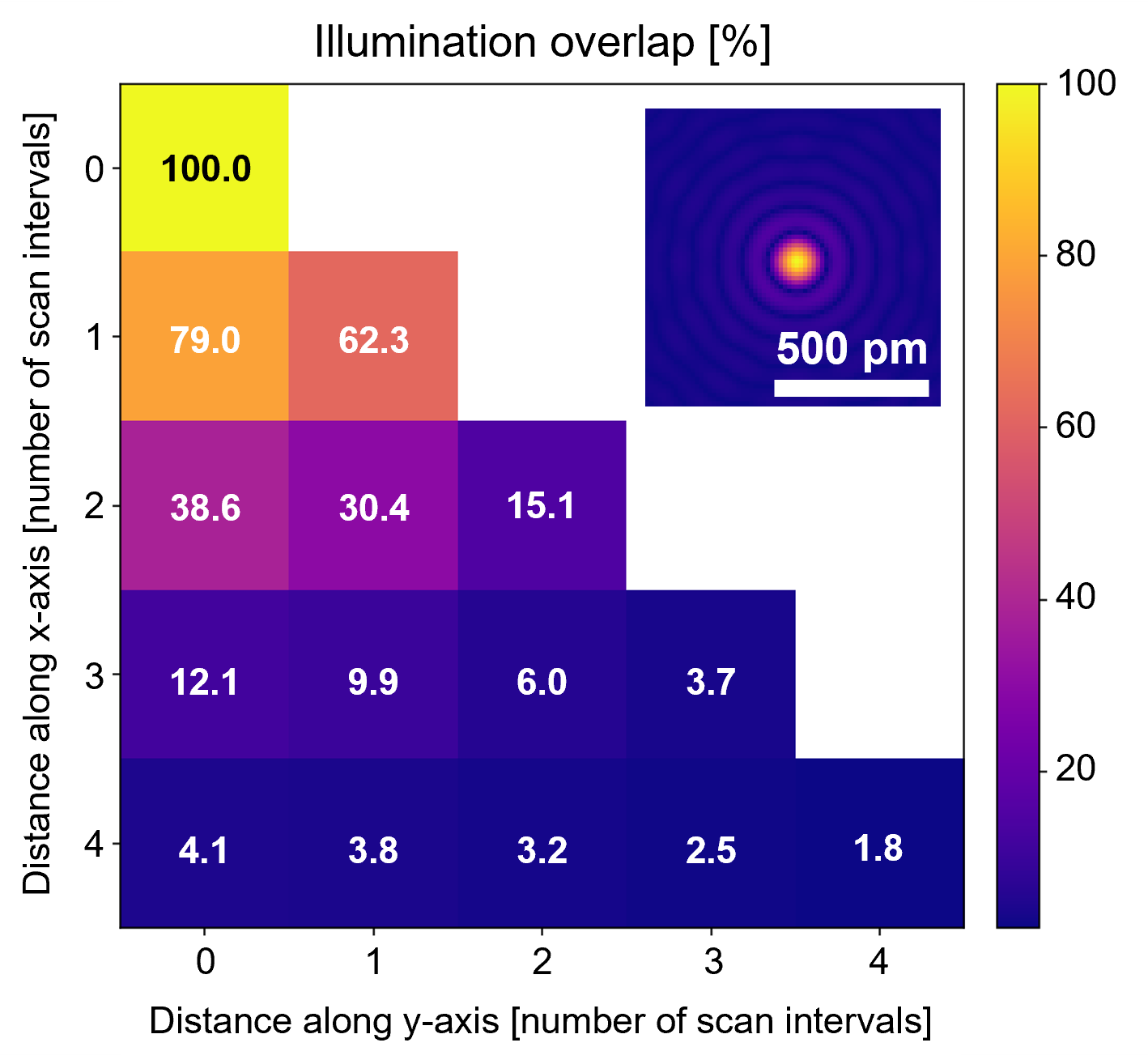}
            \caption{Depiction of the overlap ratio $\beta_{\delta \vec{r}_s}$ for a variety of scan points couple in a larger scan grid, i.e. along both scan axis and over up to four intervals in the scan grid. The scan interval is equal to about 32 pm. The electron probe is calculated given the parameters given in subsection \ref{subsec:MoS2FP}. The probe amplitude $\mid P\left(\vec{r}_0\right) \mid$, having a Rayleigh criterion of about 99 pm, is shown as an inset. More details on the calculation of this overlap ratio can be found in appendix A.}
            \label{fig:overlapratio}
        \end{figure}
        
        \begin{figure*}
            \centering
            \includegraphics[width=1.0\textwidth]{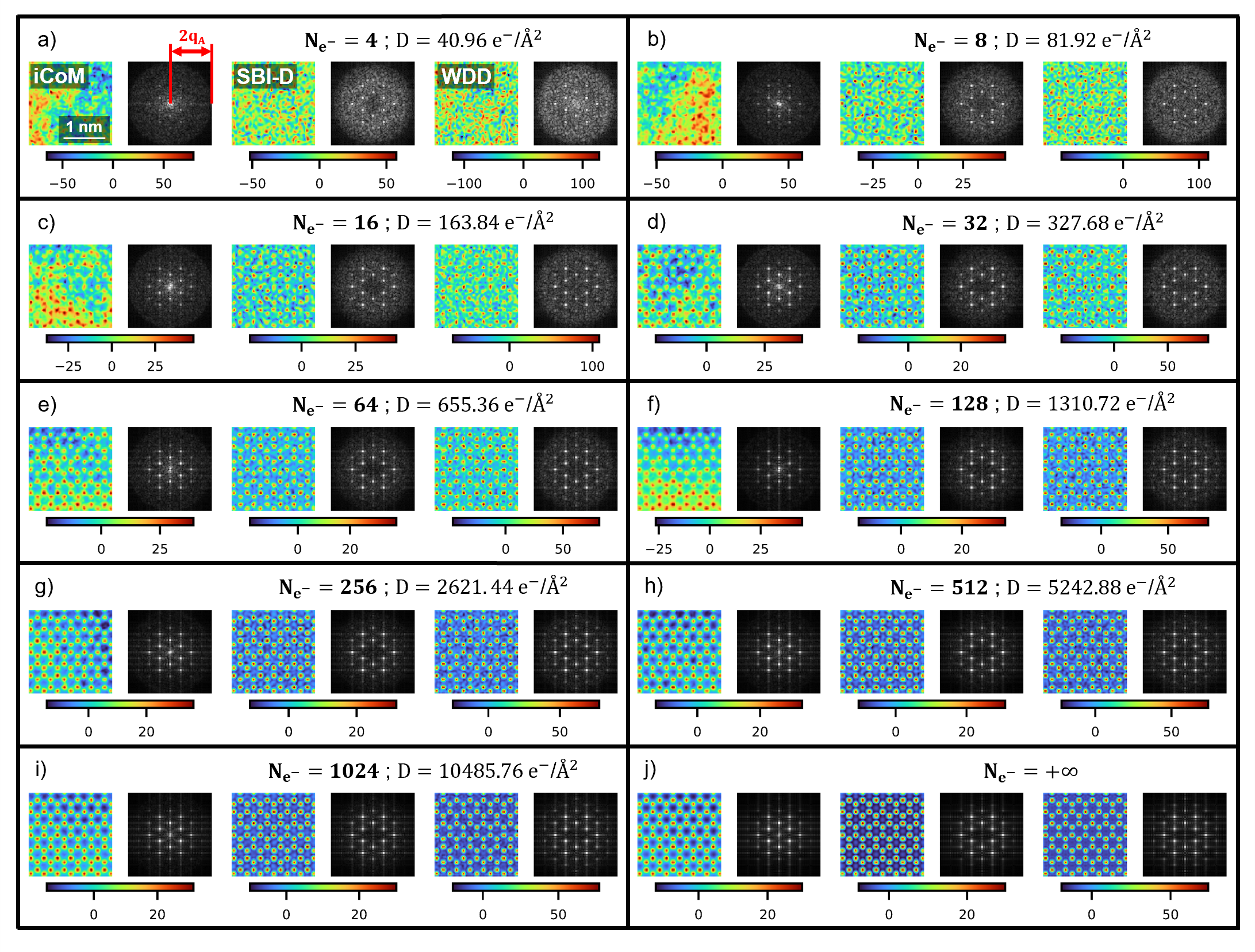}
            \caption{Results of analytical ptychography of monolayer MoS$_2$, applied on the multislice electron diffraction simulation presented in subsection \ref{subsec:MoS2FP}. Calculations are done for a variety of average numbers of electrons per pattern $N_{e^-}$, and corresponding doses $D$ given in $e^-/\text{\AA}^2$. For each case, the position-dependent measurement of the projected potential $\mu\left(\vec{r}\right)$, through the iCoM, SBI-D and WDD methods, is displayed alongside the square root of its Fourier transform's amplitude $\sqrt{\mid\tilde{\mu}\left(\vec{Q}\right)\mid}$. The colorbars reflect values of projected potential in the $\mu\left(\vec{r}\right)$ measurements, in V$\cdot$nm.}
            \label{fig:MoS2_FP}
        \end{figure*}
        
        \begin{figure}
            \centering
            \includegraphics[width=1.0\columnwidth]{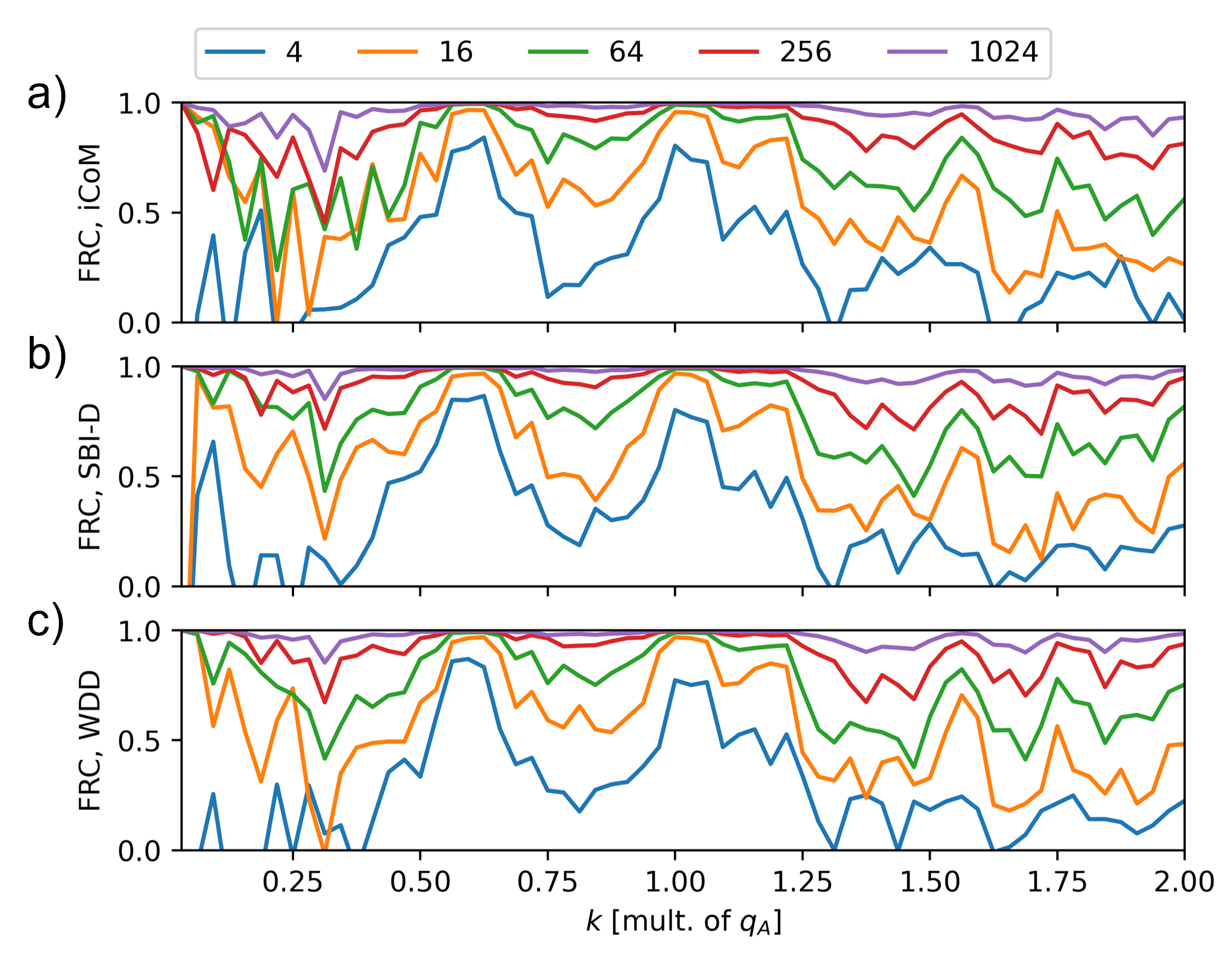}
            \caption{FRC calculated from the $\mu\left(\vec{r}\right)$ measurements presented in fig. \ref{fig:MoS2_FP}, i.e. by comparing the infinite dose cases to the various dose-limited simulations. The results are plotted as a function of the reference spatial frequency $k$, expressed as a multiple of $q_A$, and given for selected $N_{e^-}$ values. The FRC calculation is displayed in a) for iCoM, in b) for SBI-D and in c) for WDD.}
            \label{fig:FRC_MoS2_FP}
        \end{figure}
        
        \begin{figure}
            \centering
            \includegraphics[width=1.0\columnwidth]{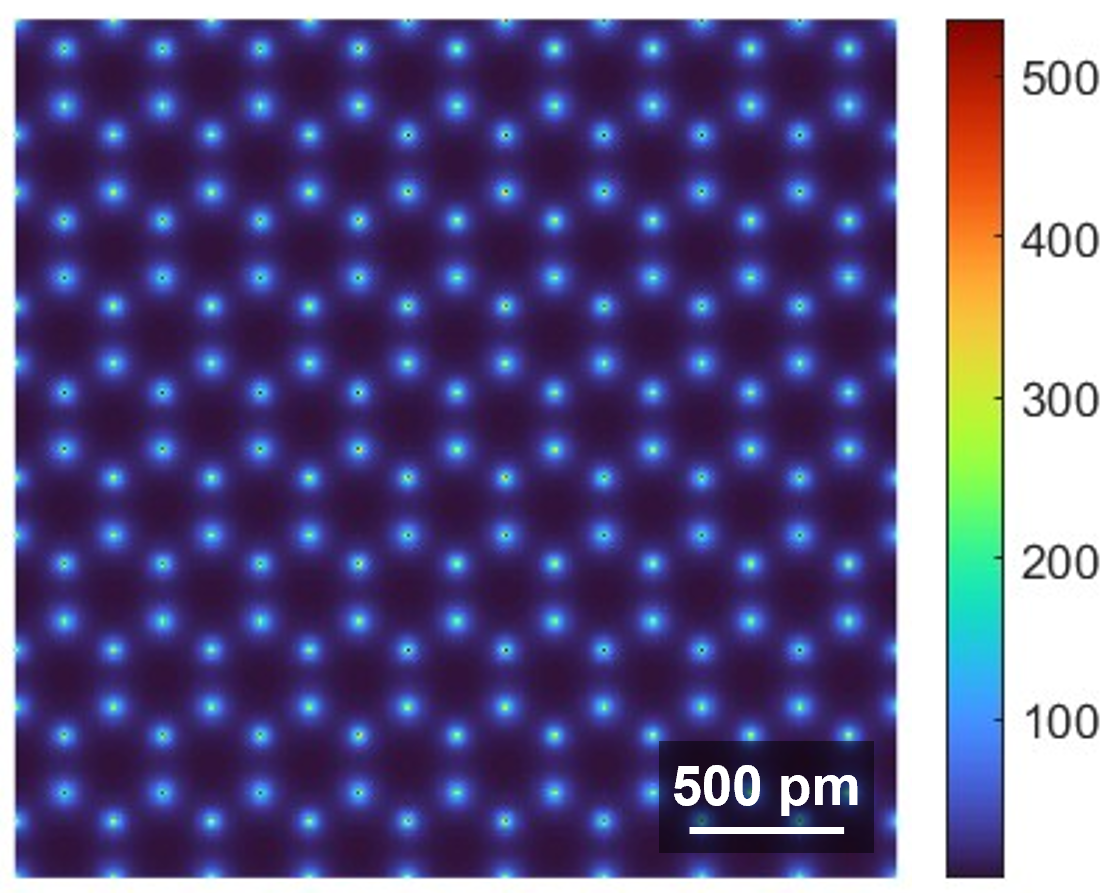}
            \caption{Laterally limited view of the vertical projection of the three-dimensional potential used for the simulation presented in subsection \ref{subsec:MoS2FP}. The quantity is expressed in V$\cdot$nm and is represented in the absence of atomic vibration, i.e. the atoms are all exactly at their rest positions.}
            \label{fig:groundtruth}
        \end{figure}
        
        \subsubsection{Simulation and processing parameters}
            
            In order to test the dose-efficiency of the iCoM, SBI-D and WDD methods in the conventional focused-probe, high-resolution, condition of electron ptychography, an MR-STEM simulation was performed based on a monolayer MoS$_2$ specimen, for which the POA can reasonably be considered fulfilled. Diffraction patterns were calculated in a scan grid of 64 by 64 points, covering an area of 2 nm by 2 nm, hence with an interval of about 32 pm. Illumination conditions were chosen as representative of the capacities of a modern aberration-corrected microscope such as e.g. a Titan Themis 60-300 (Thermo Fisher Scientific). Specifically, the acceleration voltage $U$ and the semi-convergence angle $\alpha$ were assigned values of 60 kV and 30 mrad respectively which, for reference, leads to a Rayleigh criterion $\delta r_{\text{Rayleigh}}\,=\,0.61 / q_A$ of about 99 pm. As illustrated in figure \ref{fig:overlapratio}, the optical conditions described above lead to an area overlap $\beta_{\delta \vec{r}_s}\,\approx\,79.0\text{\%}$ between two scan points neighboring each other along a scan direction, and $62.3\text{\%}$ along the diagonal. $\beta_{\delta \vec{r}_s}$ is defined for higher distances $\delta \vec{r}_s$ as well. This highlights that a degree of redundancy remains beyond immediate neighbors, which is exploited by the ptychographic process as well. More details on the calculation of the area overlap and how it differs from the conventional approach \cite{Bunk2008} can be found in appendix A.
            
            Continuing, the propagation of the electron wavefunction through the specimen was modeled based on the multislice approximation \cite{Cowley1957,Goodman1974,Ishizuka1977} and the atomic potentials were calculated using parameterized hydrogen orbitals as described in ref. \cite{Lobato2014}. The specimen potential was sliced below the atomic plane level and pixelated such that a maximum scattering vector of up to twice the range actually used could be included. Thermal motion within the lattice was accounted for by repeating the calculation for a total of 64 configurations of random lateral atomic shifts, and averaging the resulting distributions $I_{\vec{r}_s}\left(\vec{q}\right)$. The random shift vectors were determined using the frozen phonon approximation \cite{Wang1998a,Muller2001} based on the Einstein model, i.e. assuming non-correlated atomic vibrations \cite{Loane1991}. For simplicity, and also because this work aims at reproducing results obtainable with a Timepix3 chip \cite{Poikela2014,Frojdh2015} at a low acceleraton voltage $U$, thus in a condition where multiple counting is unlikely to occur \cite{Mir2017,Jannis2022,Denisov2023}, the simulation did not include an explicit MTF. As such, the values taken by the $M\left(\vec{r}_d\right)$ function, included in practice in the SBI-D and WDD calculations, only encompassed the role of the finite pixel size of the simulated camera, as implied by the kernel size.
            
            The results of the iCoM, SBI-D and WDD processes, implemented using the SFPA approach described in subsection \ref{subsec:SFPA}, are depicted in figure \ref{fig:MoS2_FP}. Specifically, the measurements of the projected potential, expressed in V$\cdot$nm, are shown alongside the square roots of the corresponding Fourier transform amplitudes, for visualization of Fourier weightings along the two-dimensional $\vec{Q}$ coordinates. The calculations were done for a variety of average numbers of electrons per pattern $N_{e^-}$ and consequent doses $D$ given in $e^-/\text{\AA}^2$. To better highlight the non-linear relation between dose and contrast, $N_{e^-}$ was given values of 2$^l$ with $l\,\in\,\left[2,3,...,10\right]$. Dose-limitation was ensured by repeated random pixel selection, with the number of repetitions being probabilistically determined across the scan window by Poisson statistics. As such, the propagation of Poisson noise from detector space to the reconstruction window can be straightforwardly reproduced, while conserving a realistic sparsity in the scattering frames \cite{OLeary2020}. This approach is described in more details in appendix B, alongside its wider interests.
            
            Furthermore, for each dose-limited case, the generated sparse diffraction patterns were individually normalized by their sum, pre-treatment. This strategy was adopted for all reconstructions presented in this work and was chosen following the suggestion of ref. \cite{Seki2018}. This is equivalent to varying the normalization of the wavefunction, scan point-wise, to match the number of counts in each corresponding pattern. Importantly, taking this normalization choice into account will be required for any theoretical estimation of measurement precision in future work, as it leads, in effect, to a change in the variance of single patterns. This solution differs from the usual quantitative STEM approach \cite{LeBeau2008,Rosenauer2009}, which would have consisted in uniformly normalizing by $N_{e^-}$. Finally, the $N_{e^-}\ =\,+\infty$ case corresponds to the direct use of the simulated $I_{\vec{r}_s}\left(\vec{q}\right)$, where the intensity is implicitly normalized. The corresponding result can thus be understood as representing the experimental situation where the best achievable dose-dependent precision is reached, and hence where the noise level is negligible.
            
            As a side-note, in the case of the WDD process, the projected electrostatic potential is obtained through a prior extraction of the phase shift map. Given that the $\left[\,-\pi\,;\,+\pi\,\right]$ range was not exceeded, no discontinuities were observed and thus no unwrapping was necessary.
        
        \subsubsection{Noise level in the micrographs}
            
            For the three methods, atomic patterns are already visible from $N_{e^-}\,=\,8$, hence with a dose below $D\,=\,81.92\,e^-/\text{\AA}^2$. Moreover, frequencies belonging to the specimen lattice are observed even in the Fourier transforms of results obtained given $N_{e^-}\,=\,4$. This first remark is particularly interesting for future applications of electron ptychography to beam-sensitive objects, as it empirically shows what is the true requirement in terms of dose, given a perfectly stable and coherent imaging system. As $N_{e^-}$ increases, the noise level in the images lessens and specimen frequencies become more dominant compared to the noise background. Such a dose-dependent precision in ptychographic computational imaging has been investigated empirically in the literature \cite{Godard2012,Katvotnik2013,DAlfonso2016,Leidl2024}, as is done here as well, and its lowest achievable value can in principle be predicted by parameter estimation theory \cite{Cederquist1987,Wei2020,Bouchet2021,Koppell2021,Dwyer2024,VegaIbanez2025}, in particular using the Cramér-Rao lower bound (CRLB) \cite{Rao1945}.
            
            In this publication, the true frequency-dependent CRLB is not provided since, unless some simplifications such as the WPOA \cite{Koppell2021,Dwyer2024} are introduced, its formulation remains specific to the specimen \cite{Wei2020,Bouchet2021}. The establishment of a general $\vec{Q}$-dependent metric, which would be dependent on the complete set of experimental parameters, is thus left for future work. Beyond that, the approximation made in ref. \cite{VegaIbanez2025}, provided below, leads to a single number $\text{CRLB}_{RS}$ representing the minimum standard deviation among distinct measurements, as induced by the propagation of Poisson noise \cite{Luczka1991}, upon retrieving the  phase shift map $\sigma \mu\left(\vec{r}\right)$ in real-space. While it was derived in ideal illumination conditions which are not met here, e.g. the total illumination is not strictly restricted to the scanned area, this metric remains useful to establish a fundamental understanding of the concept.
            \begin{equation}
                \text{CRLB}_{RS} \, = \, \sqrt{\frac{ N_{\vec{Q}} }{ 2 N_s N_{e^-} }} \, \geq \, \sqrt{ \, \frac{ \, 2\pi {q_A}^{\,\,\,2} \, - \, \frac{1}{2 S} \, }{D} \, } \quad.
                \label{eq:simpleCRLB}
            \end{equation}
            $N_s N_{e^-}$ represents the total number of probing electrons, while $N_s$ is the total number of scan positions used. $S$ is the surface covered by the scan window, necessarily smaller than the surface $S_{rec}$ as explained in subsection \ref{subsec:SFPA}. The number $N_{\vec{Q}}$ of reconstructed frequencies was otherwise described by equation \ref{eq:numbfreq}, and can here be understood as the number of useful pixels in the reconstruction. A "-1" term is added to account for the unmeasurable DC component. Given that the term $0.5 / S$ is likely to be negligible compared to $2\pi {q_A}^{\,\,\,2}$, $\text{CRLB}_{RS}$ shows rather clearly that, in order to achieve a certain goal in measurement precision, the dose $D$ has to be adapted to the aperture radius $q_A$, and thus implicitly to the spatial resolution \cite{Egerton2013}.
        
        \subsubsection{Fourier ring correlations}
            
            In order to pursue the analysis further, Fourier ring correlations $\text{FRC}^m\left(k\right)$ \cite{VanHeel1982,Saxton1982}, shown in figure \ref{fig:FRC_MoS2_FP}, were calculated from the projected potential results through
            \begin{equation}
            \begin{split}
                & \text{FRC}^m\left(k\right) \, = \, \frac{\text{X}^m\left(k\right)}{\text{Y}^m\left(k\right)} \\
                & \text{X}^m\left(k\right) \, = \, \sum\limits_{\parallel\vec{Q}\parallel \in R_k} \, \tilde{\mu}_{m}\left(\vec{Q}\right) \, {\tilde{\mu}_{+\infty}}^*\left(\vec{Q}\right) \\
                & \text{Y}^m\left(k\right) \, = \, \sqrt{ \, \sum\limits_{\parallel\vec{Q}\parallel \in R_k} \, \mid\tilde{\mu}_{m}\left(\vec{Q}\right)\mid^2 \, \sum\limits_{\parallel\vec{Q}\,'\parallel \in R_k} \, \mid\tilde{\mu}_{+\infty}\left(\vec{Q}\,'\right)\mid^2 \, } \quad.
            \end{split}
            \end{equation}
            $k$ is a spatial frequency modulus and $R_k \, = \, \left[ \, k-\delta k \, ; \, k \, \right]$ is the corresponding annular domain, with $\delta k$ a case-dependent precision. $\tilde{\mu}_{m}\left(\vec{Q}\right)$ is the Fourier transform of the measured projected potential, for the specific $N_{e^-}\,=\,m$ case. In this context, the calculated FRC can be interpreted as a frequency-wise measurement of the dose-dependent precision of each method, and thus provides a straightforward dose-efficiency metric. The closer $\text{FRC}^m\left(k\right)$ is to 1, for a given spatial frequency modulus $k$, the closer the corresponding $R_k$ range of the signal is to reaching the best achievable precision. In this example, and for better visibility, the FRC curves are provided in a reduced selection of $N_{e^-} \, = \, 4, 16, 64, 256, 1024$, for iCoM in fig. \ref{fig:FRC_MoS2_FP}.a, SBI-D in fig. \ref{fig:FRC_MoS2_FP}.b and WDD in fig. \ref{fig:FRC_MoS2_FP}.c.
            
            A few observations can immediately be done from the calculated FRC. For all imaging modes, three peaks are observed, close to 0.5, 1.0 and 1.5 times $q_A$, as well as an emerging fourth one. Those correspond to the hexagonal pattern of spatial frequencies belonging to the specimen, mirroring scattering orders of the atomic lattice, as observed in the Fourier transforms of fig. \ref{fig:MoS2_FP} as well. At the level of the peaks, perfect precision is reached at a much lower $N_{e^-}$ than in the rest of the $k$-axis. From a naive standpoint, this already tends to show that frequencies $\vec{Q}$ actually carrying information on the specimen are reconstructed much more efficiently than those containing no information, and which then end up reaching $\tilde{\mu}\left(\vec{Q}\right)\,=\,0$ at infinite dose. This is expected in a situation where the spectrum of the illuminated object is sparse. In particular, ptychographically processed electrons end up contributing only to the recovered projected potential, i.e. its associated $\vec{Q}$-coordinates. This remains true as long as the illumination characteristics are known and no artefactual features are introduced, e.g. from an inaccurate interaction model.
        
            In this context, if one considers the overall calculation as an additive inclusion of single counts' contributions to the measurement, with no question of normalization, the signal-to-noise ratio at $\vec{Q}$-coordinates belonging to the specimen is expected to directly improve for each dose increment, while the noise level at other frequencies remains the same. The reduction of the background noise is then due to the normalization, and thus occurs at a lower rate than the retrieval of specimen information in itself. Continuing, upon comparing the three frequency peaks mentioned above, it is also noticeable that $\text{FRC}^m\left(k\right)\,\approx\,1$ occurs with more difficulty as $k$ increases, i.e. higher values of $\vec{Q}$ appear more dose-expensive at first sight. The practical reason for it is that the surface covered by an $R_k$ ring increases with $k$, which thus implies a larger proportion of background noise compared to specimen frequencies. Similarly, the lower values of $k$ lead to a less visually stable value of FRC, due to the low number of actual pixels in the corresponding $R_k$.
        
        \subsubsection{Contrast transfer capacities}
            
            Arguably the most important information to draw from figure \ref{fig:FRC_MoS2_FP} is that, among the three investigated methods, the overall FRC profiles are rather similar. This is of particular interest, as it shows that, for a given frequency component, the dose-efficiencies of iCoM, SBI and WDD are more-or-less the same, in that they reach the best achievable result with comparable dose requirements. As such, what differentiates those imaging modes with regards to the measurement precision in real-space, and in particular to the noise background formed as a function of spatial frequency \cite{Seki2018}, is the existence of the CTF $\tilde{\gamma}\left(\vec{Q}\right)$ for iCoM and $\tilde{\zeta}\left(\vec{Q}\right)$ for SBI, displayed in figures \ref{fig:CTF_iCoM} and \ref{fig:CTF_SBI_S}. Whereas those CTF lead to visually different micrographs, as is directly noticeable in fig. \ref{fig:MoS2_FP}, the attenuation of frequency components also contribute to noise filtering. Consequently, $\mu^{SBI}\left(\vec{r}\right)$ appears slightly, but noticeably, less noisy than $\mu^{WDD}\left(\vec{r}\right)$, e.g. for $N_{e^-}\,\leq\,64$. This is specifically related to the existence of high-frequency noise, as observed in the Fourier transforms of the WDD results, which is otherwise eliminated by the deconvolutive SBI process.
            
            Beyond that, the underlying difference in the real-space measurement, between the two analytical ptychography methods, consists in an exaggerated dark halo around atomic sites, clearly visible at higher doses and only present in the SBI-D results. This feature is associated to $\tilde{\zeta}\left(\vec{Q}\right)$ \cite{Hofer2023}, which is thus shown to not intrinsically apply to the WDD result. As was explained in subsection \ref{subsec:SBISBI}, this then constitutes a clear indication that the specimen is not a weak scatterer, i.e. $\tilde{\zeta}\left(\vec{Q}\right)$ cannot be considered to constitute the PCTF of the experiment in general. Specifically, the WDD process only assumes the more general POA, here remaining reasonable, and should not be generally expected to show the specific frequency transfer met in the case of a weak phase object. The formulation of the SBI method, on the other hand, is still based on this assumption and will thus have $\tilde{\zeta}\left(\vec{Q}\right)$ as a CTF in any case. Note that the violation of the WPOA is here further confirmed by the range of values covered by the WDD phase shift map, which is above 1.0 rad, as well as by the ground truth of the projected potential, for reference depicted in fig. \ref{fig:groundtruth} in a limited real-space window.
            
            More fundamentally, the nonfulfillment of the WPOA means that, in the $\tilde{J}_{\vec{Q}}\left(\vec{q}_d\right)$ distribution, the general sideband-like geometry arising from a weakly scattering specimen is not met in practice. As a consequence, this is not just noise, e.g. in the triple overlap areas, that is removed by the SBI process, but also potentially useful information on the specimen itself. This can be verified in fig. \ref{fig:MoS2_FP} as well, where the Fourier components of $\mu^{WDD}\left(\vec{r}\right)$, at high $\parallel\vec{Q}\parallel$ values, are visibly higher than those of $\mu^{SBI}\left(\vec{r}\right)$. This thus leads, in addition to the absence of the artificial features mentioned above, to a slightly better resolution in the WDD measurement.
            
            In parallel, the iCoM imaging mode is affected by the OTF $\tilde{\gamma}\left(\vec{Q}\right)$, whether the PCTF $\tilde{\zeta}\left(\vec{Q}\right)$ of a weak phase object is applicable or not. As such, the higher weighting of low frequencies, with the rest of the spectrum being then attenuated, increases its susceptibility to long-range artefacts \cite{Yucelen2018,Gao2022}, as explained in subsection \ref{subsec:iCoMiCoM}. Consequently, low-frequency noise remains dominant up to e.g. $N_{e^-}\,\leq\,256$. This is verified by the Fourier transforms as well.
        
        \subsubsection{Phase shift value range}
            
            Continuing with the projected potential maps displayed in fig. \ref{fig:MoS2_FP}, one last remark remains to be made. In high dose conditions, the range of values obtained with WDD is about twice larger than it is with iCoM and SBI-D. This is not due, for instance, to a normalization issue, as the WDD result is obtained by extracting the angle of the initially retrieved transmission function. Furthermore, as is shown in section \ref{sec:ApoApo}, the opposite situation can be met as well and, as proven in subsection \ref{subsec:MoS2OF40}, the aberration function plays a role too. This thus points out the mismatch in value range as being a fundamental feature of the imaging method rather than a numerical issue. In that respect, it is also worth noting that such a mismatch was observed previously in the literature \cite{Yang2017,OLeary2021,Gao2022} as well. Finally, the effect is likely amplified by the higher resolution of the WDD reconstructions, itself due to the better transfer of high frequencies, leading to stronger atomic peaks in the image.
    
    \subsection{\textbf{Overfocused illumination conditions}}
        \label{subsec:MoS2OF40}
        
        \begin{figure}
            \centering
            \includegraphics[width=1.0\columnwidth]{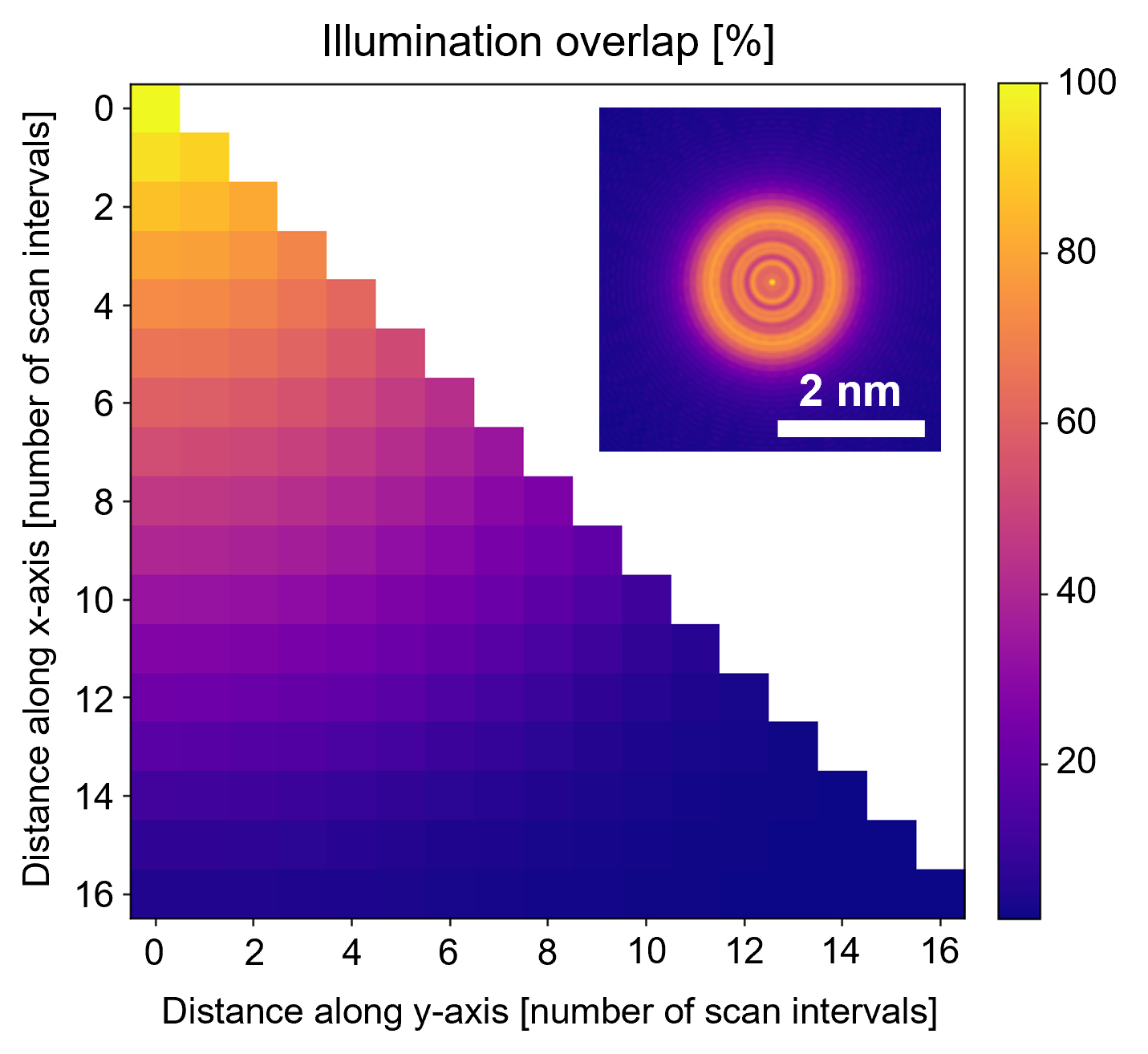}
            \caption{Depiction of the overlap ratio $\beta_{\delta \vec{r}_s}$ for a variety of scan points couple in a larger scan grid, i.e. along both scan axis and over up to 16 intervals in the scan grid. The scan interval is equal to about 133 pm. The electron probe is calculated given the parameters given in subsection \ref{subsec:MoS2OF40}. The probe amplitude $\mid P\left(\vec{r}_0\right) \mid$ is shown as an inset. More details on the calculation of this overlap ratio can be found in appendix A.}
            \label{fig:overlapratio2}
        \end{figure}
        
        \begin{figure}
            \centering
            \includegraphics[width=1.0\columnwidth]{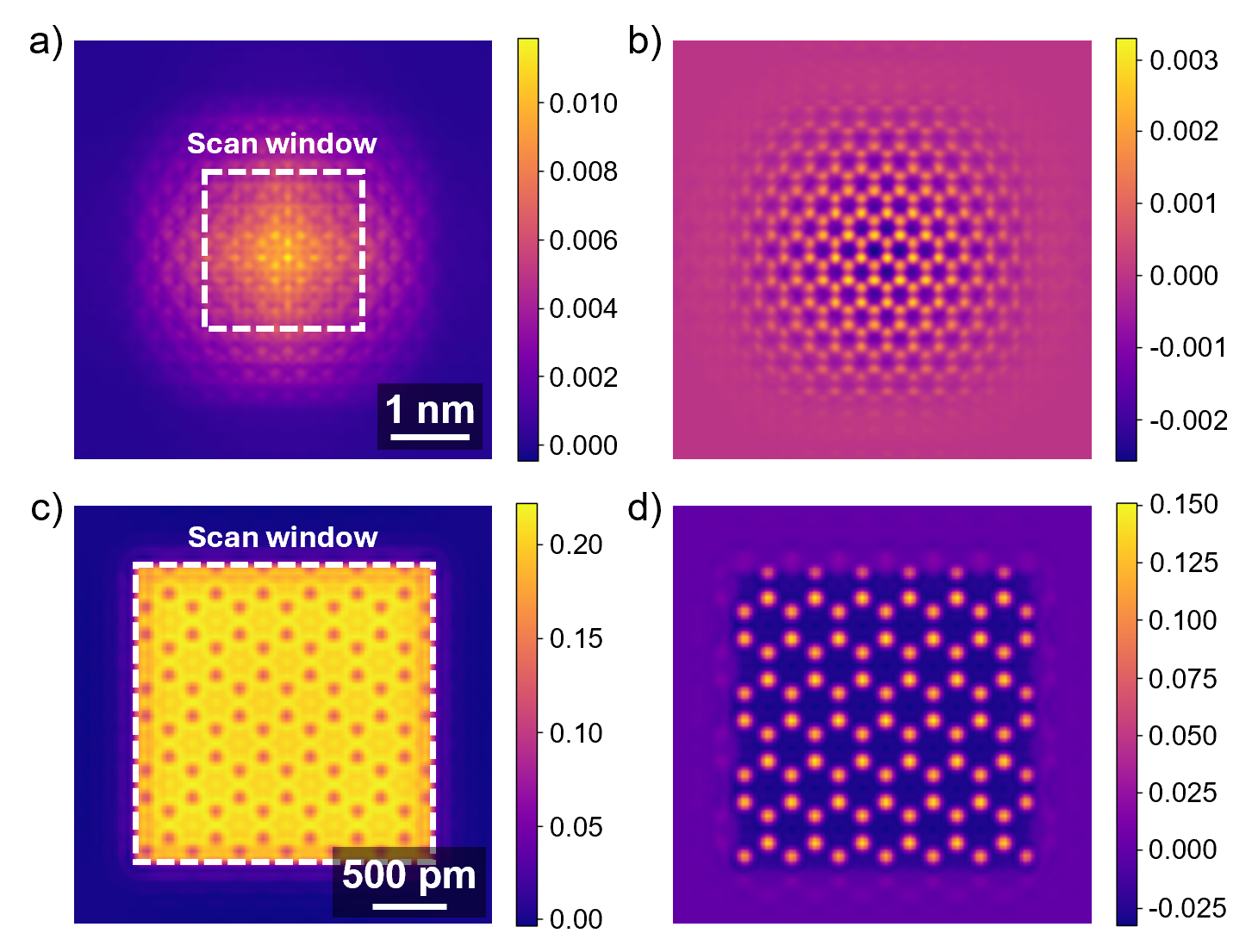}
            \caption{a) Real and b) imaginary parts of the transmission function $T^{WDD}\left(\vec{r}\right)$ retrieved from the overfocused simulation case, described in subsection \ref{subsec:MoS2OF40} and given $N_{e^-}\,=\,+\infty$. The result is here visualized in an extended field of view, reflective of the larger reconstruction window. The scanned area is highlighted as well, as a white dotted square. For comparison, the c) real and d) imaginary parts of the transmission function recovered in the focused-probe case, as described in subsection \ref{subsec:MoS2FP} and also given $N_{e^-}\,=\,+\infty$, are depicted as well.}
            \label{fig:MoS2_OF40_FoV}
        \end{figure}
        
        \begin{figure*}
            \centering
            \includegraphics[width=1.0\textwidth]{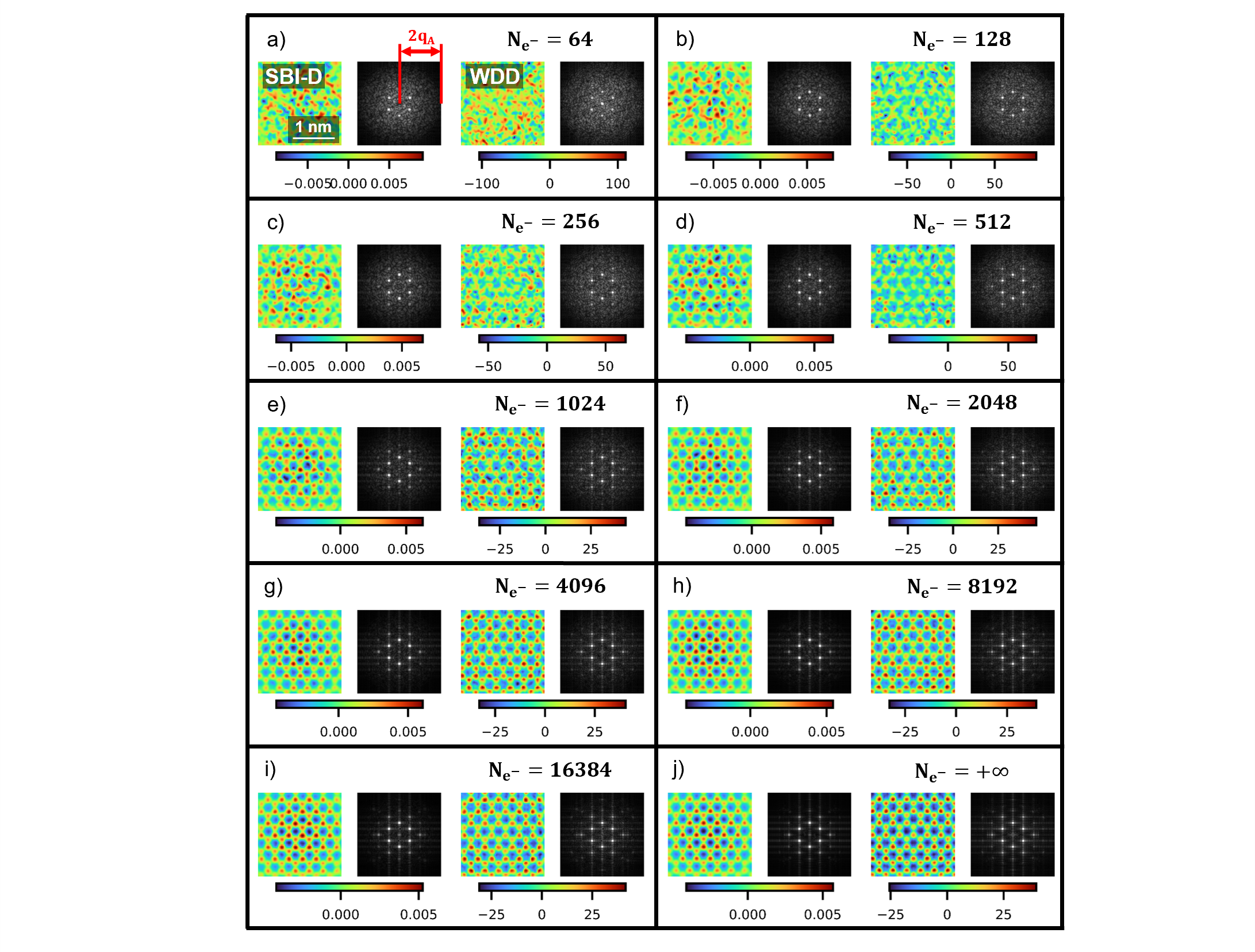}
            \caption{Results of analytical ptychography of monolayer MoS$_2$, applied on the multislice electron diffraction simulation presented in subsection \ref{subsec:MoS2OF40}. Calculations are done for a variety of average numbers of electrons per pattern $N_{e^-}$. For each case, the position-dependent measurement of the projected potential $\mu\left(\vec{r}\right)$, through the SBI-D and WDD methods, is displayed alongside the square root of its Fourier transform's amplitude $\sqrt{\mid\tilde{\mu}\left(\vec{Q}\right)\mid}$. The colorbars reflect values of projected potential in the $\mu\left(\vec{r}\right)$ measurements, in V$\cdot$nm.}
            \label{fig:MoS2_OF40}
        \end{figure*}
        
        \begin{figure}
            \centering
            \includegraphics[width=1.0\columnwidth]{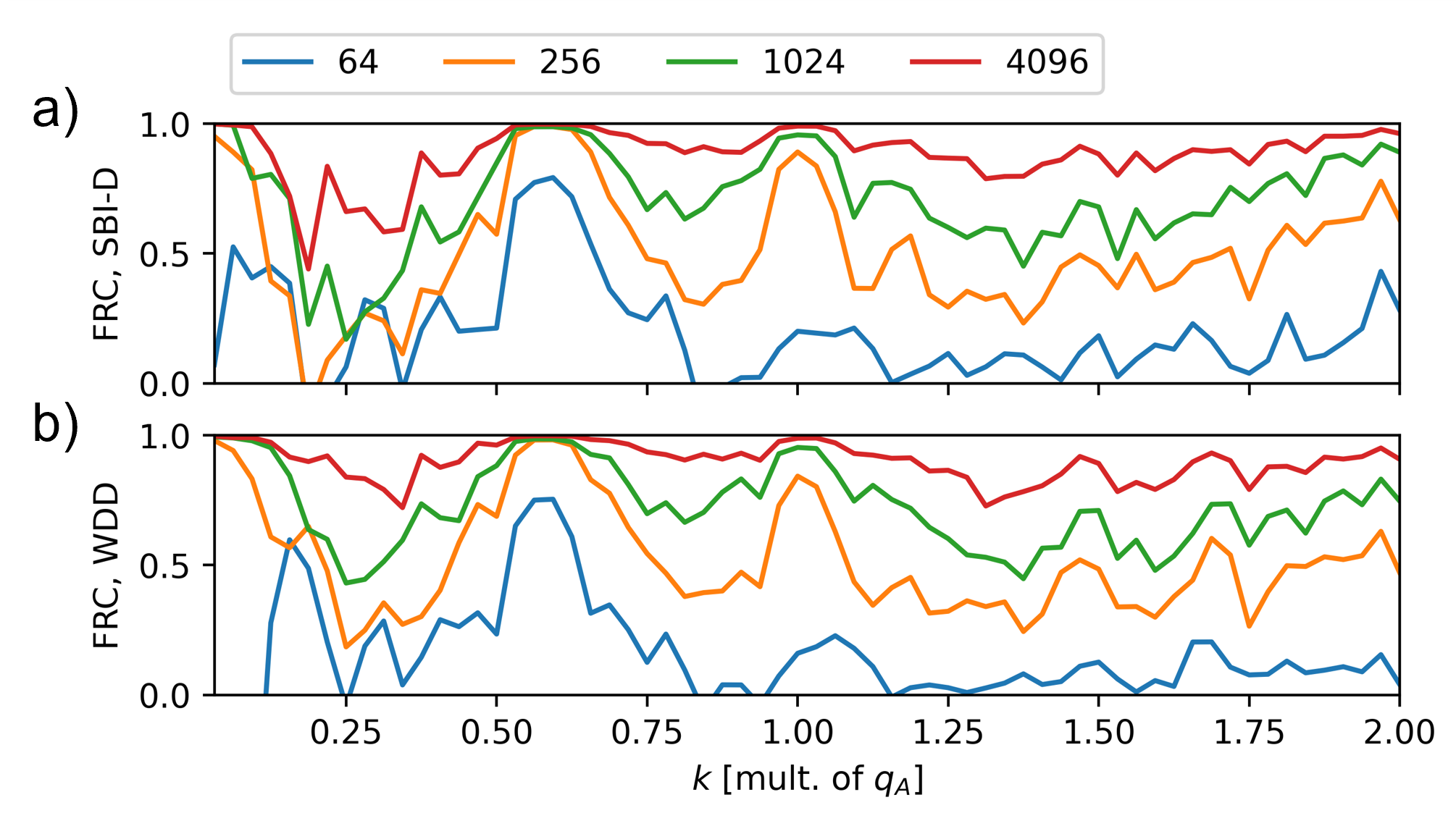}
            \caption{FRC calculated from the $\mu\left(\vec{r}\right)$ measurements presented in fig. \ref{fig:MoS2_OF40}, i.e. by comparing the infinite dose cases to the various dose-limited simulations. The results are plotted as a function of the reference spatial frequency $k$, expressed as a multiple of $q_A$, and given for selected $N_{e^-}$ values. The FRC calculation is displayed in a) for SBI-D and in b) for WDD.}
            \label{fig:FRC_MoS2_OF40}
        \end{figure}
        
        \subsubsection{Simulation and processing parameters}
            
            To test analytical ptychography methods in overfocused illumination conditions, a second simulation was performed in the same conditions as described in subsection \ref{subsec:MoS2FP}, though with an added defocus of 40 nm, and only 16 by 16 scan positions leading to an interval of about 133 pm in the scan grid. The simulation and reconstruction windows were enlarged to avoid artefactual probe self-interference. Noteworthily, the reconstruction of frequencies exceeding the maximum that is allowed, in principle, by the scan interval \cite{Li2014,Pennycook2015} is enabled by the SFPA solution described in subsection \ref{subsec:SFPA}.
            
            The illumination condition leads to $\beta_{\delta \vec{r}_s}\,\approx\,93.8\text{\%}$ between neighboring scan points, as shown in fig. \ref{fig:overlapratio2}. Whereas this somewhat high area overlap was found to be fully sufficient for the reconstruction, another attempt with only 8 by 8 points in the same region, which would have permitted 85.7 \% of overlap, was found to be insufficient. Empirically, this need for a significant value of $\beta_{\delta \vec{r}_s}$ is expected from the literature \cite{Hue2010,Humphry2012,Jiang2018,Song2019} available on the use of defocused probes.
            
            Continuing, because the correction of aberrations is not possible in the conventional framework of iCoM imaging, this method is not used in this subsection. As the number of distinct acquisitions is reduced here, the selection of $N_{e^-}$ values is adapted as well to include numbers 2$^l$ with $l\,\in\,\left[6,7,...,14\right]$. Noteworthily, the increment in the number of electrons per pattern is consistent with the higher complexity of those patterns, as they then constitute shadow images of the specimen \cite{Cowley1979a}.
            
            Given the use of a large illumination including internal features, as shown in the inset of figure \ref{fig:overlapratio2}, formula \ref{eq:dose}, used in the previous section to establish the dose, does not hold anymore. This is because a significant part of the incident intensity on the specimen surface ends up probing the area outside the scan window, hence the dose $D$ serves to recover information from an inhomogeneously sampled surface, which is larger than $S$. This is demonstrated in fig. \ref{fig:MoS2_OF40_FoV}.a,b, where the real and imaginary parts of the reconstructed transmission function $T^{WDD}\left(\vec{r}\right)$, for the $N_{e^-}\,=\,+\infty$ case, are shown in an extended field of view, though still contained in the normal reconstruction window. As can be seen directly, supplementary specimen information is obtained outside of the actual scanned area. A slight inhomogeneity may furthermore appear within the central scanned surface itself, with the regions close to the corners receiving less intensity overall. While it is not very striking, this noticeably occurs in the present case, as can be observed in fig. \ref{fig:MoS2_OF40_FoV}.a, where a cross-like pattern is visible in the real part of the retrieved transmission function.

        \subsubsection{Comparison to the focused-probe case}
            
            Projected potential measurements by WDD and SBI-D, expressed in V$\cdot$nm, are displayed in figure \ref{fig:FRC_MoS2_OF40}, for the specified values of $N_{e^-}$, alongside the corresponding square roots of Fourier transform amplitude. Similarly to the conventional focused-probe case, specimen frequencies are already detected at $N_{e^-}\,=\,64$, although a clear observation of real-space features in the scan window is, arguably, only possible for $N_{e^-}\,\geq\,128$. One further qualitative observation can be made on the noise in the micrographs, which seems more persistent than in those shown in subsection \ref{subsec:MoS2FP}. This is expected, since the use of a delocalized illumination implies an equivalent spread of retrievable information per recording, as explained in the previous paragraph.
            
            Another important difference between the overfocused and the conventional cases is a slight loss of resolution, e.g. consistent with comparisons made in ref. \cite{Jiang2018}. As the true aberration function of the illumination was included in the process, and since the number of single recordings and camera pixelisation were high enough, this cannot be attributed to an insufficiency of available scattering information \cite{Maiden2011,Edo2013} or a processing error. In particular, increasing the number of scan positions to 32 by 32 did not improve the resolution, hence showing no further need in overlap ratio. Another explanation can be found in the inherent information content of the acquired MR-STEM dataset, as determined by the CRLB \cite{Koppell2021,Dwyer2024}. In other words, different illumination conditions, including probe focus \cite{Dwyer2024}, may possess specific capacities to transfer specimen frequencies to the acquired data, hence leading to a supplementary $\vec{Q}$-dependent weighting in the result. In that context, ptychographic reconstructions, with their practical resolution limits, can be expected to remain probe-specific, even when this probe is known or refined in-process \cite{Thibault2009,Maiden2009,Yang2016,Li2025a}.
        
        \subsubsection{Fourier ring correlations}
            
            Moreover, Fourier ring correlations $\text{FRC}^m\left(k\right)$ were calculated for the overfocused probe case and are displayed in fig. \ref{fig:FRC_MoS2_OF40}.a for SBI-D and \ref{fig:FRC_MoS2_OF40}.b for WDD. They show essentially the same features as were observed in fig. \ref{fig:FRC_MoS2_FP}, in particular with four peaks at coordinates $k$ corresponding to specimen frequencies. A difference is however found in the apparently lower dose-efficiency, which can be attributed to the wider illuminated area, as explained above. Noteworthily, this effect is likely amplified by the larger amount of pixels per frequency ranges $R_k$, which is due to the greater size of the reconstruction window, leading to a more important weighting of frequency coordinates without crystal lattice information. This is particularly visible when comparing the first and second peaks to the third and fourth ones.
        
        \subsubsection{Phase shift value range}
            
            Going back to the micrographs themselves, a few more remarks can be made on the ranges of value covered by the projected potential maps. First, in the case of the WDD result, a similar, though slightly smaller, range is obtained as in the focused-probe case. In that context, the reduction can be related to the loss of resolution, and thus to less strongly peaked atomic sites. The average potential in the scan window is also greater. This should nevertheless serve to highlight that the retrieved DC component, i.e. the mean phase shift in the reconstruction window, is arbitrary and only depends, numerically, on the size of the reconstruction window and on the sampling of specimen features, including beyond the scanned area.
            
            The SBI-D result, on the other hand, shows a drastically lower range of values in comparison to fig. \ref{fig:MoS2_FP}, nearly four orders of magnitude down. Interestingly, whereas $\mu^{WDD}\left(\vec{r}\right)$ is unaffected by this problem, as mentioned above, it is not the case for the transmission function itself. To understand this, an important difference between the two analytical ptychography methods should be highlighted again, which is that WDD measures the projected potential in an indirect manner, i.e. by extracting the angle of the initially retrieved $T^{WDD}\left(\vec{r}\right)$, post-use of equation \ref{eq:transmfuncextract}. As such, it is determined by the ratio between its real and imaginary parts, irrespective of the amplitude. A comparison of the $T^{WDD}\left(\vec{r}\right)$ map obtained in the conventional focused-probe case, as shown in fig. \ref{fig:MoS2_OF40_FoV}.c,d, to the one retrieved in the overfocused case, in fig. \ref{fig:MoS2_OF40_FoV}.a,b, is sufficient to confirm the role of this indirect measurement process in avoiding a similar defocus-induced value range issue as met for the SBI-D calculation. In particular, in the overfocused case, the amplitude of the measured transmission function is found to possess values more than an order of magnitude smaller than the focused-probe reconstructions, similarly to the SBI-D case though not as strongly, while the ratio of real and imaginary parts remains roughly the same.
            
            As a supplementary note here, the slice-wise transmission functions actually used for the simulation, one set for each frozen phonon configuration, are all phase objects in the strict sense, i.e. with a constant unitary amplitude. This reflects the absence of absorption effects for the interacting electrons, which is assumed to be fully elastic. The measurement by ptychography, on the other hand, should in general not be expected to fulfill this condition, as mentioned in subsection \ref{subsec:POA}.
    
    \subsection{\textbf{Role of the numerical aperture}}
        \label{subsec:MoS2apert}
        
        \begin{figure*}
            \centering
            \includegraphics[width=1.0\textwidth]{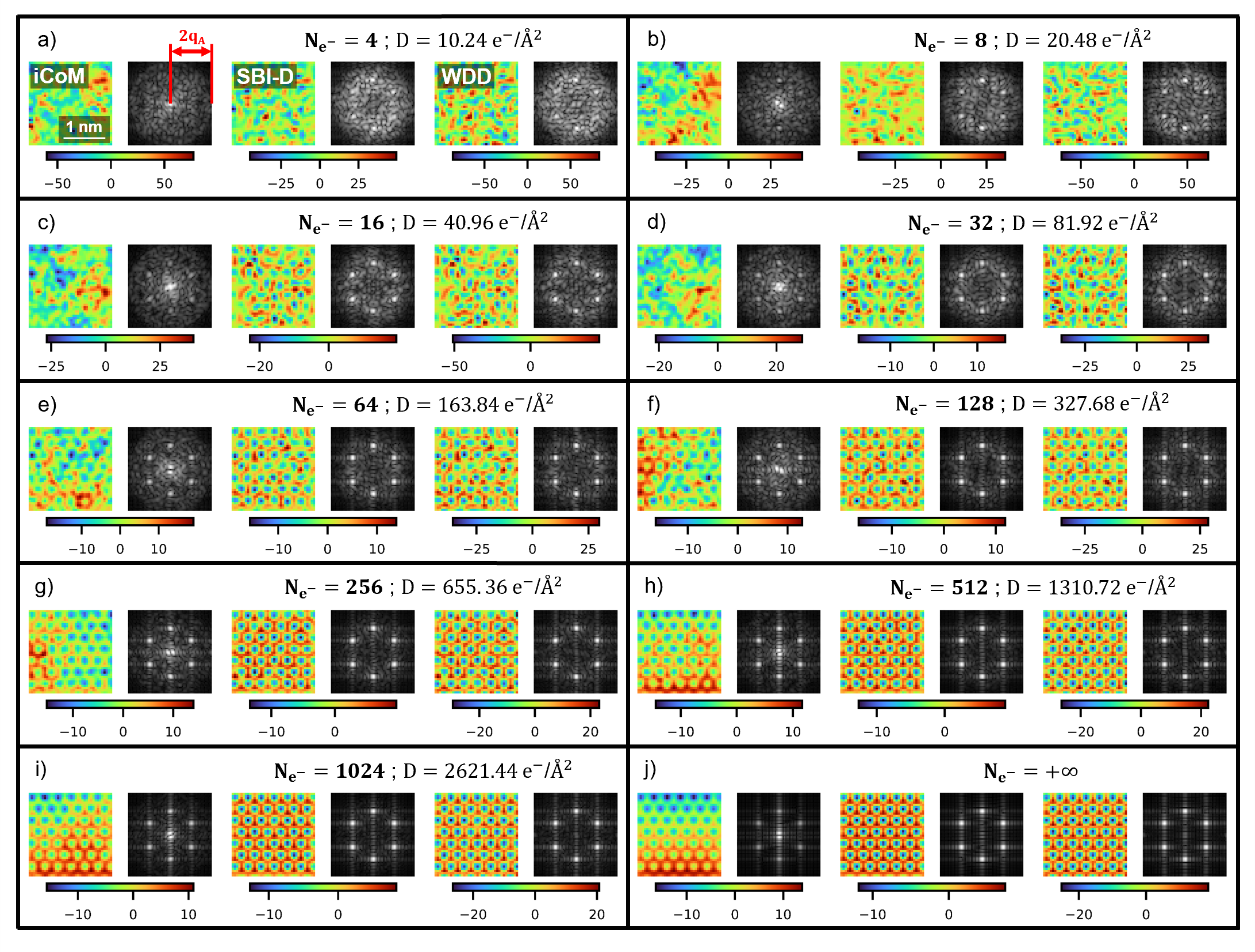}
            \caption{Results of analytical ptychography of monolayer MoS$_2$, applied on the multislice electron diffraction simulation presented in subsection \ref{subsec:MoS2apert}, given $\alpha\,=\,15\,\text{mrad}$. Calculations are done for a variety of average numbers of electrons per pattern $N_{e^-}$, and corresponding doses $D$ given in $e^-/\text{\AA}^2$. For each case, the position-dependent measurement of the projected potential $\mu\left(\vec{r}\right)$, through the iCoM, SBI-D and WDD methods, is displayed alongside the square root of its Fourier transform's amplitude $\sqrt{\mid\tilde{\mu}\left(\vec{Q}\right)\mid}$. The colorbars reflect values of projected potential in the $\mu\left(\vec{r}\right)$ measurements, in V$\cdot$nm.}
            \label{fig:MoS2_FP15}
        \end{figure*}
        
        \begin{figure*}
            \centering
            \includegraphics[width=1.0\textwidth]{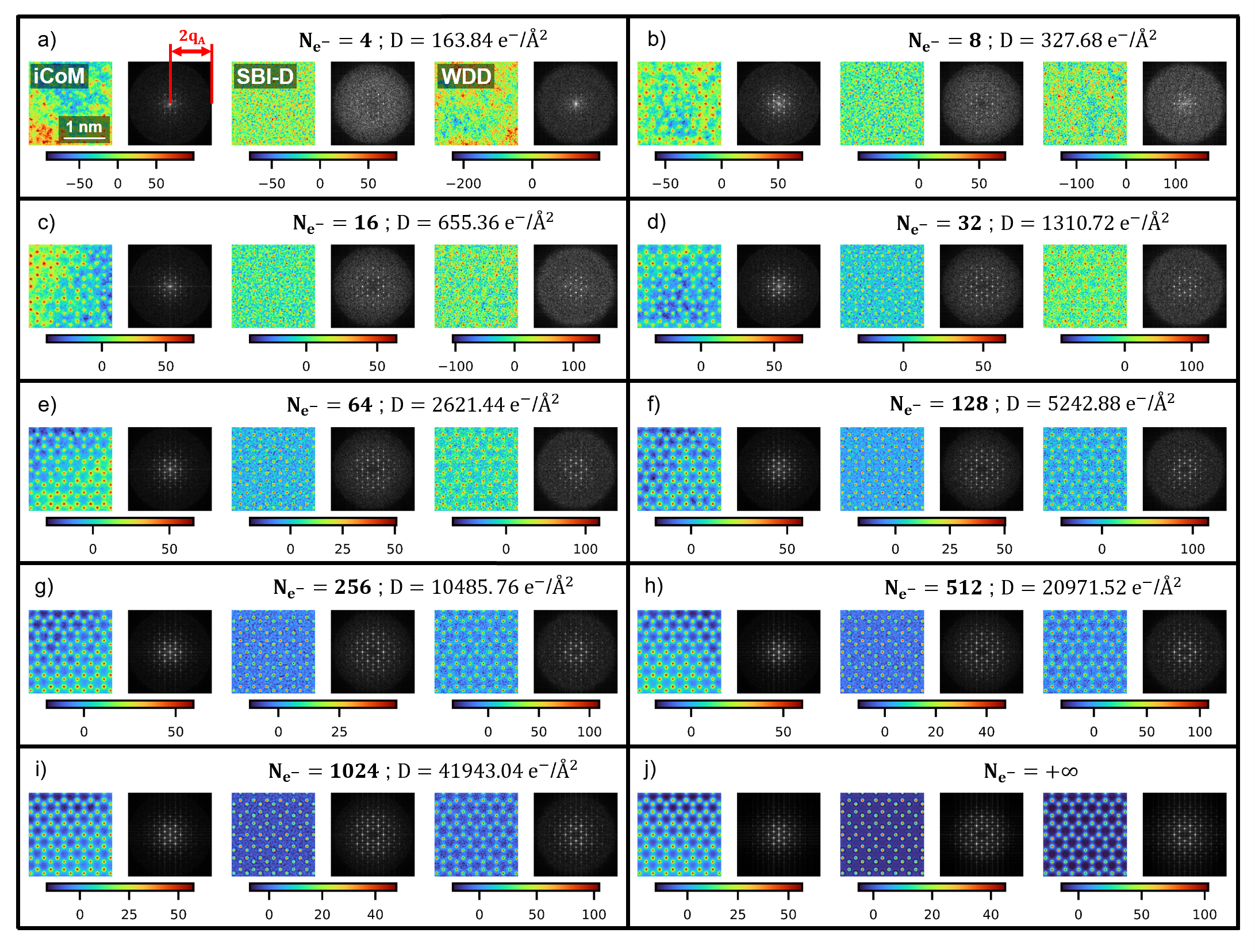}
            \caption{Results of analytical ptychography of monolayer MoS$_2$, applied on the multislice electron diffraction simulation presented in subsection \ref{subsec:MoS2apert}, given $\alpha\,=\,60\,\text{mrad}$. Calculations are done for a variety of average numbers of electrons per pattern $N_{e^-}$, and corresponding doses $D$ given in $e^-/\text{\AA}^2$. For each case, the position-dependent measurement of the projected potential $\mu\left(\vec{r}\right)$, through the iCoM, SBI-D and WDD methods, is displayed alongside the square root of its Fourier transform's amplitude $\sqrt{\mid\tilde{\mu}\left(\vec{Q}\right)\mid}$. The colorbars reflect values of projected potential in the $\mu\left(\vec{r}\right)$ measurements, in V$\cdot$nm.}
            \label{fig:MoS2_FP60}
        \end{figure*}
        
        \begin{figure}
            \centering
            \includegraphics[width=1.0\columnwidth]{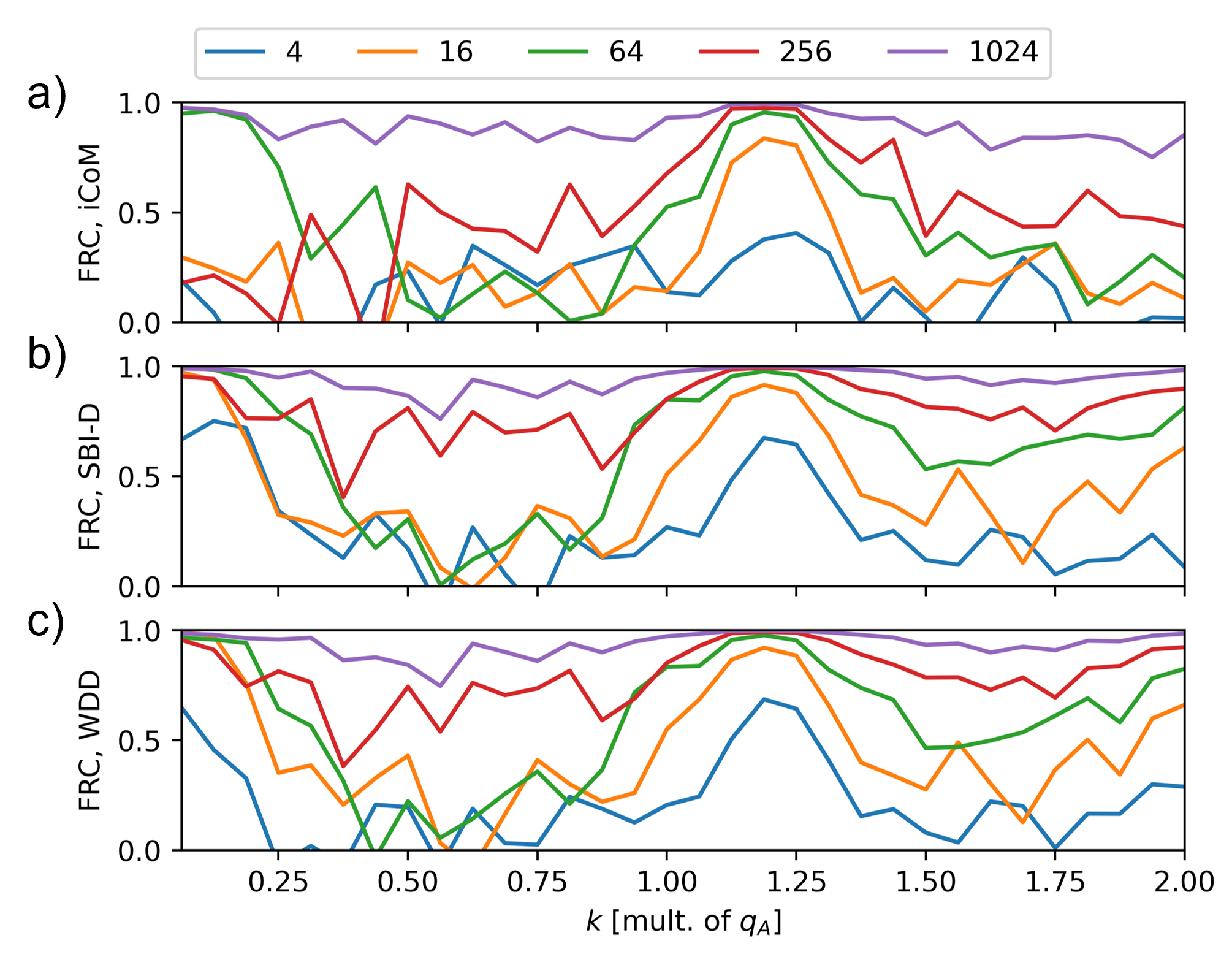}
            \caption{FRC calculated from the $\mu\left(\vec{r}\right)$ measurements presented in fig. \ref{fig:MoS2_FP15}, i.e. by comparing the infinite dose cases to the various dose-limited simulations. The results are plotted as a function of the reference spatial frequency $k$, expressed as a multiple of $q_A$, and given for selected $N_{e^-}$ values. The FRC calculation is displayed in a) for iCoM, in b) for SBI-D and in c) for WDD.}
            \label{fig:FRC_MoS2_FP15}
        \end{figure}
        
        \begin{figure}
            \centering
            \includegraphics[width=1.0\columnwidth]{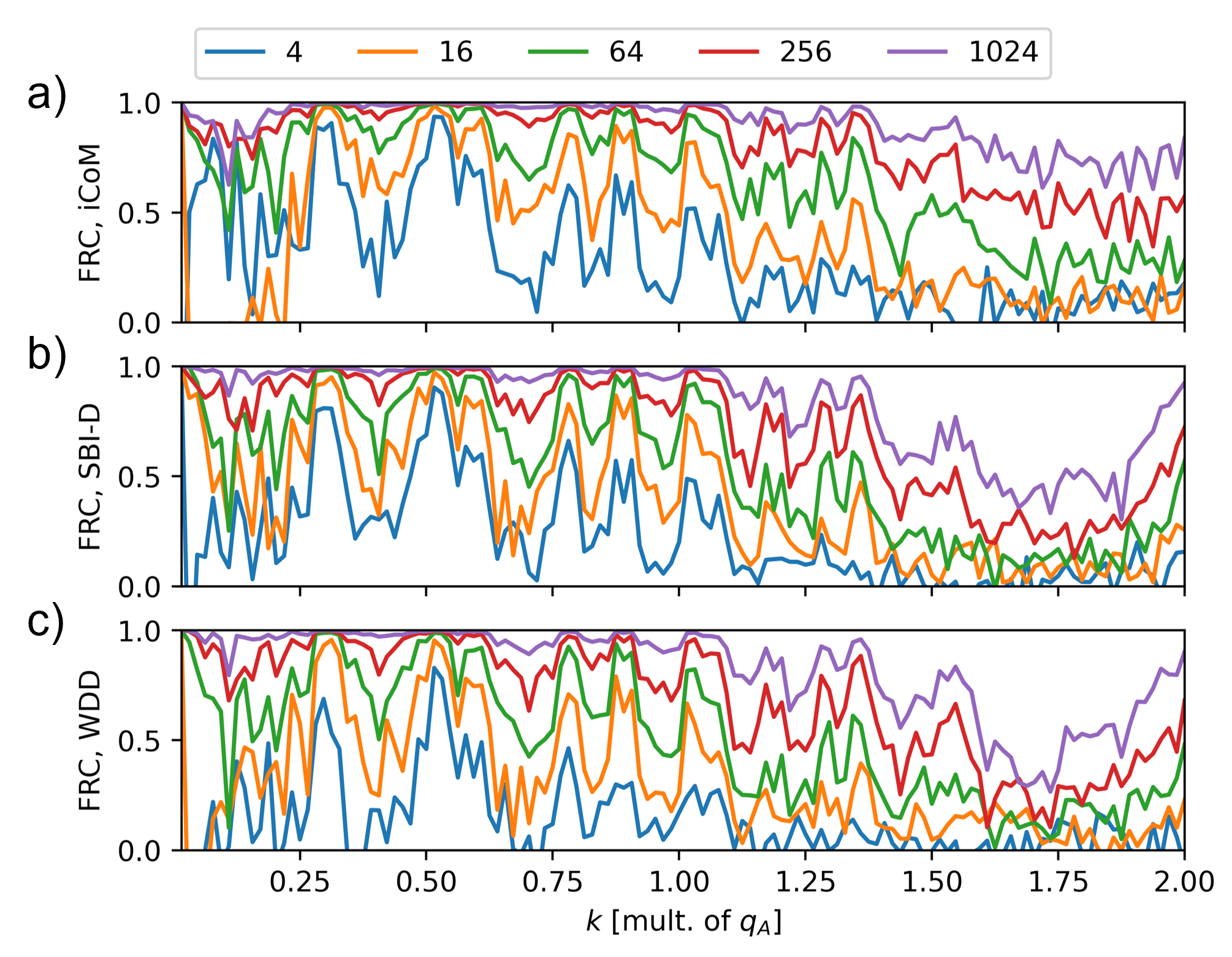}
            \caption{FRC calculated from the $\mu\left(\vec{r}\right)$ measurements presented in fig. \ref{fig:MoS2_FP60}, i.e. by comparing the infinite dose cases to the various dose-limited simulations. The results are plotted as a function of the reference spatial frequency $k$, expressed as a multiple of $q_A$, and given for selected $N_{e^-}$ values. The FRC calculation is displayed in a) for iCoM, in b) for SBI-D and in c) for WDD.}
            \label{fig:FRC_MoS2_FP60}
        \end{figure}
        
        \subsubsection{Simulation and processing parameters}
            
            In analytical ptychography, the range of accessible frequencies is, outside of super-resolution \cite{Sayre1952,Gerchberg1974,Rodenburg1992,Nellist1995,Maiden2011,Humphry2012}, determined strictly by $q_A\,=\,\sin\left(\alpha\right)/\lambda$. In this subsection, an interest is thus taken in how the numerical aperture $\sin\left(\alpha\right)$ affects the dose requirement of the reconstruction. Consequently, two supplementary simulations were performed given $\alpha\,=\,15\,\text{mrad}$, in a scan grid of 32 by 32 points, and $\alpha\,=\,60\,\text{mrad}$, with 128 by 128 scan points. For reference, the resulting unaberrated probes possess Rayleigh criterions $\delta r_{\text{Rayleigh}}$ of 198 and 49 pm, respectively. Other than that, simulation parameters were identical as those described in subsection \ref{subsec:MoS2FP}.
            
            Under those illumination conditions, the relation $\sqrt{N_s}\,\propto\,q_A$, with $N_s$ the total number of scan positions, is fulfilled, which leads to approximately the same $\beta_{\delta \vec{r}_s}$ values in all tested focused-probe cases, including in subsection \ref{subsec:MoS2FP}. The accessible frequency range is however twice smaller in the 15 mrad case, and twice larger in the 60 mrad one. Moreover, for both cases, the same selection of $N_{e^-}$ values was used as for the conventional focused-probe case, hence leading to comparable count sparsity in the exploited CBED patterns. The resulting doses nevertheless differ due to the change in the number of scan points.
            
            Here, it should furthermore be noted that, for most instruments, using a semi-convergence angle of 60 mrad is either not technically possible or leads to an excessive loss of coherence due to chromatic aberration \cite{Kabius2009}. In this publication, the use of such large numerical aperture should thus be regarded as relevant for theoretical verification rather than an immediate experimental horizon, although some work has already been performed in that direction within the last few years \cite{Sawada2015a,Ishikawa2015,Brown2019,Ma2024}.
        
        \subsubsection{Noise level in the micrographs}
            
            The results of applying the iCoM, SBI-D and WDD methods to the 15 mrad simulation are displayed in fig. \ref{fig:MoS2_FP15}, and those of the 60 mrad simulation in fig. \ref{fig:MoS2_FP60}. Owing to the different values of $\alpha$, the indicated $q_A$ differs among the two cases. An immediate consequence of the reduced frequency surface, for the 15 mrad case, is a resolution insufficient to clearly separate two neighboring atomic sites. As such, in fig. \ref{fig:MoS2_FP15}, the crystal lattice is visible only thanks to its hexagonal structure, i.e. the cavity in the middle of an given hexagon can be resolved, and no more than half of the first order of lattice-induced frequencies is transferred. In comparison, for the 30 mrad simulation presented in subsections \ref{subsec:MoS2FP} and \ref{subsec:MoS2OF40}, two orders, i.e. four hexagonal patterns of frequency peaks, were visible in the Fourier transform. Under $\alpha\,=\,60\,\text{mrad}$, the resolution is significantly improved, and up to 11 hexagons can be seen.
            
            Continuing, for both new values of $\alpha$, the specimen frequencies are visible in the Fourier transform, even with excessive noise in real-space, already from the lowest doses introduced. Beyond this, as $N_{e^-}$ increases, the observed level of noise and relative strength of specimen frequencies evolve in a rather similar manner among the three focused-probe cases tested in this section, including fig. \ref{fig:MoS2_FP}, with the reconstruction being nearly noiseless at $N_{e^-}\,=\,1024$. The same method-dependent frequency transfer capacities are furthermore observed in each case, in particular with an important presence of low-frequency artefacts in the iCoM result and slightly more persistent high-frequency noise in the WDD micrograph than for SBI, as explained previously.
        
        \subsubsection{Fourier ring correlations}
            
            Those first qualitative remarks are confirmed by the calculated Fourier ring correlations, shown in fig. \ref{fig:FRC_MoS2_FP15} for the 15 mrad case and in fig. \ref{fig:FRC_MoS2_FP60} for the 60 mrad one. The general behaviour described in subsection \ref{subsec:MoS2FP} is observed for the two newly introduced numerical apertures too. In particular, the $\vec{Q}$-dependent dose-efficiency is higher for $\vec{Q}$-coordinates that are rich in specimen information and lower for the others, where noise reduction is only due to the normalization. Moreover, as is particularly visible in fig. \ref{fig:FRC_MoS2_FP60}, the manner in which $\text{FRC}^m\left(k\right)$ is calculated for each method leads to an artificially lower dose-efficiency for higher $k$, owing to the more important weighting of noisy $\vec{Q}$-coordinates in the corresponding $R_k$ range. It should finally be noted that the micrographs shown in fig. \ref{fig:MoS2_FP15} and \ref{fig:MoS2_FP60} confirm findings from subsection \ref{subsec:MoS2FP} relating to the value range of the retrieved phase shift maps. Specifically, the WDD range is about twice as high as the SBI-D and iCoM ones. Beyond that, the resolution, in leading to more or less pronounced atomic peaks, contributes as well to this effect.
        
        \subsubsection{Dose requirement and reconstructed frequencies}
            
            At a fundamental level, the results presented in this subsection confirm, as was noted by equation \ref{eq:simpleCRLB}, that the overall dose requirement of a ptychographic reconstruction, to obtain a specific precision determined by $\text{CRLB}_{RS}$, is proportional to ${q_A}^{\,\,\,2}$ and more generally to the surface covered by the reconstructed two-dimensional frequency space. This is expected, as a larger frequency surface implies a larger number of pixels to which Poisson noise \cite{Luczka1991} is propagated from the detector plane.
            
            From a naive standpoint, as long as a well-focused probe is employed and that $\sqrt{N_s}\,\propto\,q_A$ is fulfilled, thus permitting the conservation of the same area overlap, this notion also implies that the relation between the average number of electrons per pattern $N_{e^-}$ and the noise level is not fundamentally dependent on the numerical aperture. In other words, it can be expected that, irrespective of the resolution, reconstruction can be performed with very low $N_{e^-}$, and thus with sparse CBED patterns \cite{OLeary2020}, the dose being then fixed by the number of scan points. As such, count sparsity in itself is not a limitation for the reconstruction of the electrostatic potential by analytical ptychography. Noteworthily, if the other common strategy is adopted, consisting in recording in the defocused geometry \cite{Hue2010,Song2019}, $N_{e^-}$ will need to be increased to match the information content of the focused-probe data. A revised area overlap is then also necessary, as shown in subsection \ref{subsec:MoS2OF40}.

\section{Imaging of apoferritin particles under high- and low-resolution conditions}
    \label{sec:ApoApo}
    
    \subsection{\textbf{Contrast predictions above vacuum}}
        \label{subsec:ApoVac}
        
        \begin{figure*}
            \centering
            \includegraphics[width=1.0\textwidth]{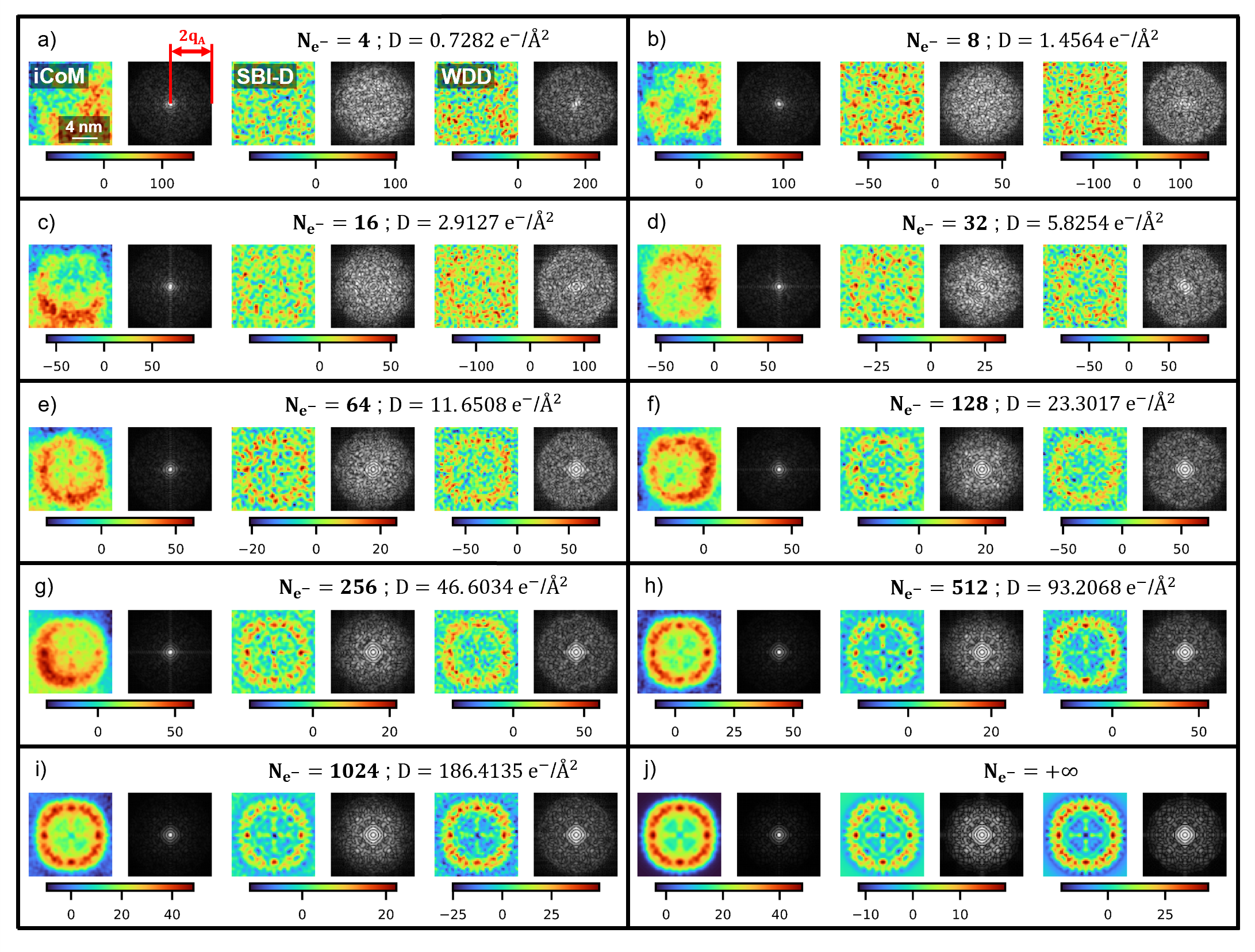}
            \caption{Results of analytical ptychography of apoferritin, applied on the multislice electron diffraction simulation presented in subsection \ref{subsec:ApoVac}, with $\alpha\,=\,1.5\,\text{mrad}$. Calculations are done for a variety of average numbers of electrons per pattern $N_{e^-}$, and corresponding doses $D$ given in $e^-/\text{\AA}^2$. For each case, the position-dependent measurement of the projected potential $\mu\left(\vec{r}\right)$, through the iCoM, SBI-D and WDD methods, is displayed alongside the square root of its Fourier transform's amplitude $\sqrt{\mid\tilde{\mu}\left(\vec{Q}\right)\mid}$. The colorbars reflect values of projected potential in the $\mu\left(\vec{r}\right)$ measurements, in V$\cdot$nm.}
            \label{fig:Apo_1.5_Vac}
        \end{figure*}
        
        \begin{figure*}
            \centering
            \includegraphics[width=1.0\textwidth]{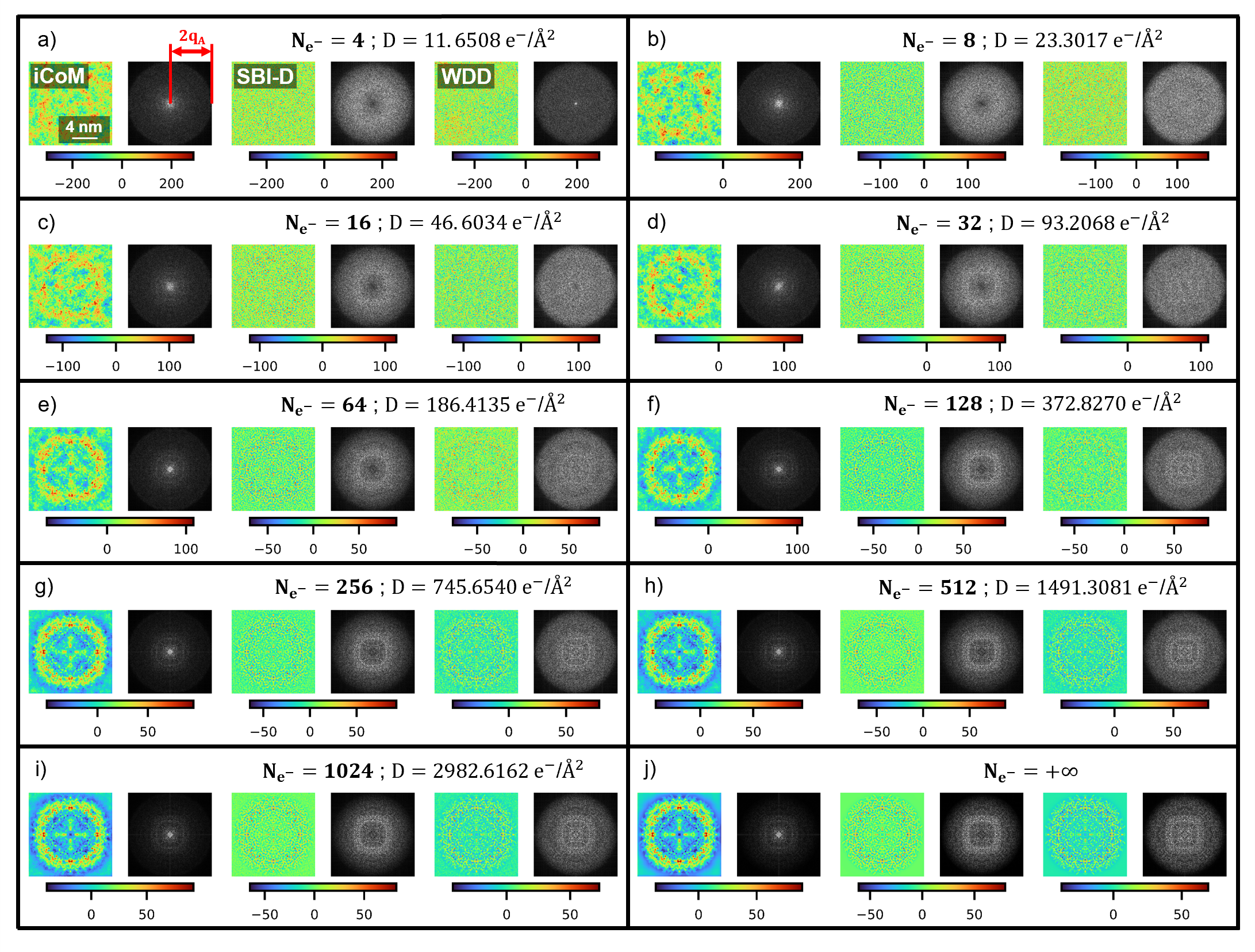}
            \caption{Results of analytical ptychography of apoferritin, applied on the multislice electron diffraction simulation presented in subsection \ref{subsec:ApoVac}, with $\alpha\,=\,6.0\,\text{mrad}$. Calculations are done for a variety of average numbers of electrons per pattern $N_{e^-}$, and corresponding doses $D$ given in $e^-/\text{\AA}^2$. For each case, the position-dependent measurement of the projected potential $\mu\left(\vec{r}\right)$, through the iCoM, SBI-D and WDD methods, is displayed alongside the square root of its Fourier transform's amplitude $\sqrt{\mid\tilde{\mu}\left(\vec{Q}\right)\mid}$. The colorbars reflect values of projected potential in the $\mu\left(\vec{r}\right)$ measurements, in V$\cdot$nm.}
            \label{fig:Apo_6.0_Vac}
        \end{figure*}
        
        \begin{figure}
            \centering
            \includegraphics[width=1.0\columnwidth]{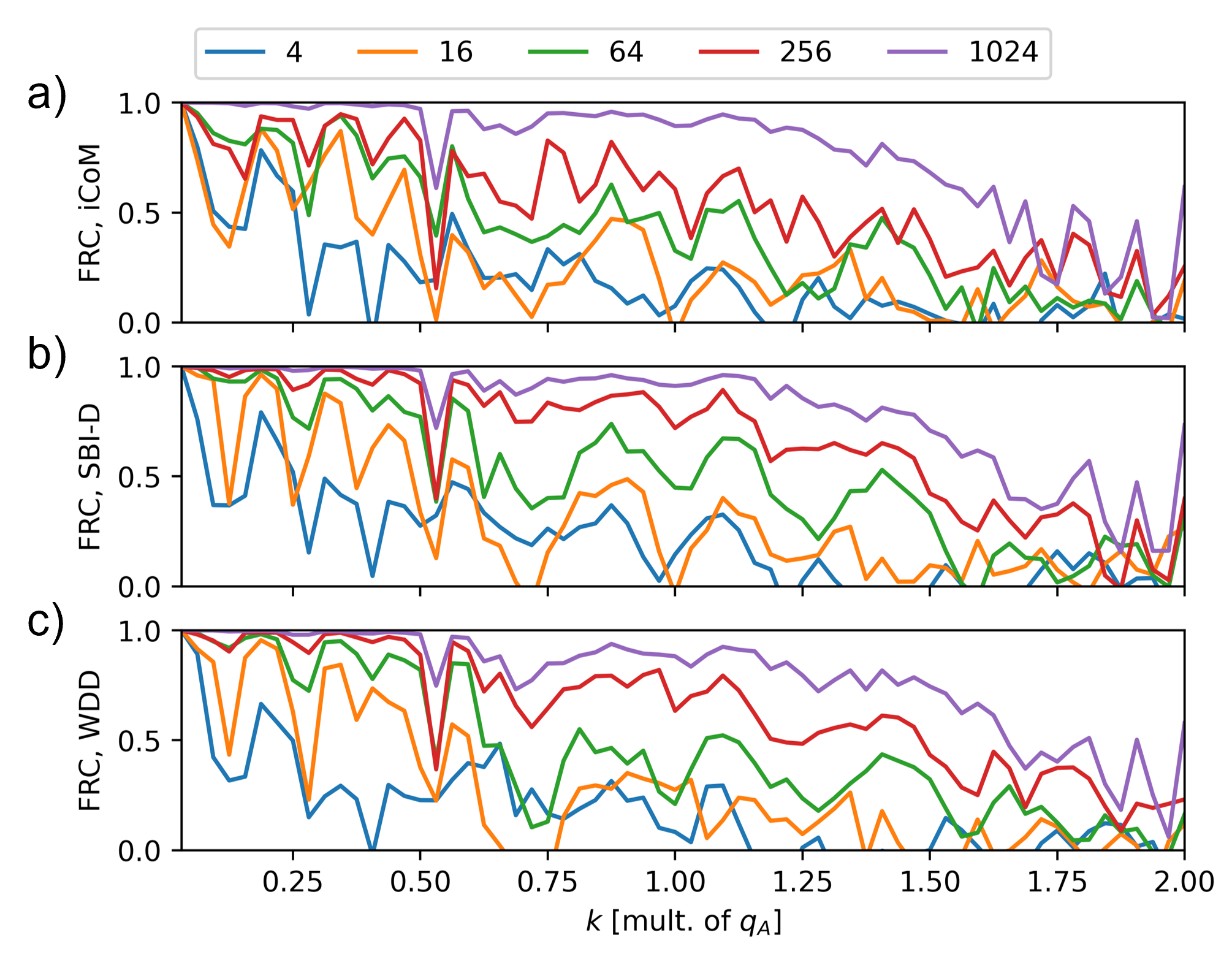}
            \caption{FRC calculated from the $\mu\left(\vec{r}\right)$ measurements presented in fig. \ref{fig:Apo_1.5_Vac}, i.e. by comparing the infinite dose cases to the various dose-limited simulations. The results are plotted as a function of the reference spatial frequency $k$, expressed as a multiple of $q_A$, and given for selected $N_{e^-}$ values. The FRC calculation is displayed in a) for iCoM, in b) for SBI-D and in c) for WDD.}
            \label{fig:FRC_Apo_1.5_Vac}
        \end{figure}
        
        \begin{figure}
            \centering
            \includegraphics[width=1.0\columnwidth]{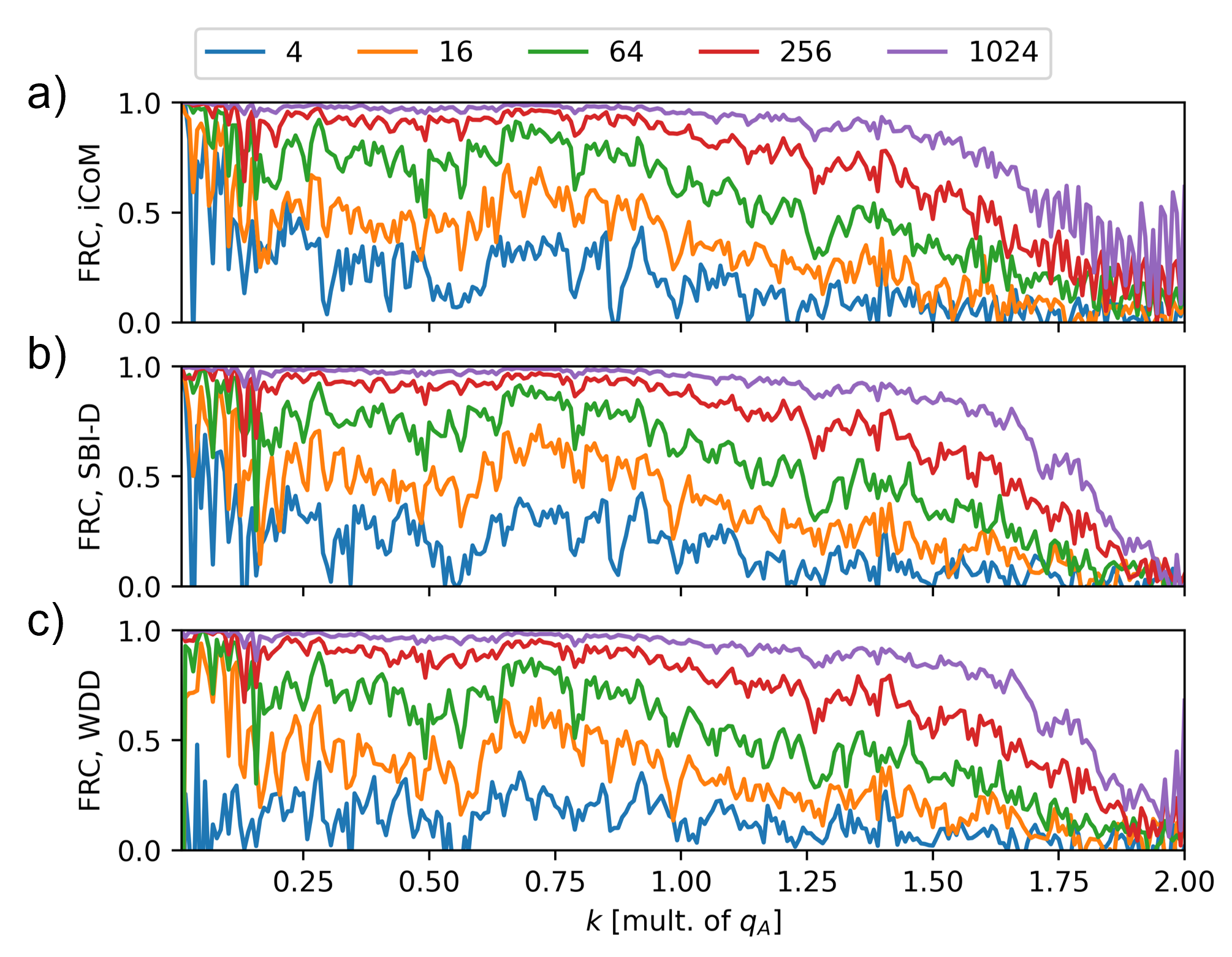}
            \caption{FRC calculated from the $\mu\left(\vec{r}\right)$ measurements presented in fig. \ref{fig:Apo_6.0_Vac}, i.e. by comparing the infinite dose cases to the various dose-limited simulations. The results are plotted as a function of the reference spatial frequency $k$, expressed as a multiple of $q_A$, and given for selected $N_{e^-}$ values. The FRC calculation is displayed in a) for iCoM, in b) for SBI-D and in c) for WDD.}
            \label{fig:FRC_Apo_6.0_Vac}
        \end{figure}
        
        \subsubsection{Difficulties in imaging light matter}
            
            In section \ref{sec:MoS2}, interest was taken in the atomically resolved measurement of the projected potential in a monolayer 2D material which, though it often requires an acceleration voltage $U$ below e.g. 80 kV to avoid excessive knock-on displacement of atoms \cite{Susi2019a,Muller2023}, remains an experimentally realistic endeavour. On the other hand, the critical dose \cite{Egerton2021a} of many beam-sensitive specimens, e.g. biological matter, is in practice too low to permit high-resolution imaging, unless done through the combination of a large number of images from identical objects, i.e. a single-particle analysis (SPA) \cite{Cheng2015,Nakane2020} procedure. In particular, the dose requirement is proportional to the surface of reconstructed frequencies, as was exemplified in subsection \ref{subsec:MoS2apert}. Consequently, electron ptychography performed on viruses and proteins \cite{Pelz2017,Zhou2020,Pei2023,Kucukoglu2024,Mao2024} has so far been focused on retrieving relatively small ranges of frequency components and limited resolutions.
            
            It should also be noted that the amount of electrons needed depends on the imaged specimen itself. Specifically, it depends on the encountered atom types and their scattering cross-section \cite{Bethe1930,Mott1930}, which determines the general amount of observed specimen-induced features in the scattering patterns. In other words, the heavier the imaged material is, the stronger the contrast ends up being in the retrieved phase shift map. This factor, as well as the high probability of radiolysis \cite{Dubochet1988,Henderson1995,Rez2021} leading to the dose limitations mentioned above, make biological specimens particularly difficult to investigate in STEM.
        
        \subsubsection{Simulation and processing parameters}
            
            In order to explore this topic further and, like in the MoS$_2$ case, empirically determine the dose requirement for the imaging of a macromolecule in the absence of further issues, e.g. the MTF of the camera, scan imperfections or an amorphous ice embedding, new simulations were performed based on an apoferritin particle in vacuum. The chosen acceleration voltage $U$ was 300 keV. Two distinct semi-convergence angles $\alpha$ were furthermore tested, specifically 1.5 and 6.0 mrad, to verify the previously observed trends on the role of the numerical aperture. In both cases, a field of view of 15 nm by 15 nm, with the specimen in the center, was employed. This field of view was filled by 64$^2$ scan positions in the 1.5 mrad simulation, and by 256$^2$ positions in the 6.0 mrad one.
            
            Importantly, those illumination conditions both permit an area overlap $\beta_{\delta \vec{r}_s}$ slightly above 82 \%, when comparing immediately neighboring scan points, and lead to Rayleigh criteria $\delta r_{\text{Rayleigh}}$ of about 801 and 200 pm, respectively. Moreover, following a suggestion made e.g. in ref. \cite{Clark2023,Gao2024}, the probe focus was placed in the middle of the vertical distance covered by the specimen. Given the large depths of focus $\delta z_{\text{DOF}} \, = \, \lambda / \left( 2 \sin( \alpha / 2 )^2 \right)$ \cite{VanBenthem2005,VanBenthem2006} of about 1750 nm, for $\alpha\,=\,1.5\,\text{mrad}$, and 109 nm, for 6.0 mrad, this is nevertheless not expected to be critical here. In particular, and also because such a light material is not expected to lead to e.g. strong channeling effects \cite{VanDyck1996}, the wave amplitude should remain sufficiently invariant throughout the propagation axis so that the POA can be considered fulfilled in any case.
            
            Like in the previous section, the multislice method \cite{Cowley1957,Goodman1974,Ishizuka1977} was used to represent the elastic propagation of the electron wavefunction through matter, while the atomic potentials were parameterized according to ref. \cite{Lobato2014}. As before, the specimen potential was pixelated in the two-dimensional plane such that a maximum scattering vector of up to twice the range actually used could be simulated. The object was furthermore sliced in the manner described in ref. \cite{Leidl2023}. To perform the calculations in a reasonable time, owing to the large size of the simulation window, thermal vibrations were accounted for by multiplying the scattering amplitudes with an isotropic Debye–Waller factor, i.e. the wave was considered to interact coherently with a time-average of the atoms in motion. In general, this approximation may lead to errors for high enough scattering angles \cite{VanDyck2009}, e.g. above 40 to 50 mrad, but was not considered to be problematic here, as the extent of $\vec{q}_d$-space available was below this limit.
            
            The results of applying reconstructions on the 1.5 mrad simulation are presented in fig. \ref{fig:Apo_1.5_Vac}, and those of the 6.0 mrad one in fig. \ref{fig:Apo_6.0_Vac}. Dose-limitation was imposed based on the procedure described in appendix B, while employing average numbers of electron per pattern $N_{e^-}$ equal to $2^l$, with $l\,\in\,\left[2,3,...,10\right]$. The resulting doses are indicated in the figures. For all cases, both the measured projected potential map, expressed in V$\cdot$nm, and the square root of its Fourier transform amplitude are displayed. For all cases, the single CBED patterns were normalized to their sum pre-treatment and the SFPA solution, described in subsection \ref{subsec:SFPA} was employed for practical implementation.
        
        \subsubsection{Noise level and contrast transfer}
            
            As a first remark, specimen frequencies are not as obviously observable in the Fourier transforms as in the MoS$_2$ case. In particular, there are no lattice-induced frequency peaks with a width dependent on the size of the scan window to be observed, but rather a complex specimen pattern corresponding to this specific projection of the potential. The overall shape of the particle is also easy to notice in real-space, e.g. from $N_{e^-}\,=\,16$. This is especially true in the iCoM result, where the higher weighting of low-frequencies, high-frequency information having then been reduced, permits an easy detection of the edges \cite{Wu2017,Mahr2022}. This is nevertheless accompanied by a prevalence of low-frequency noise, as explained in the previous section. For the other imaging modes, the two distinct numerical apertures used permit the visibility of a varying degree of details in the inner structure of the specimen.
            
            Furthermore, the CTF $\tilde{\zeta}\left(\vec{Q}\right)$, for SBI, and $\tilde{\gamma}\left(\vec{Q}\right)$, for iCoM, lead to clear differences between the different micrographs, as those two methods highlight specific information in the projected potential map. In contrast, in the case of an atomically resolved crystal where specimen frequencies are sparse, as mentioned above, those effects are not as striking. Upon comparing the two analytical ptychography approaches, it can thus be noticed that the presence of $\tilde{\zeta}\left(\vec{Q}\right)$ leads to an exaggeration of the intermediary frequencies, e.g. close to $\parallel\vec{Q}\parallel\,=\,q_A$, as is especially visible in the high-dose micrographs.
            
            Relating to arguments given in the MoS$_2$ case, as well as in subsection \ref{subsec:SBISBI}, such a clear difference between the SBI-D and WDD results indicates that the imaged specimen cannot be strictly defined as a weak scatterer, i.e. $\tilde{\zeta}\left(\vec{Q}\right)$ does not intrinsically apply while, under the WPOA, it should represent the information content of the scattering data itself and thus occur in all ptychographic imaging modes. This is furthermore confirmed by the value ranges of the WDD phase shift maps themselves, which are found above 0.4 rad for $\alpha\,=\,1.5\,\text{mrad}$ and 0.7 rad for $\alpha\,=\,6.0\,\text{mrad}$, at infinite dose. Such a finding is of particular interest here, as it demonstrates that considering biological specimens as weak phase objects, even with a low value of $\sigma$ following equation \ref{eq:sigma}, may not be correct in the general case.
            
            Outside of those aspects, the $N_{e^-}$-dependent measurement precision is found to be comparable among the two $q_A$ cases, as explained in subsection \ref{subsec:MoS2apert}. This thus implies, following the necessary adaptation of $N_s$ for the conservation of the area overlap, a proportionality between the required dose and the frequency surface being reconstructed, which here extends to ${q_A}^{\,\,\,2}$ directly as was illustrated by equation \ref{eq:simpleCRLB}. The count sparsity of the CBED patterns used is furthermore not a limitation for the reconstruction itself, as expected from previous results. While $N_{e^-}$ increases, internal features of the particle become better resolved, thus here providing a direct empirical verification of the dose requirement of specifically targeted structural information.
        
        \subsubsection{Fourier ring correlations}
            
            Those observations are confirmed by the calculated $\text{FRC}^m\left(k\right)$, as shown in fig. \ref{fig:FRC_Apo_1.5_Vac}, for the 1.5 mrad case, and in fig. \ref{fig:FRC_Apo_6.0_Vac}, for the 6.0 mrad one. Like in section \ref{sec:MoS2}, no striking differences of dose-efficiency are observed among the three imaging modes used in this work, which again is reflective of the comparable needed dose to reach the best achievable precision for a given frequency component. This furthermore illustrate the preponderant role of the CTF in making high- and low-frequency noise more persistent in the WDD and iCoM micrographs, respectively.
            
            As was highlighted previously as well, the FRC shows much more difficulty in reaching high dose-dependent precision for higher frequencies $k$ in general, which can be related to the higher number of pixels in the corresponding $R_k$ range. Moreover, the absence of frequency peaks, owing to the difference of structure in the Fourier transform, observed in the previous paragraph, can be noted here as well and leads to a more homogeneous variation of the FRC, though a specific frequency response is still visible. Relating to figures \ref{fig:Apo_1.5_Vac} and \ref{fig:Apo_6.0_Vac}, it is noteworthy that this specific frequency response, upon comparing the dose-limited cases to the infinite dose reconstruction, can arguably be noticed from e.g. $N_{e^-}\,=\,64$.
        
        \subsubsection{Phase shift value range}
            
            Finally, going back to the micrographs themselves, it should be noted that the different imaging modes, much like for the MoS$_2$ simulations, do not lead to the same general value ranges, though this time the effect depends on the semi-convergence angle $\alpha$ as well. This confirms the hypothesis made in subsection \ref{subsec:MoS2FP} that disagreements in the range of values covered by the retrieved phase shift map, outside of the role of the resolution and of the reconstructed frequency surface, are related to the methods themselves.
    
    \subsection{\textbf{Influence of protective amorphous ice}}
        \label{subsec:ApoIce}
        
        \begin{figure*}
            \centering
            \includegraphics[width=1.0\textwidth]{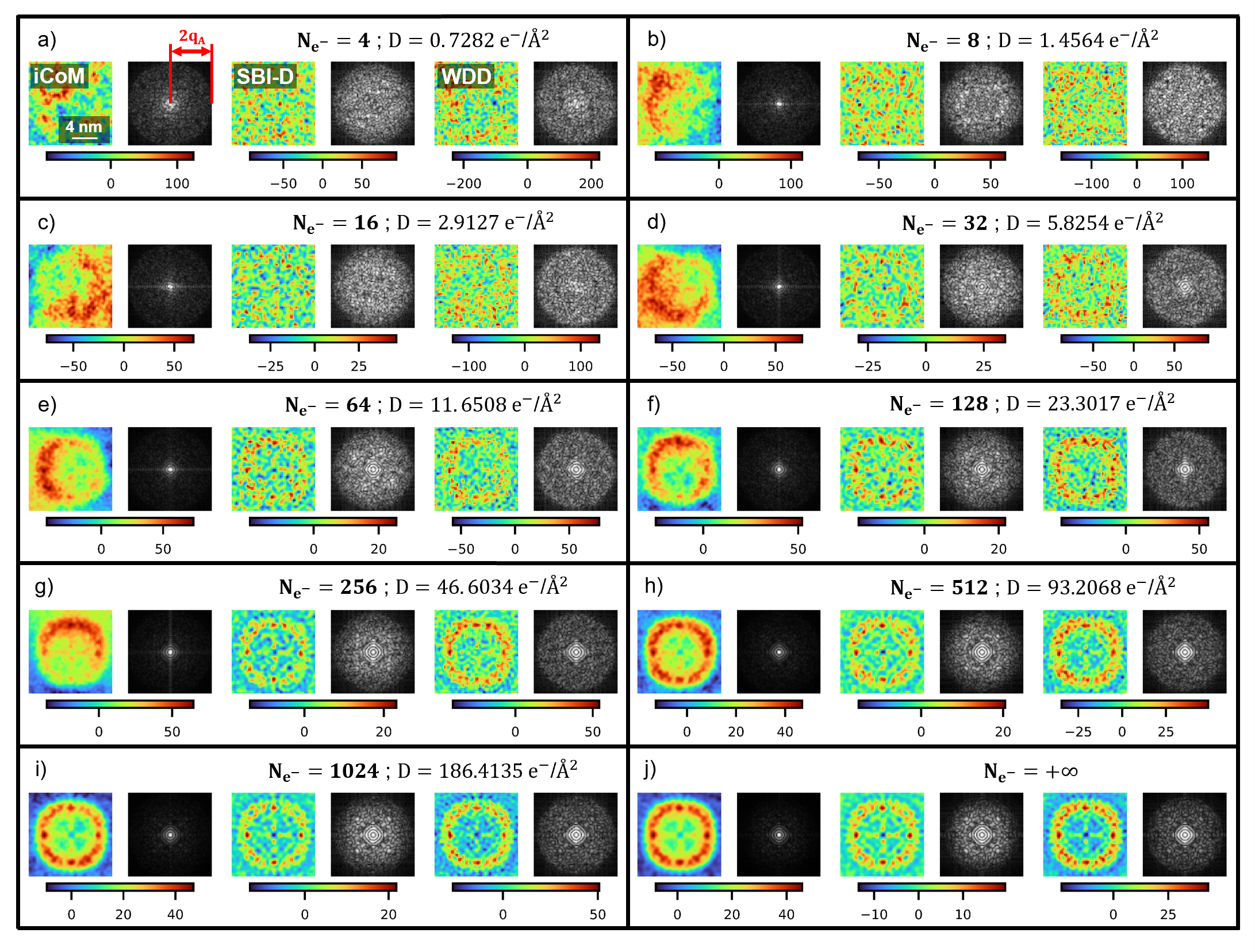}
            \caption{Results of analytical ptychography of apoferritin, applied on the multislice electron diffraction simulation presented in subsection \ref{subsec:ApoIce}, with $\alpha\,=\,1.5\,\text{mrad}$. Calculations are done for a variety of average numbers of electrons per pattern $N_{e^-}$, and corresponding doses $D$ given in $e^-/\text{\AA}^2$. For each case, the position-dependent measurement of the projected potential $\mu\left(\vec{r}\right)$, through the iCoM, SBI-D and WDD methods, is displayed alongside the square root of its Fourier transform's amplitude $\sqrt{\mid\tilde{\mu}\left(\vec{Q}\right)\mid}$. The colorbars reflect values of projected potential in the $\mu\left(\vec{r}\right)$ measurements, in V$\cdot$nm.}
            \label{fig:Apo_1.5_Ice}
        \end{figure*}
        
        \begin{figure*}
            \centering
            \includegraphics[width=1.0\textwidth]{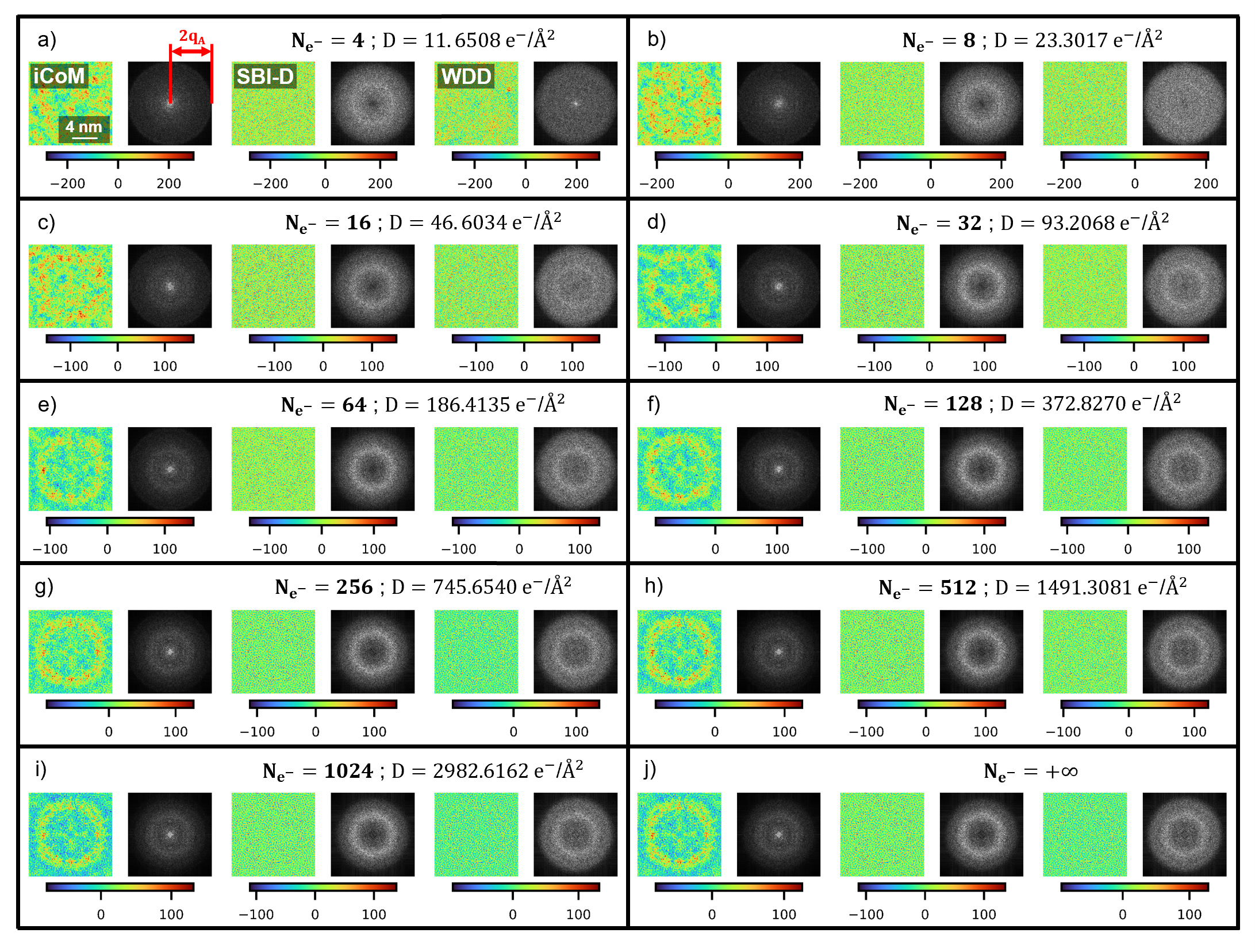}
            \caption{Results of analytical ptychography of apoferritin, applied on the multislice electron diffraction simulation presented in subsection \ref{subsec:ApoIce}, with $\alpha\,=\,6.0\,\text{mrad}$. Calculations are done for a variety of average numbers of electrons per pattern $N_{e^-}$, and corresponding doses $D$ given in $e^-/\text{\AA}^2$. For each case, the position-dependent measurement of the projected potential $\mu\left(\vec{r}\right)$, through the iCoM, SBI-D and WDD methods, is displayed alongside the square root of its Fourier transform's amplitude $\sqrt{\mid\tilde{\mu}\left(\vec{Q}\right)\mid}$. The colorbars reflect values of projected potential in the $\mu\left(\vec{r}\right)$ measurements, in V$\cdot$nm.}
            \label{fig:Apo_6.0_Ice}
        \end{figure*}
        
        \begin{figure}
            \centering
            \includegraphics[width=1.0\columnwidth]{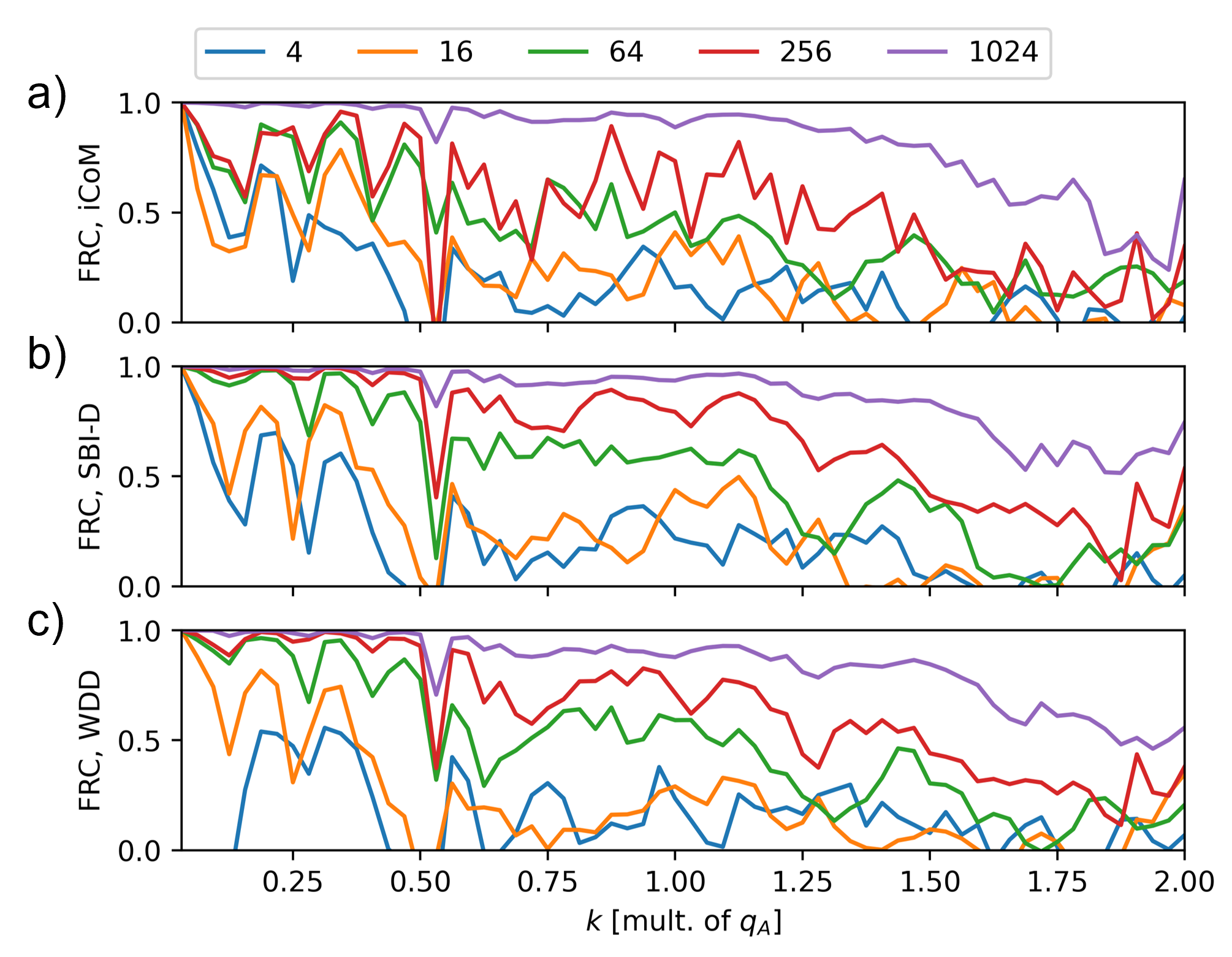}
            \caption{FRC calculated from the $\mu\left(\vec{r}\right)$ measurements presented in fig. \ref{fig:Apo_1.5_Ice}, i.e. by comparing the infinite dose cases to the various dose-limited simulations. The results are plotted as a function of the reference spatial frequency $k$, expressed as a multiple of $q_A$, and given for selected $N_{e^-}$ values. The FRC calculation is displayed in a) for iCoM, in b) for SBI-D and in c) for WDD.}
            \label{fig:FRC_Apo_1.5_Ice}
        \end{figure}
        
        \begin{figure}
            \centering
            \includegraphics[width=1.0\columnwidth]{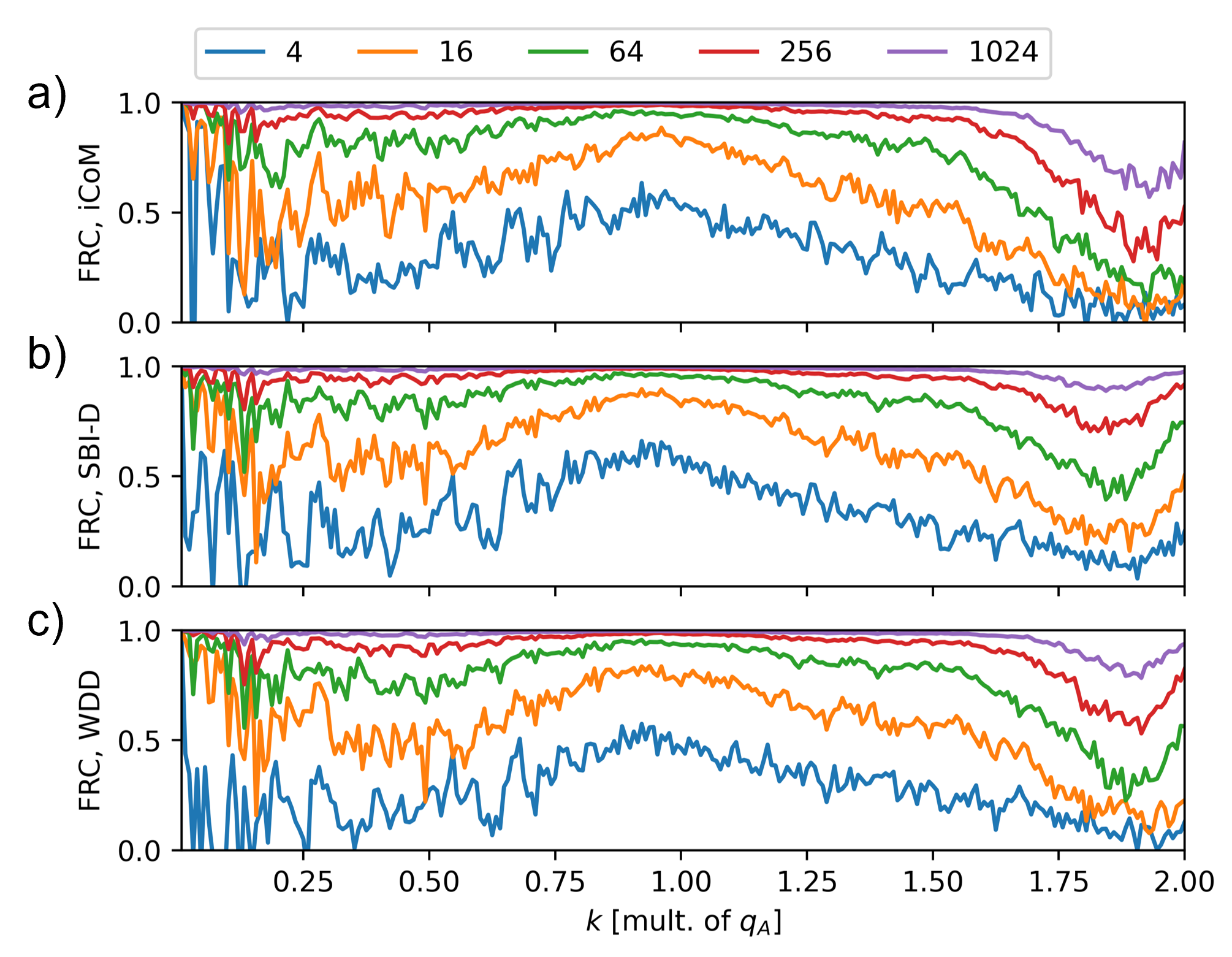}
            \caption{FRC calculated from the $\mu\left(\vec{r}\right)$ measurements presented in fig. \ref{fig:Apo_6.0_Ice}, i.e. by comparing the infinite dose cases to the various dose-limited simulations. The results are plotted as a function of the reference spatial frequency $k$, expressed as a multiple of $q_A$, and given for selected $N_{e^-}$ values. The FRC calculation is displayed in a) for iCoM, in b) for SBI-D and in c) for WDD.}
            \label{fig:FRC_Apo_6.0_Ice}
        \end{figure}
        
        \subsubsection{Simulation and processing parameters}
            
            While subsection \ref{subsec:ApoVac} was sufficient to provide a general prospect for the imaging of biological objects via analytical ptychography, it left out an important practical aspect of such experiments. When performing TEM imaging on this type of specimens, it is common to first embed them in a relatively thick layer of amorphous ice, in order to permit stability in a vacuum environment \cite{Dubochet1988}. This experimental protocol is known for leading to a so-called structural noise \cite{Baxter2009} effect in the result. In particular, as the micrograph should represent a vertical projection of the illuminated object \cite{Vulovic2013}, the frequency distribution of the ice directly adds up to the image spectrum, potentially creating difficulties of interpretation.
            
            In order to investigate this effect further in the case of analytical ptychography, the simulations described in subsection \ref{subsec:ApoVac} were repeated with a specimen consisting of the same apoferritin particle, though this time embedded within a representative amorphous ice layer. The ice and particle ensemble, having a total thickness of about 50 nm, was relaxed via molecular dynamics, as described in ref. \cite{Leidl2023}, before performing the actual multislice calculation. The focus point of the probe was placed in the middle of the object, and other simulation parameters were chosen identically to the in-vacuum simulation cases.
        
        \subsubsection{Noise level and contrast transfer}
            
            Results of applying the imaging methods to the new simulations are displayed in fig. \ref{fig:Apo_1.5_Ice}, for $\alpha\,=\,1.5\,\text{mrad}$, and in fig. \ref{fig:Apo_6.0_Ice}, for 6.0 mrad. In general, the remarks made in subsection \ref{subsec:ApoVac} can be transferred to this second situation as well. In particular, the visualization of the overall specimen structure in real-space is possible from about $N_{e^-}\,=\,16$ and the CTF of iCoM and SBI lead to very different final images among the methods. The main difference is the presence of the projected potential of the ice, superposed to the contribution of the apoferritin particle and thus leading to the deterministic noise-like effect \cite{Baxter2009} mentioned in the last paragraph.
            
            An important difference in its overall influence should furthermore be noted between the two semi-convergence angles employed here. Specifically, while the 1.5 mrad case does not show a very striking loss of contrast due to the ice, as can be observed by comparing it to fig. \ref{fig:Apo_1.5_Vac}, the 6.0 mrad one is affected much more strongly. In particular, in the SBI-D and WDD images, the particle is nearly not visible at all anymore, in contrast to fig. \ref{fig:Apo_6.0_Vac} where its inner structure was well-resolved even at relatively low doses.
            
            As can be directly noticed in the Fourier transforms, the frequency spectrum of the amorphous ice, which possesses a ring-like shape owing to its amorphous structure \cite{Vulovic2013,Parkhurst2024}, is added to the specimen frequencies, thus obstructing them in the resulting image. It is then clear that, while the value of 1.5 mrad is sufficiently small to mostly cut off the affected $\vec{Q}$-coordinates, then found beyond the $2 q_A$ limit, it is not so in the 6.0 mrad case. In this context, due to its large thickness compared to the apoferritin itself, the amorphous ice furthermore ends up dominating the projected potential measurement, and thus preventing a direct interpretation of the micrograph. Noteworthily, the iCoM method, in fig. \ref{fig:Apo_6.0_Ice} is the least affected of the three, which is related to its CTF attenuating higher frequency components.

            In general, those findings show that, for the imaging of a single ice-embedded biological object, not only does the frequency distribution of the amorphous ice has to be known beforehand \cite{Vulovic2013}, but the numerical aperture may need to be adapted as well in order to obtain an interpretable micrograph. That is, unless further post-processing is employed like in the context of SPA \cite{Cheng2015,Nakane2020}. It is moreover clear that, if the WPOA does not strictly apply to a protein particle standing in vacuum, it will be the case as well for its ice-embedded version.
        
        \subsubsection{Fourier ring correlations}
            
            For completeness of the arguments, FRC calculations were performed for both cases and are displayed in figures \ref{fig:FRC_Apo_1.5_Ice} and \ref{fig:FRC_Apo_6.0_Ice}, respectively for 1.5 and 6.0 mrad. As expected, the results displayed in fig. \ref{fig:FRC_Apo_1.5_Ice} do not show obvious differences from those in fig. \ref{fig:FRC_Apo_1.5_Vac}, while the FRC profiles in fig. \ref{fig:FRC_Apo_6.0_Ice} have completely changed from their in-vacuum version, as displayed in fig. \ref{fig:FRC_Apo_6.0_Vac}. Specifically, at the coordinates $k$ where the ice-induced ring of frequencies is highest, the fine structure of specimen-related information has been largely replaced by a near-homogeneous response.

\section{Discussion}
    
    \subsection{\textbf{Comparison of frequency transfer capacities and role of the interaction model}}
        \label{subsec:freqtranscap}
        
        \subsubsection{Generalizable contrast transfer function for a weak scatterer}
            
            The three STEM-based phase retrieval techniques used in this work, encompassing the iCoM approach and the two existing analytical ptychography methods, can be distinguished in how well they recover specimen information at reconstructible spatial frequencies $\vec{Q}$. In particular, they may be attributed contrast transfer functions, denoting a $\vec{Q}$-dependent attenuation of signal-to-noise ratio. In principle, such frequency-wise reductions of the object spectrum can be solved by deconvolving the result with the predicted point-spread function. This is however difficult in the low-dose case, as the concerned $\vec{Q}$-coordinates may then have been brought below the noise level, hence resulting in significant noise amplification upon deconvolution. As a result, the dose-efficiency is then expected to be much worse for frequency components where the CTF has a low value.
            
            A first step towards determining the CTF is the derivation of the intrinsic phase contrast transfer function $\tilde{\zeta}\left(\vec{Q}\right)$ \cite{Yang2015b}, depicted in figure \ref{fig:CTF_SBI_S}, occurring when the illuminated specimen is a weak phase object and being then reflective of a sideband-like geometry in the acquired scattering data. Owing to this geometry, the SBI method \cite{Pennycook2015,Yang2015b} constitutes an optimized approach for the reconstruction of the phase shift map, where parts of the data containing only noise are excluded as explained in subsection \ref{subsec:SBISBI}.
            
            More generally, when the WPOA is fulfilled, the PCTF $\tilde{\zeta}\left(\vec{Q}\right)$ is applicable to all methods investigated in this work, which implies the equality of the WDD and SBI results, given a high enough dose. The known form of the PCTF, and the resulting $\vec{Q}$-dependent noise level in SBI \cite{Seki2018}, furthermore make it possible to establish a noise normalization strategy \cite{Seki2018,OLeary2021}, rendering the noise level homogeneous across the full frequency spectrum of the retrieved object.
            
            In the particular case of iCoM imaging, and as explained in subsection \ref{subsec:iCoMiCoM}, a supplementary frequency weighting is imposed, following the optical transfer function $\tilde{\gamma}\left(\vec{Q}\right)$ \cite{Black1957,Muller2014} shown in fig. \ref{fig:CTF_iCoM}. In contrast to the PCTF arising in the case of a weak scatterer, this OTF is due to the much simpler measurement method based on the prior calculation of the average momentum transfer at each scan position, and does not represent the information content of the experiment. As such, under the WPOA, both $\tilde{\zeta}\left(\vec{Q}\right)$ and $\tilde{\gamma}\left(\vec{Q}\right)$ can be expected to apply to the iCoM result.
        
        \subsubsection{Specimen-dependent frequency transfer}
            
            Continuing, from the results presented in this work, it should nevertheless be clear that the WPOA is inappropriate in the general case. Specifically, it was shown, for both the monolayer MoS$_2$ and the apoferritin model-objects, that not only did the SBI and WDD results differed in a clear manner, but also that the retrieved ranges of phase shift exceeded those of weak scatterers. What this then implies is the inapplicability of the derived $\tilde{\zeta}\left(\vec{Q}\right)$ as a general PCTF in focused-probe ptychography. In this context, it is noteworthy that, due to its process still being based on the assumption of a weak phase object, the SBI method may then remove useful specimen information from the available scattering data rather than just noise, and still imposes $\tilde{\zeta}\left(\vec{Q}\right)$ as a CTF \cite{OLeary2021,Clark2023}.
            
            This is not the case of the WDD method, which only assumes the more general POA, and consists in a complete deconvolution of the four-dimensional Wigner distribution $\Gamma\left(\vec{Q}\,;\,\vec{R}\right)$ of the illumination from the scattering data. As such, and in contrast to iCoM and SBI, its process may be expected to lead to no supplementary $\vec{Q}$-wise reduction in signal-to-noise ratio, at least outside of the frequency-dependent CRLB \cite{Koppell2021,Dwyer2024}, then representing the fundamental information content of the data.
            
            At a deeper level, what this means is that, while the WPOA allowed for the extraction of a specimen-independent frequency transfer capacity in ptychography, the POA alone does not permit this simplification. This then implies that the empirical CTF of the WDD method should always be expected to depend on the illuminated specimen and its scattering power \cite{Wei2020,Bouchet2021}. In general, more work on the CRLB as a theoretical precision metric will be needed in the future, as it can be used to derive inhomogeneities along $\vec{Q}$-space, for different specific cases. Noteworthily, this encompasses the role of the aberration function which, even when included in the process, still affects the reconstruction \cite{Dwyer2024}, as exemplified in subsection \ref{subsec:MoS2OF40}.
        
        \subsubsection{Further limitations}
            
            Continuing on the WDD method, the absence of a truly process-induced frequency attenuation effect may remain true only as long as the assumed interaction model, i.e. the fully coherent POA, is fulfilled. In this condition, equation \ref{eq:truthofWigner} is correct and may be used as an accurate basis for the treatment of the scattering data. A first practical limit is the partial coherence of the illumination, which imposes an envelope effect in $\Gamma\left(\vec{Q}\,;\,\vec{R}\right)$ \cite{Nellist1994}.
            
            Furthermore, in the case where the specimen is too thick to be accurately described as a phase object, but a ptychographic reconstruction is still performed on the basis of a single transmission function, artificial features may then be introduced in the result \cite{Plamann1998,Yang2017,Clark2023}. In this context, it is also noteworthy that, even with a thicker specimen, using an optimally focused illumination has been shown to partly alleviate the artificial features mentioned above \cite{Clark2023,Gao2024}, as is also well-known in the case of CoM and DPC \cite{Close2015,Addiego2020,Burger2020,Robert2021,Liang2023}. In iterative ptychography, another increasingly popular solution is the inclusion of a multislice propagation within the process \cite{Maiden2012,Tsai2016,Gao2017,Chen2021}, i.e. the use of a more accurate interaction model.
        
        \subsubsection{Use of dark field electrons and super-resolution}
            
            One more advantage of the WDD method, in comparison to SBI, is its ability to exploit the dark field electrons, which are otherwise neglected under the WPOA, as shown by equations \ref{eq:FTalongscanWPOA} and \ref{eq:sidebandsideband}.
            
            Importantly, in focused-probe ptychography, the use of the scattering vectors above $q_A$ is also necessary to achieve super-resolution \cite{Sayre1952,Gerchberg1974,Rodenburg1992,Nellist1995,Maiden2011,Humphry2012}, i.e. the ability to enhance resolution in real-space by accessing spatial frequencies that extend beyond the diffraction limit. The spectrum of the object is then completed outside of the conventional $2 q_A$ range, by exploiting the relation of the intensity scattered outside the primary beam with those initially missing frequencies. In practice, for analytical ptychography, this is done through the stepping out approach \cite{Rodenburg1992,Li2014}.
            
            Because of its ability for super-resolution, WDD has the potential to access much larger frequency ranges than SBI and iCoM, though this requires a significant amount of electrons to be present in the dark field, which makes this approach very expensive in terms of dose \cite{Jiang2018}. As the present work focuses on the imaging of beam-sensitive specimens, that topic was left out of it.
    
    \subsection{\textbf{Other aspects of the reconstruction strategy for low-dose imaging}}
        
        \subsubsection{Dose requirement}
            
            Overarchingly, this publication verifies that the dose requirement of ptychography is proportional to the frequency surface to be reconstructed. In the case where no super-resolution \cite{Sayre1952,Gerchberg1974,Rodenburg1992,Nellist1995,Maiden2011,Humphry2012} is sought, this proportionality extends directly to ${q_A}^{\,\,\,2}$.
            
            As illustrated by equation \ref{eq:simpleCRLB}, this furthermore implies that the total number of detected electrons, needed to reach a certain accuracy in the real-space measurement, should be expected to be proportional to the quantity of reconstructed frequency pixels $N_{\vec{Q}}$. What this then means is that, in general, to achieve a pre-defined signal-to-noise ratio in the micrograph, the numerical aperture has to be adapted to the critical dose of the imaged specimen \cite{Egerton2013,Egerton2021a}.
            
            Furthermore, while super-resolution may be of interest for many other applications, it should realistically not be relied on for the low-dose imaging of beam-sensitive objects, as it is based on exploiting the least intense scattering vectors across the far-field. Consequently, enhancing the resolution when imaging such specimens should rather be done by enlarging the numerical aperture itself, as this then represents the least dose-expensive option.
        
        \subsubsection{Normalization}
            
            An appropriate normalization choice is also important in analytical ptychography, especially in the case where the acquired data is sparse \cite{OLeary2020}, and thus where large changes in the variance of single patterns occur across the scan window. For this purpose, the strategy proposed in ref. \cite{Seki2018}, consisting in dividing the acquisitions by their individual sums pre-treatment, was adopted in this work.
            
            Noteworthily, this pattern-wise approach is equivalent to performing the calculation while adapting the normalization of the electron wavefunction itself to the number of counts in each corresponding pattern. Further investigations on the normalization strategy may otherwise be relevant in the future, which will also need to be correctly accounted for in any estimation of the theoretical measurement precision, e.g. using the CRLB \cite{Rao1945}.
        
        \subsubsection{Use of a Wiener filter}
            
            Finally, in the case of the WDD and SBI-D processes, the use of a Wiener filter \cite{Bates1986} as a deconvolution method implies the introduction of a parameter $\epsilon$ to avoid divisions by zero, as included in equations \ref{eq:oldstyleWDD} and \ref{eq:SBIdeconv}. Whereas, at infinite dose, this number may be considered as a simple numerical precision term, it in practice needs to be adapted to the noise level in the distributions $\tilde{J}_{\vec{Q}}\left(\vec{q}_d\right)$ and $J_{\vec{Q}}\left(\vec{R}\right)$ to avoid its amplification in the final result. This is however at the cost of accuracy for the reconstruction itself, i.e. the range of phase shift values and the transfer of higher frequencies are affected, as noted in more details e.g. in ref. \cite{Susi2025}.
            
            In this work, a single $\epsilon\,=\,10^{-6}$ was consistently used in all deconvolutions, which was found sufficient to avoid noise amplification, including in the condition of highest count sparsity, or an unwanted modification of the value range in the retrieved phase shift maps, as verified in the infinite dose case.
            
            High values of $\epsilon$, up to 10$^{-3}$, were tested as well in the WDD reconstruction, with no further reductions in the noise level, but leading to excessive modifications of the result in high-dose conditions. In the case of SBI, comparisons were performed between the deconvolutive form, using the Wiener parameter of 10$^{-6}$, and the summative form. This was done for the few lowest doses considered and did not show a better signal-to-noise ratio in the SBI-S result, hence confirming the stability of the deconvolutive process in this case. There too, higher values did not permit an improved noise level compared to the 10$^{-6}$ case. For both the WDD and SBI-D processes, $\epsilon\,=\,10^{-7}$ furthermore led to a slight, but clear, amplification of the noise.
            
            Importantly, the stability of both the WDD and SBI-D calculations, even with very sparse scattering data and given no adaptation of the parameter to the dose, can be related in large part to the scan position-wise normalization strategy chosen here. In particular, it ensures that each treated CBED pattern has a total value of one, hence leading to a reduction in their individual variances \cite{Seki2018}. As such, similar value ranges are consistently found in the amplitudes of the distributions $\tilde{J}_{\vec{Q}}\left(\vec{q}_d\right)$ and $J_{\vec{Q}}\left(\vec{R}\right)$. Other practical choices included the use of orthonormal Fourier transforms. In this context, the precise selection of the Wiener parameter $\epsilon$ becomes less critical for the reconstruction, which then permits more reproducible performances for the analytical ptychography procedures.
            
            As a final note on this topic, reaching a correspondence between the ranges of phase shift covered by the SBI and the WDD results, which was an argument used in ref. \cite{Susi2025}, should not be an objective in choosing a value for the Wiener parameter $\epsilon$. As was extensively discussed in subsection \ref{subsec:freqtranscap} and noted e.g. in ref. \cite{Yang2017}, the two methods are generally expected to lead to different results, since their fundamental assumptions on the specimen differ. On the other hand, once $\epsilon$ has been elevated sufficiently above zero to make sure that numerical divergence and noise amplification are avoided even with very sparse data, there is little justification in further increasing it. In fact, the higher this parameter becomes, the farther away the numerical operation goes from an actual inversion, and the less representative of the real interaction the deconvolved Wigner distribution $\Upsilon\left(\vec{Q}\,;\,\vec{R}\right)$ becomes. As such, the change in values for the retrieved WDD phase shift, induced by an increment of $\epsilon$, should be seen as artefactual rather than a possible validation of the SSB/SBI result.

\section*{Conclusion}
    
    Analytical ptychography methods present several advantages for the imaging of beam-sensitive materials. Specifically, they are direct, fast and relatively easy to implement. Their requirements in terms of computer memory can furthermore be reduced to allow efficient parallelization and GPU implementation, for instance through the scan-frequency partitioning algorithm introduced in this publication. They also do not require a specific choice in reconstruction parameters for particular cases, such as e.g. a coupling of loss and regularization functions, the batch size or an update strength, which could otherwise be needed to obtain a satisfying output. Finally, they permit the treatment of sparse scattering data \cite{OLeary2020} with no risk of numerical divergence.
    
    Overall, those advantages make analytical ptychography especially relevant for the low-dose investigation of beam-sensitive objects, where the same measurement often has to be repeated multiple times to reach an accurate result, thus creating a need for streamlined acquisition and reconstruction procedures, even encompassing live processing \cite{Strauch2021,Bangun2023}. In that manner, the direct form of analytical ptychography would also facilitate the inclusion in a more complex experimental protocol, such as e.g. three-dimensional structure retrieval based on a single-particle analysis \cite{Cheng2015,Nakane2020,Pei2023}. Moreover, in this context, potential reproducibility issues could be prevented, given the uniqueness of the processing compared to the high variability among iterative algorithms and their parameter sets.

\section*{Appendices}
    
    \subsection*{\textbf{Appendix A: redundancy condition and illuminated area overlap}}
        
        For a successful ptychographic reconstruction, given a well-focused and unaberrated probe, the overlap between successively illuminated regions is usually required to be 70 to 80 \%. In particular, this condition leads to the necessary degree of redundancy in the four-dimensional STEM dataset $I^{det}_{\vec{r}_s}\left(\vec{q}_d\right)$, i.e. specific object locations are probed as part of multiple recordings, thus creating common recognizable features among neighboring scan positions and making the correlative measurement of $\mu\left(\vec{r}\right)$ possible.
        
        The illumination overlap $\beta_{\delta \vec{r}_s}$ between two scan positions distant from a vectorial distance $\delta \vec{r}_s$ can be determined numerically using a normalized autocorrelation metric, e.g. given by
        \begin{equation}
            \beta_{\delta \vec{r}_s} \, = \, \frac{ \sum\limits_{\vec{r}_0} \, \mid P\left( \vec{r}_0 - \delta \vec{r}_s \right) \mid^2 \,\, \mid P\left(\vec{r}_0\right) \mid^2 }{ \sum\limits_{\vec{r}_0\,'} \, \mid P\left(\vec{r}_0\,'\right) \mid^4 } \quad.
            \label{eq:overlapratio}
        \end{equation}
        In comparison to simpler approaches \cite{Bunk2008}, the ratio described by equation \ref{eq:overlapratio} has the advantage of being calculated in two dimensions rather than just one, which can be expected to make it more accurate. $\beta_{\delta \vec{r}_s}$ is furthermore defined for an arbitrary aberration function or aperture shape, which would be useful e.g. for future work involving phase plates \cite{Yang2016a,Verbeeck2018,VegaIbanez2023,Yu2023}. In the case of a strongly overfocused probe, it encompasses the influence of near-field propagation on the precise incident intensity distribution in real-space, rather than simply assuming it to be an homogeneous disk. Examples of calculation are provided in section \ref{sec:MoS2}.
        
        Continuing, as long as the probe remains well-focused and if the area overlap $\beta_{\delta \vec{r}_s}$ is sufficient, it can be assumed that the scanned area is homogeneously illuminated and that the region outside does not receive any electrons \cite{Egerton2021a}. The dose is then simply estimated by
        \begin{equation}
            D \, = \, \frac{N_s \, N_{e^-}}{S} \quad,
            \label{eq:dose}
        \end{equation}
        where $N_s$ is the total number of scan points and $S$ is the surface of the square scan window. Importantly, equation \ref{eq:dose} assumes a perfect detection probability, such that all electrons sent to the specimen, and thus contributing to the dose $D$, end up being measured. In practice, this is not necessarily the case, as higher energy thresholds for electron detection may be imposed when the acceleration voltage $U$ is above e.g. 200 kV \cite{Ballabriga2018}. This is then done to prevent multiple counting \cite{Jannis2022,Denisov2023}, which tends to lower the effective DQE of the camera. This effect nevertheless cannot be represented in the presented dose-limitation process, except by correcting the assumed dose value post-calculation. It may also be that the maximum collection angle of the camera is too low to include every strongly scattered electrons, though this is easily prevented by a correct choice of camera length.
    
    \subsection*{\textbf{Appendix B: inclusion of dose-limitation in simulated scattering patterns}}
        
        In order to include dose-limitation in the simulations of scattering data while reproducing the single electron sensitivity, and thus the resulting count sparsity, of a hybrid-pixels DED \cite{Ballabriga2018}, this publication proposes the following approach. First, the user defines an average number $N_{e^-}$ of electrons sent on the specimen at a given scan position, i.e. a $\vec{r}_s$-wise expectancy of the incident intensity. Each CBED pattern is then attributed a random number of counts $n\left(\vec{r}_s\right)$ following the Poisson probability
        \begin{equation}
            p\left[ \, n\left(\vec{r}_s\right) \, \mid \, N_{e^-} \, \right] \, = \, \frac{ {N_{e^-}}^{n\left(\vec{r}_s\right)} \, e^{-N_{e^-}} }{ n\left(\vec{r}_s\right)\,! } \quad.
        \end{equation}
        A dose-limited intensity $I^{N_{e^-}}_{\vec{r}_s}\left(\vec{q}\right)$ is then obtained, for each scan point $\vec{r}_s$. This is done through the random selection of a single pixel, with probability weighted by the underlying pre-calculated $I_{\vec{r}_s}\left(\vec{q}\right)$ and repeated $n\left(\vec{r}_s\right)$ times, thus generating a new count at every step.
        
        In that manner, the simulation results can be made to encompass Poisson statistics \cite{Luczka1991} in the amount of counts per scan points, while providing a faithful representation of the detection process involved in devices such as e.g. the Medipix3 \cite{Ballabriga2011}, Timepix3 \cite{Poikela2014} and Timepix4 \cite{Llopart2022} chips, which is itself represented by a multinomial distribution \cite{Seki2018}. This then constitutes an alternative to the more conventional approach, which would simply consist in adding noise over the simulated CBED patterns, hence with no direct representation of sparsity other than rounding pixel values to the closest integer.
        
        The newly introduced dose-limitation procedure possesses an additional advantage, in that it provides an opportunity to make predictions on multiple counting \cite{Milazzo2010,Mir2017,Paton2021,Jannis2022,Denisov2023} in hybrid-pixels DED \cite{Ballabriga2018}, which is normally due to single electrons depositing an amount of energy above the detection threshold in more than one location. As each wavefront collapse on the camera is here represented individually, it becomes in principle possible to model the stochastic travel among pixels, e.g. through a Monte-Carlo calculation informed on the varying velocity of the incident electron \cite{Pennicard2011} and encompassing a choice of threshold energy. This would in turn lead to a more realistic representation of the resulting information spread effect than a direct convolution of $I_{\vec{r}_s}\left(\vec{q}\right)$ with a known isotropic $M\left(\vec{r}_d\right)$, either post-pixel selection or pre-noise supplementation. In particular, multiple counting manifests as non-isotropic clusters unique to each incident electrons \cite{Mir2017,VanSchayck2020,VanSchayck2023,Kuttruff2024}, whose sizes and shapes depend on the acceleration voltage and which maintain a constant value of 1 among activated pixels. Hence, for a ptychographic calculation based on a collection of sparse diffraction patterns such as the ones generated in this work, it can be expected that those subtleties become important. This topic will thus be critical for future work on low-dose ptychography making use of the Timepix3 \cite{Poikela2014} or the Timepix4 \cite{Llopart2022} chips.

\section*{Acknowledgments}
    
    The authors acknowledge financial support from the Horizon 2020 research and innovation programme, as well as from the European Research Council within the Horizon Europe innovation funding programme.

\section*{Funding}
    
    \textbf{H.L.L.R.} and \textbf{J.V.} acknowledge funding from the Horizon 2020 research and innovation programme (European Union), under grant agreement No 101017720 (FET-Proactive EBEAM). \textbf{M.L.L.} and \textbf{K.M.-C.} acknowledge funding from the European Research Council within the Horizon Europe innovation funding programme under Grant Agreement 101118656 (ERC Synergy project 4D-BioSTEM).

\section*{Conflicts of interest}
    
    The authors declare no conflicts of interest.

\section*{Data availability statement}
    
    No experimental data was required in this publication.

\section*{Author contribution statement}
    
    \textbf{H.L.L.R.}: conceptualization, methodology, software, simulation, data treatment, writing. \textbf{M.L.L.}: simulation, draft review. \textbf{K.M.-C.}: draft review, supervision. \textbf{J.V.}: conceptualization, draft review, supervision.

\bibliography{Sources_HLLR}

\begin{thebibliography}{100}
\expandafter\ifx\csname url\endcsname\relax
  \def\url#1{\texttt{#1}}\fi
\expandafter\ifx\csname urlprefix\endcsname\relax\def\urlprefix{URL }\fi
\expandafter\ifx\csname href\endcsname\relax
  \def\href#1#2{#2} \def\path#1{#1}\fi

\bibitem{Zhou2020}
L.~Zhou, J.~Song, J.~S. Kim, X.~Pei, C.~Huang, M.~Boyce, L.~Mendon{\c{c}}a, D.~Clare, A.~Siebert, C.~S. Allen, E.~Liberti, D.~Stuart, X.~Pan, P.~D. Nellist, P.~Zhang, A.~I. Kirkland, P.~Wang, \href{http://dx.doi.org/10.1038/s41467-020-16391-6 https://www.nature.com/articles/s41467-020-16391-6}{{Low-dose phase retrieval of biological specimens using cryo-electron ptychography}}, Nature Communications 11~(1) (2020) 2773.
\newblock \href {https://doi.org/10.1038/s41467-020-16391-6} {\path{doi:10.1038/s41467-020-16391-6}}.
\newline\urlprefix\url{http://dx.doi.org/10.1038/s41467-020-16391-6 https://www.nature.com/articles/s41467-020-16391-6}

\bibitem{Lazic2022}
I.~Lazi{\'{c}}, M.~Wirix, M.~L. Leidl, F.~de~Haas, D.~Mann, M.~Beckers, E.~V. Pechnikova, K.~M{\"{u}}ller-Caspary, R.~Egoavil, E.~G. Bosch, C.~Sachse, {Single-particle cryo-EM structures from iDPC–STEM at near-atomic resolution}, Nature Methods 19~(9) (2022) 1126--1136.
\newblock \href {https://doi.org/10.1038/s41592-022-01586-0} {\path{doi:10.1038/s41592-022-01586-0}}.

\bibitem{Muller-Caspary2018}
K.~M{\"{u}}ller-Caspary, M.~Duchamp, M.~R{\"{o}}sner, V.~Migunov, F.~Winkler, H.~Yang, M.~Huth, R.~Ritz, M.~Simson, S.~Ihle, H.~Soltau, T.~Wehling, R.~E. Dunin-Borkowski, S.~{Van Aert}, A.~Rosenauer, {Atomic-scale quantification of charge densities in two-dimensional materials}, Physical Review B 98~(12) (sep 2018).
\newblock \href {https://doi.org/10.1103/PhysRevB.98.121408} {\path{doi:10.1103/PhysRevB.98.121408}}.

\bibitem{Wen2019}
Y.~Wen, C.~Ophus, C.~S. Allen, S.~Fang, J.~Chen, E.~Kaxiras, A.~I. Kirkland, J.~H. Warner, {Simultaneous Identification of Low and High Atomic Number Atoms in Monolayer 2D Materials Using 4D Scanning Transmission Electron Microscopy}, Nano Letters 19~(9) (2019) 6482--6491.
\newblock \href {https://doi.org/10.1021/acs.nanolett.9b02717} {\path{doi:10.1021/acs.nanolett.9b02717}}.

\bibitem{Chen2024c}
Y.~Chen, T.-C. Chou, C.-H. Fang, C.-Y. Lu, C.-N. Hsiao, W.-T. Hsu, C.-C. Chen, \href{https://doi.org/10.1038/s41598-023-50784-z https://www.nature.com/articles/s41598-023-50784-z}{{Direct observation of single-atom defects in monolayer two-dimensional materials by using electron ptychography at 200 kV acceleration voltage}}, Scientific Reports 14~(1) (2024) 277.
\newblock \href {https://doi.org/10.1038/s41598-023-50784-z} {\path{doi:10.1038/s41598-023-50784-z}}.
\newline\urlprefix\url{https://doi.org/10.1038/s41598-023-50784-z https://www.nature.com/articles/s41598-023-50784-z}

\bibitem{Liu2020b}
L.~Liu, N.~Wang, C.~Zhu, X.~Liu, Y.~Zhu, P.~Guo, L.~Alfilfil, X.~Dong, D.~Zhang, Y.~Han, {Direct Imaging of Atomically Dispersed Molybdenum that Enables Location of Aluminum in the Framework of Zeolite ZSM-5}, Angewandte Chemie - International Edition 59~(2) (2020) 819--825.
\newblock \href {https://doi.org/10.1002/anie.201909834} {\path{doi:10.1002/anie.201909834}}.

\bibitem{Sha2023a}
H.~Sha, J.~Cui, J.~Li, Y.~Zhang, W.~Yang, Y.~Li, R.~Yu, \href{https://doi.org/10.1126/sciadv.adf1151}{{Ptychographic measurements of varying size and shape along zeolite channels}}, Science Advances 9~(11) (2023) 1--8.
\newblock \href {https://doi.org/10.1126/sciadv.adf1151} {\path{doi:10.1126/sciadv.adf1151}}.
\newline\urlprefix\url{https://doi.org/10.1126/sciadv.adf1151}

\bibitem{Zhang2023}
H.~Zhang, G.~Li, J.~Zhang, D.~Zhang, Z.~Chen, X.~Liu, P.~Guo, Y.~Zhu, C.~Chen, L.~Liu, X.~Guo, Y.~Han, \href{https://www.science.org/doi/10.1126/science.adg3183}{{Three-dimensional inhomogeneity of zeolite structure and composition revealed by electron ptychography}}, Science 380~(6645) (2023) 633--638.
\newblock \href {https://doi.org/10.1126/science.adg3183} {\path{doi:10.1126/science.adg3183}}.
\newline\urlprefix\url{https://www.science.org/doi/10.1126/science.adg3183}

\bibitem{Dong2023}
Z.~Dong, E.~Zhang, Y.~Jiang, Q.~Zhang, A.~Mayoral, H.~Jiang, Y.~Ma, \href{https://pubs.acs.org/doi/10.1021/jacs.2c12673}{{Atomic-Level Imaging of Zeolite Local Structures Using Electron Ptychography}}, Journal of the American Chemical Society 145~(12) (2023) 6628--6632.
\newblock \href {https://doi.org/10.1021/jacs.2c12673} {\path{doi:10.1021/jacs.2c12673}}.
\newline\urlprefix\url{https://pubs.acs.org/doi/10.1021/jacs.2c12673}

\bibitem{Mitsuishi2023}
K.~Mitsuishi, K.~Nakazawa, R.~Sagawa, M.~Shimizu, H.~Matsumoto, H.~Shima, T.~Takewaki, \href{https://doi.org/10.1038/s41598-023-27452-3 https://www.nature.com/articles/s41598-023-27452-3}{{Direct observation of Cu in high-silica chabazite zeolite by electron ptychography using Wigner distribution deconvolution}}, Scientific Reports 13~(1) (2023) 316.
\newblock \href {https://doi.org/10.1038/s41598-023-27452-3} {\path{doi:10.1038/s41598-023-27452-3}}.
\newline\urlprefix\url{https://doi.org/10.1038/s41598-023-27452-3 https://www.nature.com/articles/s41598-023-27452-3}

\bibitem{Lozano2018}
J.~G. Lozano, G.~T. Martinez, L.~Jin, P.~D. Nellist, P.~G. Bruce, {Low-Dose Aberration-Free Imaging of Li-Rich Cathode Materials at Various States of Charge Using Electron Ptychography}, Nano Letters 18~(11) (2018) 6850--6855.
\newblock \href {https://doi.org/10.1021/acs.nanolett.8b02718} {\path{doi:10.1021/acs.nanolett.8b02718}}.

\bibitem{Song2022}
W.~Song, M.~A. P{\'{e}}rez-Osorio, J.-J. Marie, E.~Liberti, X.~Luo, C.~O'Leary, R.~A. House, P.~G. Bruce, P.~D. Nellist, \href{https://linkinghub.elsevier.com/retrieve/pii/S2542435122001465}{{Direct imaging of oxygen shifts associated with the oxygen redox of Li-rich layered oxides}}, Joule 6~(5) (2022) 1049--1065.
\newblock \href {https://doi.org/10.1016/j.joule.2022.04.008} {\path{doi:10.1016/j.joule.2022.04.008}}.
\newline\urlprefix\url{https://linkinghub.elsevier.com/retrieve/pii/S2542435122001465}

\bibitem{Song2024c}
W.~Song, M.~A. P{\'{e}}rez-Osorio, J.~Chen, Z.~Ding, J.-J. Marie, M.~Juelsholt, R.~A. House, P.~G. Bruce, P.~D. Nellist, \href{https://pubs.acs.org/doi/10.1021/jacs.4c05556}{{Visualization of Tetrahedral Li in the Alkali Layers of Li-Rich Layered Metal Oxides}}, Journal of the American Chemical Society 146~(34) (2024) 23814--23824.
\newblock \href {https://doi.org/10.1021/jacs.4c05556} {\path{doi:10.1021/jacs.4c05556}}.
\newline\urlprefix\url{https://pubs.acs.org/doi/10.1021/jacs.4c05556}

\bibitem{Hao2023}
B.~Hao, Z.~Ding, X.~Tao, P.~D. Nellist, H.~E. Assender, \href{https://doi.org/10.1016/j.polymer.2023.126305 https://linkinghub.elsevier.com/retrieve/pii/S0032386123006353}{{Atomic-scale imaging of polyvinyl alcohol crystallinity using electron ptychography}}, Polymer 284~(August) (2023) 126305.
\newblock \href {https://doi.org/10.1016/j.polymer.2023.126305} {\path{doi:10.1016/j.polymer.2023.126305}}.
\newline\urlprefix\url{https://doi.org/10.1016/j.polymer.2023.126305 https://linkinghub.elsevier.com/retrieve/pii/S0032386123006353}

\bibitem{Ma2023}
M.~Ma, X.~Zhang, X.~Chen, H.~Xiong, L.~Xu, T.~Cheng, J.~Yuan, F.~Wei, B.~Shen, \href{https://www.nature.com/articles/s41467-023-42999-5}{{In situ imaging of the atomic phase transition dynamics in metal halide perovskites}}, Nature Communications 14~(1) (2023) 7142.
\newblock \href {https://doi.org/10.1038/s41467-023-42999-5} {\path{doi:10.1038/s41467-023-42999-5}}.
\newline\urlprefix\url{https://www.nature.com/articles/s41467-023-42999-5}

\bibitem{Scheid2023}
A.~Scheid, Y.~Wang, M.~Jung, T.~Heil, D.~Moia, J.~Maier, P.~A. van Aken, \href{https://doi.org/10.1093/micmic/ozad017 https://academic.oup.com/mam/article/29/3/869/7131445}{{Electron Ptychographic Phase Imaging of Beam-sensitive All-inorganic Halide Perovskites Using Four-dimensional Scanning Transmission Electron Microscopy}}, Microscopy and Microanalysis 29~(3) (2023) 869--878.
\newblock \href {https://doi.org/10.1093/micmic/ozad017} {\path{doi:10.1093/micmic/ozad017}}.
\newline\urlprefix\url{https://doi.org/10.1093/micmic/ozad017 https://academic.oup.com/mam/article/29/3/869/7131445}

\bibitem{Schrenker2024}
N.~J. Schrenker, T.~Braeckevelt, A.~{De Backer}, N.~Livakas, C.-P. Yu, T.~Friedrich, M.~B.~J. Roeffaers, J.~Hofkens, J.~Verbeeck, L.~Manna, V.~{Van Speybroeck}, S.~{Van Aert}, S.~Bals, \href{https://pubs.acs.org/doi/10.1021/acs.nanolett.4c02811}{{Investigation of the Octahedral Network Structure in Formamidinium Lead Bromide Nanocrystals by Low-Dose Scanning Transmission Electron Microscopy}}, Nano Letters 24~(35) (2024) 10936--10942.
\newblock \href {https://doi.org/10.1021/acs.nanolett.4c02811} {\path{doi:10.1021/acs.nanolett.4c02811}}.
\newline\urlprefix\url{https://pubs.acs.org/doi/10.1021/acs.nanolett.4c02811}

\bibitem{Li2019b}
X.~Li, J.~Wang, X.~Liu, L.~Liu, D.~Cha, X.~Zheng, A.~A. Yousef, K.~Song, Y.~Zhu, D.~Zhang, Y.~Han, \href{https://pubs.acs.org/doi/10.1021/jacs.9b04896}{{Direct Imaging of Tunable Crystal Surface Structures of MOF MIL-101 Using High-Resolution Electron Microscopy}}, Journal of the American Chemical Society 141~(30) (2019) 12021--12028.
\newblock \href {https://doi.org/10.1021/jacs.9b04896} {\path{doi:10.1021/jacs.9b04896}}.
\newline\urlprefix\url{https://pubs.acs.org/doi/10.1021/jacs.9b04896}

\bibitem{Li2025}
G.~Li, M.~Xu, W.-q. Tang, Y.~Liu, C.~Chen, D.~Zhang, L.~Liu, S.~Ning, H.~Zhang, Z.-y. Gu, Z.~Lai, D.~A. Muller, Y.~Han, \href{http://dx.doi.org/10.1038/s41467-025-56215-z https://www.nature.com/articles/s41467-025-56215-z}{{Atomically resolved imaging of radiation-sensitive metal-organic frameworks via electron ptychography}}, Nature Communications 16~(1) (2025) 914.
\newblock \href {https://doi.org/10.1038/s41467-025-56215-z} {\path{doi:10.1038/s41467-025-56215-z}}.
\newline\urlprefix\url{http://dx.doi.org/10.1038/s41467-025-56215-z https://www.nature.com/articles/s41467-025-56215-z}

\bibitem{Egerton2019}
R.~Egerton, \href{https://doi.org/10.1016/j.micron.2019.01.005 https://linkinghub.elsevier.com/retrieve/pii/S0968432818304359}{{Radiation damage to organic and inorganic specimens in the TEM}}, Micron 119~(January) (2019) 72--87.
\newblock \href {https://doi.org/10.1016/j.micron.2019.01.005} {\path{doi:10.1016/j.micron.2019.01.005}}.
\newline\urlprefix\url{https://doi.org/10.1016/j.micron.2019.01.005 https://linkinghub.elsevier.com/retrieve/pii/S0968432818304359}

\bibitem{King1987}
W.~E. King, R.~Benedek, K.~Merkle, M.~Meshii, \href{https://linkinghub.elsevier.com/retrieve/pii/0304399187902452}{{Damage effects of high energy electrons on metals}}, Ultramicroscopy 23~(3-4) (1987) 345--353.
\newblock \href {https://doi.org/10.1016/0304-3991(87)90245-2} {\path{doi:10.1016/0304-3991(87)90245-2}}.
\newline\urlprefix\url{https://linkinghub.elsevier.com/retrieve/pii/0304399187902452}

\bibitem{Hobbs1994}
L.~W. Hobbs, F.~W. Clinard, S.~J. Zinkle, R.~C. Ewing, {Radiation effects in ceramics}, Journal of Nuclear Materials 216~(C) (1994) 291--321.
\newblock \href {https://doi.org/10.1016/0022-3115(94)90017-5} {\path{doi:10.1016/0022-3115(94)90017-5}}.

\bibitem{Bornes1944}
B.~Bornes, W.~Glaser, \href{https://doi.org/10.1007/BF01502110 http://link.springer.com/10.1007/BF01502110}{{{\"{U}}ber die Temperaturerh{\"{o}}hung der Objekte im {\"{U}}bermikroskop}}, Kolloid-Zeitschrift 106~(2) (1944) 123--128.
\newblock \href {https://doi.org/10.1007/BF01502110} {\path{doi:10.1007/BF01502110}}.
\newline\urlprefix\url{https://doi.org/10.1007/BF01502110 http://link.springer.com/10.1007/BF01502110}

\bibitem{Grubb1974}
D.~T. Grubb, \href{https://www.scopus.com/inward/record.uri?eid=2-s2.0-0016116491&doi=10.1007%2FBF00540772&partnerID=40&md5=c3f4a818cf7b5566d631eec9343ec5cb http://link.springer.com/10.1007/BF00540772}{{Radiation damage and electron microscopy of organic polymers}}, Journal of Materials Science 9~(10) (1974) 1715--1736.
\newblock \href {https://doi.org/10.1007/BF00540772} {\path{doi:10.1007/BF00540772}}.
\newline\urlprefix\url{https://www.scopus.com/inward/record.uri?eid=2-s2.0-0016116491&doi=10.1007%2FBF00540772&partnerID=40&md5=c3f4a818cf7b5566d631eec9343ec5cb http://link.springer.com/10.1007/BF00540772}

\bibitem{Egerton2021a}
R.~F. Egerton, \href{https://doi.org/10.1016/j.ultramic.2021.113363}{{Dose measurement in the TEM and STEM}}, Ultramicroscopy 229~(June) (2021) 113363.
\newblock \href {https://doi.org/10.1016/j.ultramic.2021.113363} {\path{doi:10.1016/j.ultramic.2021.113363}}.
\newline\urlprefix\url{https://doi.org/10.1016/j.ultramic.2021.113363}

\bibitem{Egerton2007}
R.~Egerton, \href{https://linkinghub.elsevier.com/retrieve/pii/S0304399106002245}{{Limits to the spatial, energy and momentum resolution of electron energy-loss spectroscopy}}, Ultramicroscopy 107~(8) (2007) 575--586.
\newblock \href {https://doi.org/10.1016/j.ultramic.2006.11.005} {\path{doi:10.1016/j.ultramic.2006.11.005}}.
\newline\urlprefix\url{https://linkinghub.elsevier.com/retrieve/pii/S0304399106002245}

\bibitem{Rez2021}
P.~Rez, \href{https://doi.org/10.1016/j.ultramic.2021.113301}{{Coherent and incoherent imaging of biological specimens with electrons and X-rays}}, Ultramicroscopy 231 (2021) 113301.
\newblock \href {https://doi.org/10.1016/j.ultramic.2021.113301} {\path{doi:10.1016/j.ultramic.2021.113301}}.
\newline\urlprefix\url{https://doi.org/10.1016/j.ultramic.2021.113301}

\bibitem{McMullan2007}
G.~McMullan, D.~Cattermole, S.~Chen, R.~Henderson, X.~Llopart, C.~Summerfield, L.~Tlustos, A.~Faruqi, \href{https://linkinghub.elsevier.com/retrieve/pii/S0304399106001963}{{Electron imaging with Medipix2 hybrid pixel detector}}, Ultramicroscopy 107~(4-5) (2007) 401--413.
\newblock \href {https://doi.org/10.1016/j.ultramic.2006.10.005} {\path{doi:10.1016/j.ultramic.2006.10.005}}.
\newline\urlprefix\url{https://linkinghub.elsevier.com/retrieve/pii/S0304399106001963}

\bibitem{Llopart2007}
X.~Llopart, R.~Ballabriga, M.~Campbell, L.~Tlustos, W.~Wong, \href{https://linkinghub.elsevier.com/retrieve/pii/S0168900207017020}{{Timepix, a 65k programmable pixel readout chip for arrival time, energy and/or photon counting measurements}}, Nuclear Instruments and Methods in Physics Research Section A: Accelerators, Spectrometers, Detectors and Associated Equipment 581~(1-2) (2007) 485--494.
\newblock \href {https://doi.org/10.1016/j.nima.2007.08.079} {\path{doi:10.1016/j.nima.2007.08.079}}.
\newline\urlprefix\url{https://linkinghub.elsevier.com/retrieve/pii/S0168900207017020}

\bibitem{Ballabriga2011}
R.~Ballabriga, M.~Campbell, E.~Heijne, X.~Llopart, L.~Tlustos, W.~Wong, \href{https://linkinghub.elsevier.com/retrieve/pii/S0168900210012982}{{Medipix3: A 64k pixel detector readout chip working in single photon counting mode with improved spectrometric performance}}, Nuclear Instruments and Methods in Physics Research Section A: Accelerators, Spectrometers, Detectors and Associated Equipment 633~(SUPPL. 1) (2011) S15--S18.
\newblock \href {https://doi.org/10.1016/j.nima.2010.06.108} {\path{doi:10.1016/j.nima.2010.06.108}}.
\newline\urlprefix\url{https://linkinghub.elsevier.com/retrieve/pii/S0168900210012982}

\bibitem{Plackett2013}
R.~Plackett, I.~Horswell, E.~N. Gimenez, J.~Marchal, D.~Omar, N.~Tartoni, \href{http://iopscience.iop.org/article/10.1088/1748-0221/8/01/C01038/meta}{{Merlin: a fast versatile readout system for Medipix3}}, Journal of Instrumentation 8~(1) (2013) C01038.
\newblock \href {https://doi.org/10.1088/1748-0221/8/01/C01038} {\path{doi:10.1088/1748-0221/8/01/C01038}}.
\newline\urlprefix\url{http://iopscience.iop.org/article/10.1088/1748-0221/8/01/C01038/meta}

\bibitem{Poikela2014}
T.~Poikela, J.~Plosila, T.~Westerlund, M.~Campbell, M.~D. Gaspari, X.~Llopart, V.~Gromov, R.~Kluit, M.~V. Beuzekom, F.~Zappon, V.~Zivkovic, C.~Brezina, K.~Desch, Y.~Fu, A.~Kruth, \href{https://iopscience.iop.org/article/10.1088/1748-0221/9/05/C05013}{{Timepix3: a 65K channel hybrid pixel readout chip with simultaneous ToA/ToT and sparse readout}}, Journal of Instrumentation 9~(05) (2014) C05013--C05013.
\newblock \href {https://doi.org/10.1088/1748-0221/9/05/C05013} {\path{doi:10.1088/1748-0221/9/05/C05013}}.
\newline\urlprefix\url{https://iopscience.iop.org/article/10.1088/1748-0221/9/05/C05013}

\bibitem{Ryll2016}
H.~Ryll, M.~Simson, R.~Hartmann, P.~Holl, M.~Huth, S.~Ihle, Y.~Kondo, P.~Kotula, A.~Liebel, K.~M{\"{u}}ller-Caspary, A.~Rosenauer, R.~Sagawa, J.~Schmidt, H.~Soltau, L.~Str{\"{u}}der, \href{http://stacks.iop.org/1748-0221/11/i=04/a=P04006}{{A pnCCD-based, fast direct single electron imaging camera for TEM and STEM}}, J. Instrum. 11~(04) (2016) P04006.
\newblock \href {https://doi.org/10.1088/1748-0221/11/04/P04006} {\path{doi:10.1088/1748-0221/11/04/P04006}}.
\newline\urlprefix\url{http://stacks.iop.org/1748-0221/11/i=04/a=P04006}

\bibitem{Tate2016}
M.~W. Tate, P.~Purohit, D.~Chamberlain, K.~X. Nguyen, R.~Hovden, C.~S. Chang, P.~Deb, E.~Turgut, J.~T. Heron, D.~G. Schlom, D.~C. Ralph, G.~D. Fuchs, K.~S. Shanks, H.~T. Philipp, D.~A. Muller, S.~M. Gruner, \href{http://journals.cambridge.org/article_S1431927615015664}{{High Dynamic Range Pixel Array Detector for Scanning Transmission Electron Microscopy}}, Microsc. Microanal. 22~(1) (2016) 237--249.
\newblock \href {http://arxiv.org/abs/1511.03539} {\path{arXiv:1511.03539}}, \href {https://doi.org/10.1017/S1431927615015664} {\path{doi:10.1017/S1431927615015664}}.
\newline\urlprefix\url{http://journals.cambridge.org/article_S1431927615015664}

\bibitem{Philipp2022}
H.~T. Philipp, M.~W. Tate, K.~S. Shanks, L.~Mele, M.~Peemen, P.~Dona, R.~Hartong, G.~van Veen, Y.-T. Shao, Z.~Chen, J.~Thom-Levy, D.~A. Muller, S.~M. Gruner, \href{https://academic.oup.com/mam/article/28/2/425/6889399}{{Very-High Dynamic Range, 10,000 Frames/Second Pixel Array Detector for Electron Microscopy}}, Microscopy and Microanalysis 28~(2) (2022) 425--440.
\newblock \href {http://arxiv.org/abs/2111.05889} {\path{arXiv:2111.05889}}, \href {https://doi.org/10.1017/S1431927622000174} {\path{doi:10.1017/S1431927622000174}}.
\newline\urlprefix\url{https://academic.oup.com/mam/article/28/2/425/6889399}

\bibitem{Llopart2022}
X.~Llopart, J.~Alozy, R.~Ballabriga, M.~Campbell, R.~Casanova, V.~Gromov, E.~Heijne, T.~Poikela, E.~Santin, V.~Sriskaran, L.~Tlustos, A.~Vitkovskiy, \href{https://iopscience.iop.org/article/10.1088/1748-0221/17/01/C01044}{{Timepix4, a large area pixel detector readout chip which can be tiled on 4 sides providing sub-200 ps timestamp binning}}, Journal of Instrumentation 17~(01) (2022) C01044.
\newblock \href {https://doi.org/10.1088/1748-0221/17/01/C01044} {\path{doi:10.1088/1748-0221/17/01/C01044}}.
\newline\urlprefix\url{https://iopscience.iop.org/article/10.1088/1748-0221/17/01/C01044}

\bibitem{Zambon2023}
P.~Zambon, S.~Bottinelli, R.~Schnyder, D.~Musarra, D.~Boye, A.~Dudina, N.~Lehmann, S.~{De Carlo}, M.~Rissi, C.~Schulze-Briese, M.~Meffert, M.~Campanini, R.~Erni, L.~Piazza, \href{https://doi.org/10.1016/j.nima.2022.167888 https://linkinghub.elsevier.com/retrieve/pii/S0168900222011809}{{KITE: High frame rate, high count rate pixelated electron counting ASIC for 4D STEM applications featuring high-Z sensor}}, Nuclear Instruments and Methods in Physics Research Section A: Accelerators, Spectrometers, Detectors and Associated Equipment 1048~(November 2022) (2023) 167888.
\newblock \href {https://doi.org/10.1016/j.nima.2022.167888} {\path{doi:10.1016/j.nima.2022.167888}}.
\newline\urlprefix\url{https://doi.org/10.1016/j.nima.2022.167888 https://linkinghub.elsevier.com/retrieve/pii/S0168900222011809}

\bibitem{Ercius2024}
P.~Ercius, I.~J. Johnson, P.~Pelz, B.~H. Savitzky, L.~Hughes, H.~G. Brown, S.~E. Zeltmann, S.-L. Hsu, C.~C.~S. Pedroso, B.~E. Cohen, R.~Ramesh, D.~Paul, J.~M. Joseph, T.~Stezelberger, C.~Czarnik, M.~Lent, E.~Fong, J.~Ciston, M.~C. Scott, C.~Ophus, A.~M. Minor, P.~Denes, \href{http://arxiv.org/abs/2305.11961 https://academic.oup.com/mam/advance-article/doi/10.1093/mam/ozae086/7762045}{{The 4D Camera: An 87 kHz Direct Electron Detector for Scanning/Transmission Electron Microscopy}}, Microscopy and Microanalysis (2024) 1--10\href {http://arxiv.org/abs/2305.11961} {\path{arXiv:2305.11961}}, \href {https://doi.org/10.1093/mam/ozae086} {\path{doi:10.1093/mam/ozae086}}.
\newline\urlprefix\url{http://arxiv.org/abs/2305.11961 https://academic.oup.com/mam/advance-article/doi/10.1093/mam/ozae086/7762045}

\bibitem{Ruskin2013}
R.~S. Ruskin, Z.~Yu, N.~Grigorieff, \href{http://dx.doi.org/10.1016/j.jsb.2013.10.016 https://linkinghub.elsevier.com/retrieve/pii/S1047847713002815}{{Quantitative characterization of electron detectors for transmission electron microscopy}}, Journal of Structural Biology 184~(3) (2013) 385--393.
\newblock \href {https://doi.org/10.1016/j.jsb.2013.10.016} {\path{doi:10.1016/j.jsb.2013.10.016}}.
\newline\urlprefix\url{http://dx.doi.org/10.1016/j.jsb.2013.10.016 https://linkinghub.elsevier.com/retrieve/pii/S1047847713002815}

\bibitem{Milazzo2010}
A.-C. Milazzo, G.~Moldovan, J.~Lanman, L.~Jin, J.~C. Bouwer, S.~Klienfelder, S.~T. Peltier, M.~H. Ellisman, A.~I. Kirkland, N.-H. Xuong, \href{http://www.sciencedirect.com/science/article/pii/S0304399110000859}{{Characterization of a direct detection device imaging camera for transmission electron microscopy}}, Ultramicroscopy 110~(7) (2010) 741--744.
\newblock \href {https://doi.org/10.1016/j.ultramic.2010.03.007} {\path{doi:10.1016/j.ultramic.2010.03.007}}.
\newline\urlprefix\url{http://www.sciencedirect.com/science/article/pii/S0304399110000859}

\bibitem{Mir2017}
J.~A. Mir, R.~Clough, R.~MacInnes, C.~Gough, R.~Plackett, I.~Shipsey, H.~Sawada, I.~MacLaren, R.~Ballabriga, D.~Maneuski, V.~O'Shea, D.~McGrouther, A.~I. Kirkland, \href{http://dx.doi.org/10.1016/j.ultramic.2017.06.010}{{Characterisation of the Medipix3 detector for 60 and 80 keV electrons}}, Ultramicroscopy 182 (2017) 44--53.
\newblock \href {https://doi.org/10.1016/j.ultramic.2017.06.010} {\path{doi:10.1016/j.ultramic.2017.06.010}}.
\newline\urlprefix\url{http://dx.doi.org/10.1016/j.ultramic.2017.06.010}

\bibitem{Paton2021}
K.~A. Paton, M.~C. Veale, X.~Mu, C.~S. Allen, D.~Maneuski, C.~K{\"{u}}bel, V.~O'Shea, A.~I. Kirkland, D.~McGrouther, \href{https://doi.org/10.1016/j.ultramic.2021.113298}{{Quantifying the performance of a hybrid pixel detector with GaAs:Cr sensor for transmission electron microscopy}}, Ultramicroscopy 227~(April) (2021) 113298.
\newblock \href {http://arxiv.org/abs/2009.14565} {\path{arXiv:2009.14565}}, \href {https://doi.org/10.1016/j.ultramic.2021.113298} {\path{doi:10.1016/j.ultramic.2021.113298}}.
\newline\urlprefix\url{https://doi.org/10.1016/j.ultramic.2021.113298}

\bibitem{Muller2012a}
K.~M{\"{u}}ller, H.~Ryll, I.~Ordavo, S.~Ihle, L.~Str{\"{u}}der, K.~Volz, J.~Zweck, H.~Soltau, A.~Rosenauer, {Scanning transmission electron microscopy strain measurement from millisecond frames of a direct electron charge coupled device}, Applied Physics Letters 101~(21) (2012).
\newblock \href {https://doi.org/10.1063/1.4767655} {\path{doi:10.1063/1.4767655}}.

\bibitem{Muller-Caspary2018a}
K.~M{\"{u}}ller-Caspary, F.~F. Krause, F.~Winkler, A.~B{\'{e}}ch{\'{e}}, J.~Verbeeck, S.~{Van Aert}, A.~Rosenauer, S.~VanAert, A.~Rosenauer, \href{http://www.sciencedirect.com/science/article/pii/S0304399118302730 https://doi.org/10.1016/j.ultramic.2018.12.018}{{Comparison of first moment STEM with conventional differential phase contrast and the dependence on electron dose}}, Ultramicroscopy accepted~(August 2018) (2018) in print.
\newblock \href {https://doi.org/10.1016/j.ultramic.2018.12.018} {\path{doi:10.1016/j.ultramic.2018.12.018}}.
\newline\urlprefix\url{http://www.sciencedirect.com/science/article/pii/S0304399118302730 https://doi.org/10.1016/j.ultramic.2018.12.018}

\bibitem{Yang2015a}
H.~Yang, L.~Jones, H.~Ryll, M.~Simson, H.~Soltau, Y.~Kondo, R.~Sagawa, H.~Banba, I.~MacLaren, P.~D. Nellist, \href{http://stacks.iop.org/1742-6596/644/i=1/a=012032}{{4D STEM: High efficiency phase contrast imaging using a fast pixelated detector}}, J. Phys.: Conf. Ser. 644~(1) (2015) 12032.
\newblock \href {https://doi.org/10.1088/1742-6596/644/1/012032} {\path{doi:10.1088/1742-6596/644/1/012032}}.
\newline\urlprefix\url{http://stacks.iop.org/1742-6596/644/i=1/a=012032}

\bibitem{Fan1998}
G.~Fan, P.~Datte, E.~Beuville, J.-F. Beche, J.~Millaud, K.~Downing, F.~Burkard, M.~Ellisman, N.-H. Xuong, \href{https://linkinghub.elsevier.com/retrieve/pii/S0304399197001095}{{ASIC-based event-driven 2D digital electron counter for TEM imaging}}, Ultramicroscopy 70~(3) (1998) 107--113.
\newblock \href {https://doi.org/10.1016/S0304-3991(97)00109-5} {\path{doi:10.1016/S0304-3991(97)00109-5}}.
\newline\urlprefix\url{https://linkinghub.elsevier.com/retrieve/pii/S0304399197001095}

\bibitem{Frojdh2015}
E.~Frojdh, M.~Campbell, M.~D. Gaspari, S.~Kulis, X.~Llopart, T.~Poikela, L.~Tlustos, \href{https://iopscience.iop.org/article/10.1088/1748-0221/10/01/C01039}{{Timepix3: first measurements and characterization of a hybrid-pixel detector working in event driven mode}}, Journal of Instrumentation 10~(01) (2015) C01039--C01039.
\newblock \href {https://doi.org/10.1088/1748-0221/10/01/C01039} {\path{doi:10.1088/1748-0221/10/01/C01039}}.
\newline\urlprefix\url{https://iopscience.iop.org/article/10.1088/1748-0221/10/01/C01039}

\bibitem{Jannis2022}
D.~Jannis, C.~Hofer, C.~Gao, X.~Xie, A.~B{\'{e}}ch{\'{e}}, T.~Pennycook, J.~Verbeeck, \href{https://linkinghub.elsevier.com/retrieve/pii/S0304399121001996 https://doi.org/10.1016/j.ultramic.2021.113423}{{Event driven 4D STEM acquisition with a Timepix3 detector: Microsecond dwell time and faster scans for high precision and low dose applications}}, Ultramicroscopy 233~(October 2021) (2022) 113423.
\newblock \href {http://arxiv.org/abs/2107.02864} {\path{arXiv:2107.02864}}, \href {https://doi.org/10.1016/j.ultramic.2021.113423} {\path{doi:10.1016/j.ultramic.2021.113423}}.
\newline\urlprefix\url{https://linkinghub.elsevier.com/retrieve/pii/S0304399121001996 https://doi.org/10.1016/j.ultramic.2021.113423}

\bibitem{Auad2023a}
Y.~Auad, J.~Baaboura, J.~D. Blazit, M.~Tenc{\'{e}}, O.~St{\'{e}}phan, M.~Kociak, L.~H. Tizei, \href{https://doi.org/10.1016/j.ultramic.2023.113889}{{Time calibration studies for the Timepix3 hybrid pixel detector in electron microscopy}}, Ultramicroscopy 257~(November 2023) (2024) 113889.
\newblock \href {https://doi.org/10.1016/j.ultramic.2023.113889} {\path{doi:10.1016/j.ultramic.2023.113889}}.
\newline\urlprefix\url{https://doi.org/10.1016/j.ultramic.2023.113889}

\bibitem{Kuttruff2024}
J.~Kuttruff, J.~Holder, Y.~Meng, P.~Baum, \href{https://doi.org/10.1016/j.ultramic.2023.113864}{{Real-time electron clustering in an event-driven hybrid pixel detector}}, Ultramicroscopy 255~(September 2023) (2024) 113864.
\newblock \href {https://doi.org/10.1016/j.ultramic.2023.113864} {\path{doi:10.1016/j.ultramic.2023.113864}}.
\newline\urlprefix\url{https://doi.org/10.1016/j.ultramic.2023.113864}

\bibitem{Hoppe1969}
W.~Hoppe, \href{https://scripts.iucr.org/cgi-bin/paper?S0567739469001045}{{Beugung im inhomogenen Prim{\"{a}}rstrahlwellenfeld. I. Prinzip einer Phasenmessung von Elektronenbeungungsinterferenzen}}, Acta Crystallographica Section A 25~(4) (1969) 495--501.
\newblock \href {https://doi.org/10.1107/S0567739469001045} {\path{doi:10.1107/S0567739469001045}}.
\newline\urlprefix\url{https://scripts.iucr.org/cgi-bin/paper?S0567739469001045}

\bibitem{Hoppe1969a}
W.~Hoppe, G.~Strube, \href{https://scripts.iucr.org/cgi-bin/paper?S0567739469001057}{{Beugung in inhomogenen Prim{\"{a}}rstrahlenwellenfeld. II. Lichtoptische Analogieversuche zur Phasenmessung von Gitterinterferenzen}}, Acta Crystallographica Section A 25~(4) (1969) 502--507.
\newblock \href {https://doi.org/10.1107/S0567739469001057} {\path{doi:10.1107/S0567739469001057}}.
\newline\urlprefix\url{https://scripts.iucr.org/cgi-bin/paper?S0567739469001057}

\bibitem{Hoppe1969b}
W.~Hoppe, \href{https://scripts.iucr.org/cgi-bin/paper?S0567739469001069}{{Beugung im inhomogenen Prim{\"{a}}rstrahlwellenfeld. III. Amplituden- und Phasenbestimmung bei unperiodischen Objekten}}, Acta Crystallographica Section A 25~(4) (1969) 508--514.
\newblock \href {https://doi.org/10.1107/S0567739469001069} {\path{doi:10.1107/S0567739469001069}}.
\newline\urlprefix\url{https://scripts.iucr.org/cgi-bin/paper?S0567739469001069}

\bibitem{Drenth1975}
A.~Drenth, A.~Huiser, H.~Ferwerda, \href{https://doi.org/10.1080/713819083 https://www.tandfonline.com/doi/full/10.1080/713819083}{{The Problem of Phase Retrieval in Light and Electron Microscopy of Strong Objects}}, Optica Acta: International Journal of Optics 22~(7) (1975) 615--628.
\newblock \href {https://doi.org/10.1080/713819083} {\path{doi:10.1080/713819083}}.
\newline\urlprefix\url{https://doi.org/10.1080/713819083 https://www.tandfonline.com/doi/full/10.1080/713819083}

\bibitem{Fienup1978}
J.~R. Fienup, \href{https://opg.optica.org/abstract.cfm?URI=ol-3-1-27}{{Reconstruction of an object from the modulus of its Fourier transform}}, Optics Letters 3~(1) (1978) 27.
\newblock \href {https://doi.org/10.1364/OL.3.000027} {\path{doi:10.1364/OL.3.000027}}.
\newline\urlprefix\url{https://opg.optica.org/abstract.cfm?URI=ol-3-1-27}

\bibitem{Miao1998}
J.~Miao, D.~Sayre, H.~N. Chapman, \href{https://opg.optica.org/abstract.cfm?URI=josaa-15-6-1662}{{Phase retrieval from the magnitude of the Fourier transforms of nonperiodic objects}}, Journal of the Optical Society of America A 15~(6) (1998) 1662.
\newblock \href {https://doi.org/10.1364/JOSAA.15.001662} {\path{doi:10.1364/JOSAA.15.001662}}.
\newline\urlprefix\url{https://opg.optica.org/abstract.cfm?URI=josaa-15-6-1662}

\bibitem{Miao1999}
J.~Miao, P.~Charalambous, J.~Kirz, D.~Sayre, \href{http://dx.doi.org/10.1038/22498 https://www.nature.com/articles/22498}{{Extending the methodology of X-ray crystallography to allow imaging of micrometre-sized non-crystalline specimens}}, Nature 400~(6742) (1999) 342--344.
\newblock \href {https://doi.org/10.1038/22498} {\path{doi:10.1038/22498}}.
\newline\urlprefix\url{http://dx.doi.org/10.1038/22498 https://www.nature.com/articles/22498}

\bibitem{Weierstall2002}
U.~Weierstall, Q.~Chen, J.~Spence, M.~Howells, M.~Isaacson, R.~Panepucci, \href{https://linkinghub.elsevier.com/retrieve/pii/S0304399101001346}{{Image reconstruction from electron and X-ray diffraction patterns using iterative algorithms: experiment and simulation}}, Ultramicroscopy 90~(2-3) (2002) 171--195.
\newblock \href {https://doi.org/10.1016/S0304-3991(01)00134-6} {\path{doi:10.1016/S0304-3991(01)00134-6}}.
\newline\urlprefix\url{https://linkinghub.elsevier.com/retrieve/pii/S0304399101001346}

\bibitem{Williams2006}
G.~J. Williams, H.~M. Quiney, B.~B. Dhal, C.~Q. Tran, K.~A. Nugent, A.~G. Peele, D.~Paterson, M.~D. de~Jonge, \href{https://link.aps.org/doi/10.1103/PhysRevLett.97.025506}{{Fresnel Coherent Diffractive Imaging}}, Physical Review Letters 97~(2) (2006) 025506.
\newblock \href {https://doi.org/10.1103/PhysRevLett.97.025506} {\path{doi:10.1103/PhysRevLett.97.025506}}.
\newline\urlprefix\url{https://link.aps.org/doi/10.1103/PhysRevLett.97.025506}

\bibitem{Fienup1982}
J.~R. Fienup, \href{http://ieeexplore.ieee.org/document/1169765/ https://opg.optica.org/abstract.cfm?URI=ao-21-15-2758}{{Phase retrieval algorithms: a comparison}}, Applied Optics 21~(15) (1982) 2758.
\newblock \href {https://doi.org/10.1364/AO.21.002758} {\path{doi:10.1364/AO.21.002758}}.
\newline\urlprefix\url{http://ieeexplore.ieee.org/document/1169765/ https://opg.optica.org/abstract.cfm?URI=ao-21-15-2758}

\bibitem{Pelz2017}
P.~M. Pelz, W.~X. Qiu, R.~B{\"{u}}cker, G.~Kassier, R.~J. Miller, {Low-dose cryo electron ptychography via non-convex Bayesian optimization}, Scientific Reports 7~(1) (2017) 1--13.
\newblock \href {http://arxiv.org/abs/1702.05732} {\path{arXiv:1702.05732}}, \href {https://doi.org/10.1038/s41598-017-07488-y} {\path{doi:10.1038/s41598-017-07488-y}}.

\bibitem{Pei2023}
X.~Pei, L.~Zhou, C.~Huang, M.~Boyce, J.~S. Kim, E.~Liberti, Y.~Hu, T.~Sasaki, P.~D. Nellist, P.~Zhang, D.~I. Stuart, A.~I. Kirkland, P.~Wang, \href{https://www.nature.com/articles/s41467-023-38268-0}{{Cryogenic electron ptychographic single particle analysis with wide bandwidth information transfer}}, Nature Communications 14~(1) (2023) 3027.
\newblock \href {https://doi.org/10.1038/s41467-023-38268-0} {\path{doi:10.1038/s41467-023-38268-0}}.
\newline\urlprefix\url{https://www.nature.com/articles/s41467-023-38268-0}

\bibitem{Kucukoglu2024}
B.~K{\"{u}}{\c{c}}{\"{u}}koğlu, I.~Mohammed, R.~C. Guerrero-Ferreira, S.~M. Ribet, G.~Varnavides, M.~L. Leidl, K.~Lau, S.~Nazarov, A.~Myasnikov, M.~Kube, J.~Radecke, C.~Sachse, K.~M{\"{u}}ller-Caspary, C.~Ophus, H.~Stahlberg, \href{https://www.nature.com/articles/s41467-024-52403-5}{{Low-dose cryo-electron ptychography of proteins at sub-nanometer resolution}}, Nature Communications 15~(1) (2024) 8062.
\newblock \href {https://doi.org/10.1038/s41467-024-52403-5} {\path{doi:10.1038/s41467-024-52403-5}}.
\newline\urlprefix\url{https://www.nature.com/articles/s41467-024-52403-5}

\bibitem{Mao2024}
W.~Mao, W.~Zhang, C.~Huang, L.~Zhou, J.~S. Kim, S.~Gao, Y.~Lei, X.~Wu, Y.~Hu, X.~Pei, W.~Fang, X.~Liu, J.~Song, C.~Fan, Y.~Nie, A.~I. Kirkland, P.~Wang, \href{http://arxiv.org/abs/2403.16902}{{Multi-Convergence-Angle Ptychography with Simultaneous Strong Contrast and High Resolution}} (2024) 1--25\href {http://arxiv.org/abs/2403.16902} {\path{arXiv:2403.16902}}.
\newline\urlprefix\url{http://arxiv.org/abs/2403.16902}

\bibitem{Elser2003}
V.~Elser, \href{https://opg.optica.org/abstract.cfm?URI=josaa-20-1-40}{{Phase retrieval by iterated projections}}, Journal of the Optical Society of America A 20~(1) (2003) 40.
\newblock \href {https://doi.org/10.1364/JOSAA.20.000040} {\path{doi:10.1364/JOSAA.20.000040}}.
\newline\urlprefix\url{https://opg.optica.org/abstract.cfm?URI=josaa-20-1-40}

\bibitem{Faulkner2004}
H.~M. Faulkner, J.~M. Rodenburg, {Movable aperture lensless transmission microscopy: A novel phase retrieval algorithm}, Physical Review Letters 93~(2) (2004) 2--5.
\newblock \href {https://doi.org/10.1103/PhysRevLett.93.023903} {\path{doi:10.1103/PhysRevLett.93.023903}}.

\bibitem{Rodenburg2004}
J.~M. Rodenburg, H.~M. Faulkner, {A phase retrieval algorithm for shifting illumination}, Applied Physics Letters 85~(20) (2004) 4795--4797.
\newblock \href {https://doi.org/10.1063/1.1823034} {\path{doi:10.1063/1.1823034}}.

\bibitem{Thibault2008}
P.~Thibault, M.~Dierolf, A.~Menzel, O.~Bunk, C.~David, F.~Pfeiffer, {High-resolution scanning X-ray diffraction microscopy}, Science 321~(5887) (2008) 379--382.
\newblock \href {https://doi.org/10.1126/science.1158573} {\path{doi:10.1126/science.1158573}}.

\bibitem{Thibault2009}
P.~Thibault, M.~Dierolf, O.~Bunk, A.~Menzel, F.~Pfeiffer, {Probe retrieval in ptychographic coherent diffractive imaging}, Ultramicroscopy 109~(4) (2009) 338--343.
\newblock \href {https://doi.org/10.1016/j.ultramic.2008.12.011} {\path{doi:10.1016/j.ultramic.2008.12.011}}.

\bibitem{Maiden2009}
A.~M. Maiden, J.~M. Rodenburg, \href{https://linkinghub.elsevier.com/retrieve/pii/S0304399109001284}{{An improved ptychographical phase retrieval algorithm for diffractive imaging}}, Ultramicroscopy 109~(10) (2009) 1256--1262.
\newblock \href {https://doi.org/10.1016/j.ultramic.2009.05.012} {\path{doi:10.1016/j.ultramic.2009.05.012}}.
\newline\urlprefix\url{https://linkinghub.elsevier.com/retrieve/pii/S0304399109001284}

\bibitem{Maiden2017}
A.~Maiden, D.~Johnson, P.~Li, {Further improvements to the ptychographical iterative engine}, Optica 4~(7) (2017) 736.
\newblock \href {https://doi.org/10.1364/optica.4.000736} {\path{doi:10.1364/optica.4.000736}}.

\bibitem{Maiden2024}
A.~Maiden, W.~Mei, P.~Li, {WASP: Weighted Average of Sequential Projections for ptychographic phase retrieval}, Optics Express 32~(12) (2024) 21327--21344.
\newblock \href {https://doi.org/10.1364/oe.516946} {\path{doi:10.1364/oe.516946}}.

\bibitem{Gerchberg1972}
R.~Gerchberg, W.~Saxton, \href{http://ci.nii.ac.jp/naid/10010556614/%0Ahttp://stacks.iop.org/1063-7818/39/i=6/a=A06?key=crossref.08ac46f1c5d5b70d17b930a813765f44}{{A Practical Algorithm for the Determination of Phase from Image and Diffraction Plane Pictures}}, Optik 35~(2) (1972).
\newline\urlprefix\url{http://ci.nii.ac.jp/naid/10010556614/%0Ahttp://stacks.iop.org/1063-7818/39/i=6/a=A06?key=crossref.08ac46f1c5d5b70d17b930a813765f44}

\bibitem{Gerchberg1974}
R.~Gerchberg, \href{https://www.tandfonline.com/doi/full/10.1080/713818946}{{Super-resolution through Error Energy Reduction}}, Optica Acta: International Journal of Optics 21~(9) (1974) 709--720.
\newblock \href {https://doi.org/10.1080/713818946} {\path{doi:10.1080/713818946}}.
\newline\urlprefix\url{https://www.tandfonline.com/doi/full/10.1080/713818946}

\bibitem{Guizar-Sicairos2008}
M.~Guizar-Sicairos, J.~R. Fienup, {Phase retrieval with transverse translation diversity: a nonlinear optimization approach}, Optics Express 16~(10) (2008) 7264.
\newblock \href {https://doi.org/10.1364/oe.16.007264} {\path{doi:10.1364/oe.16.007264}}.

\bibitem{Godard2012}
P.~Godard, M.~Allain, V.~Chamard, J.~Rodenburg, \href{https://opg.optica.org/abstract.cfm?URI=oe-20-23-25914}{{Noise models for low counting rate coherent diffraction imaging}}, Optics Express 20~(23) (2012) 25914.
\newblock \href {https://doi.org/10.1364/OE.20.025914} {\path{doi:10.1364/OE.20.025914}}.
\newline\urlprefix\url{https://opg.optica.org/abstract.cfm?URI=oe-20-23-25914}

\bibitem{Bian2016}
L.~Bian, J.~Suo, J.~Chung, X.~Ou, C.~Yang, F.~Chen, Q.~Dai, {Fourier ptychographic reconstruction using Poisson maximum likelihood and truncated Wirtinger gradient}, Scientific Reports 6~(June) (2016) 1--10.
\newblock \href {http://arxiv.org/abs/1603.04746} {\path{arXiv:1603.04746}}, \href {https://doi.org/10.1038/srep27384} {\path{doi:10.1038/srep27384}}.

\bibitem{Odstrcil2018}
M.~Odstr{\v{c}}il, A.~Menzel, M.~Guizar-Sicairos, {Iterative least-squares solver for generalized maximum-likelihood ptychography}, Optics Express 26~(3) (2018) 3108.
\newblock \href {https://doi.org/10.1364/oe.26.003108} {\path{doi:10.1364/oe.26.003108}}.

\bibitem{Thibault2012}
P.~Thibault, M.~Guizar-Sicairos, {Maximum-likelihood refinement for coherent diffractive imaging}, New Journal of Physics 14 (2012) 1--20.
\newblock \href {https://doi.org/10.1088/1367-2630/14/6/063004} {\path{doi:10.1088/1367-2630/14/6/063004}}.

\bibitem{Pham2019a}
M.~Pham, P.~Yin, A.~Rana, S.~Osher, J.~Miao, \href{https://opg.optica.org/abstract.cfm?URI=oe-27-3-2792}{{Generalized proximal smoothing (GPS) for phase retrieval}}, Optics Express 27~(3) (2019) 2792.
\newblock \href {http://arxiv.org/abs/1803.05610} {\path{arXiv:1803.05610}}, \href {https://doi.org/10.1364/OE.27.002792} {\path{doi:10.1364/OE.27.002792}}.
\newline\urlprefix\url{https://opg.optica.org/abstract.cfm?URI=oe-27-3-2792}

\bibitem{Schloz2020}
M.~Schloz, T.~C. Pekin, Z.~Chen, W.~{Van den Broek}, D.~A. Muller, C.~T. Koch, {Overcoming information reduced data and experimentally uncertain parameters in ptychography with regularized optimization}, Optics Express 28~(19) (2020) 28306.
\newblock \href {http://arxiv.org/abs/2005.01530} {\path{arXiv:2005.01530}}, \href {https://doi.org/10.1364/oe.396925} {\path{doi:10.1364/oe.396925}}.

\bibitem{Lee2024}
K.~C. Lee, H.~Chae, S.~Xu, K.~Lee, R.~Horstmeyer, S.~A. Lee, B.-W. Hong, \href{https://opg.optica.org/abstract.cfm?URI=oe-32-14-25343}{{Anisotropic regularization for sparsely sampled and noise-robust Fourier ptychography}}, Optics Express 32~(14) (2024) 25343.
\newblock \href {https://doi.org/10.1364/OE.529023} {\path{doi:10.1364/OE.529023}}.
\newline\urlprefix\url{https://opg.optica.org/abstract.cfm?URI=oe-32-14-25343}

\bibitem{Herdegen2024}
Z.~Herdegen, B.~Diederichs, K.~M{\"{u}}ller-Caspary, \href{https://link.aps.org/doi/10.1103/PhysRevB.110.064102}{{Thermal vibrations in the inversion of dynamical electron scattering}}, Physical Review B 110~(6) (2024) 064102.
\newblock \href {https://doi.org/10.1103/PhysRevB.110.064102} {\path{doi:10.1103/PhysRevB.110.064102}}.
\newline\urlprefix\url{https://link.aps.org/doi/10.1103/PhysRevB.110.064102}

\bibitem{Diederichs2024}
B.~Diederichs, Z.~Herdegen, A.~Strauch, F.~Filbir, K.~M{\"{u}}ller-Caspary, {Exact inversion of partially coherent dynamical electron scattering for picometric structure retrieval}, Nature Communications 15~(1) (2024).
\newblock \href {https://doi.org/10.1038/s41467-023-44268-x} {\path{doi:10.1038/s41467-023-44268-x}}.

\bibitem{Yang2024b}
W.~Yang, H.~Sha, J.~Cui, L.~Mao, R.~Yu, \href{https://www.nature.com/articles/s41565-023-01595-w}{{Local-orbital ptychography for ultrahigh-resolution imaging}}, Nature Nanotechnology (jan 2024).
\newblock \href {https://doi.org/10.1038/s41565-023-01595-w} {\path{doi:10.1038/s41565-023-01595-w}}.
\newline\urlprefix\url{https://www.nature.com/articles/s41565-023-01595-w}

\bibitem{Yang2025a}
W.~Yang, H.~Sha, J.~Cui, R.~Yu, \href{http://arxiv.org/abs/2502.18294}{{Imaging thick objects with deep-sub-angstrom resolution and deep-sub-picometer precision}} (feb 2025).
\newblock \href {http://arxiv.org/abs/2502.18294} {\path{arXiv:2502.18294}}.
\newline\urlprefix\url{http://arxiv.org/abs/2502.18294}

\bibitem{Chang2019b}
H.~Chang, P.~Enfedaque, J.~Zhang, J.~Reinhardt, B.~Enders, Y.-S. Yu, D.~Shapiro, C.~G. Schroer, T.~Zeng, S.~Marchesini, \href{https://opg.optica.org/abstract.cfm?URI=oe-27-8-10395}{{Advanced denoising for X-ray ptychography}}, Optics Express 27~(8) (2019) 10395.
\newblock \href {http://arxiv.org/abs/1811.02081} {\path{arXiv:1811.02081}}, \href {https://doi.org/10.1364/OE.27.010395} {\path{doi:10.1364/OE.27.010395}}.
\newline\urlprefix\url{https://opg.optica.org/abstract.cfm?URI=oe-27-8-10395}

\bibitem{Leidl2024}
M.~L. Leidl, B.~Diederichs, C.~Sachse, K.~M{\"{u}}ller-Caspary, \href{https://doi.org/10.1016/j.micron.2024.103688 https://linkinghub.elsevier.com/retrieve/pii/S0968432824001057}{{Influence of loss function and electron dose on ptychography of 2D materials using the Wirtinger flow}}, Micron 185~(July) (2024) 103688.
\newblock \href {https://doi.org/10.1016/j.micron.2024.103688} {\path{doi:10.1016/j.micron.2024.103688}}.
\newline\urlprefix\url{https://doi.org/10.1016/j.micron.2024.103688 https://linkinghub.elsevier.com/retrieve/pii/S0968432824001057}

\bibitem{Seifert2024}
J.~Seifert, Y.~Shao, R.~van Dam, D.~Bouchet, T.~van Leeuwen, A.~P. Mosk, \href{http://arxiv.org/abs/2308.02436 http://dx.doi.org/10.1364/OL.502344}{{Maximum-likelihood estimation in ptychography in the presence of Poisson-Gaussian noise statistics}} (2023) 1--11\href {http://arxiv.org/abs/2308.02436} {\path{arXiv:2308.02436}}, \href {https://doi.org/10.1364/OL.502344} {\path{doi:10.1364/OL.502344}}.
\newline\urlprefix\url{http://arxiv.org/abs/2308.02436 http://dx.doi.org/10.1364/OL.502344}

\bibitem{Melnyk2023}
O.~Melnyk, \href{http://arxiv.org/abs/2306.08750}{{Convergence properties of gradient methods for blind ptychography}} (jun 2023).
\newblock \href {http://arxiv.org/abs/2306.08750} {\path{arXiv:2306.08750}}.
\newline\urlprefix\url{http://arxiv.org/abs/2306.08750}

\bibitem{Katvotnik2013}
V.~Katkovnik, J.~Astola, \href{https://opg.optica.org/abstract.cfm?URI=josaa-30-3-367}{{Sparse ptychographical coherent diffractive imaging from noisy measurements}}, Journal of the Optical Society of America A 30~(3) (2013) 367.
\newblock \href {https://doi.org/10.1364/JOSAA.30.000367} {\path{doi:10.1364/JOSAA.30.000367}}.
\newline\urlprefix\url{https://opg.optica.org/abstract.cfm?URI=josaa-30-3-367}

\bibitem{Marchesini2016}
S.~Marchesini, H.~Krishnan, B.~J. Daurer, D.~A. Shapiro, T.~Perciano, J.~A. Sethian, F.~R. Maia, {SHARP: A distributed GPU-based ptychographic solver}, Journal of Applied Crystallography 49~(4) (2016) 1245--1252.
\newblock \href {http://arxiv.org/abs/1602.01448} {\path{arXiv:1602.01448}}, \href {https://doi.org/10.1107/S1600576716008074} {\path{doi:10.1107/S1600576716008074}}.

\bibitem{Yu2022}
X.~Yu, V.~Nikitin, D.~J. Ching, S.~Aslan, D.~G{\"{u}}rsoy, T.~Bi{\c{c}}er, \href{https://doi.org/10.1038/s41598-022-09430-3}{{Scalable and accurate multi-GPU-based image reconstruction of large-scale ptychography data}}, Scientific Reports 12~(1) (2022) 1--16.
\newblock \href {http://arxiv.org/abs/2106.07575} {\path{arXiv:2106.07575}}, \href {https://doi.org/10.1038/s41598-022-09430-3} {\path{doi:10.1038/s41598-022-09430-3}}.
\newline\urlprefix\url{https://doi.org/10.1038/s41598-022-09430-3}

\bibitem{Wang2022c}
X.~Wang, A.~Tsaris, D.~Mukherjee, M.~Wahib, P.~Chen, M.~Oxley, O.~Ovchinnikova, J.~Hinkle, \href{https://ieeexplore.ieee.org/document/10045785/}{{Image Gradient Decomposition for Parallel and Memory-Efficient Ptychographic Reconstruction}}, in: SC22: International Conference for High Performance Computing, Networking, Storage and Analysis, Vol. 2022-Novem, IEEE, 2022, pp. 1--13.
\newblock \href {http://arxiv.org/abs/2205.06327} {\path{arXiv:2205.06327}}, \href {https://doi.org/10.1109/SC41404.2022.00013} {\path{doi:10.1109/SC41404.2022.00013}}.
\newline\urlprefix\url{https://ieeexplore.ieee.org/document/10045785/}

\bibitem{Mukherjee2022}
D.~Mukherjee, K.~M. Roccapriore, A.~Al-Najjar, A.~Ghosh, J.~D. Hinkle, A.~R. Lupini, R.~K. Vasudevan, S.~V. Kalinin, O.~S. Ovchinnikova, M.~A. Ziatdinov, N.~S. Rao, \href{https://academic.oup.com/mt/article/30/6/10/6995490}{{A Roadmap for Edge Computing Enabled Automated Multidimensional Transmission Electron Microscopy}}, Microscopy Today 30~(6) (2022) 10--19.
\newblock \href {http://arxiv.org/abs/2210.02538} {\path{arXiv:2210.02538}}, \href {https://doi.org/10.1017/S1551929522001286} {\path{doi:10.1017/S1551929522001286}}.
\newline\urlprefix\url{https://academic.oup.com/mt/article/30/6/10/6995490}

\bibitem{Welborn2024}
S.~S. Welborn, C.~Harris, S.~M. Ribet, G.~Varnavides, C.~Ophus, B.~Enders, P.~Ercius, \href{http://arxiv.org/abs/2407.03215 https://academic.oup.com/mam/advance-article/doi/10.1093/mam/ozae109/7900426}{{Streaming Large-Scale Microscopy Data to a Supercomputing Facility}}, Microscopy and Microanalysis (2024) 1--9\href {http://arxiv.org/abs/2407.03215} {\path{arXiv:2407.03215}}, \href {https://doi.org/10.1093/mam/ozae109} {\path{doi:10.1093/mam/ozae109}}.
\newline\urlprefix\url{http://arxiv.org/abs/2407.03215 https://academic.oup.com/mam/advance-article/doi/10.1093/mam/ozae109/7900426}

\bibitem{Bates1989}
R.~Bates, J.~Rodenburg, \href{https://linkinghub.elsevier.com/retrieve/pii/0304399189900521}{{Sub-{\aa}ngstr{\"{o}}m transmission microscopy: A fourier transform algorithm for microdiffraction plane intensity information}}, Ultramicroscopy 31~(3) (1989) 303--307.
\newblock \href {https://doi.org/10.1016/0304-3991(89)90052-1} {\path{doi:10.1016/0304-3991(89)90052-1}}.
\newline\urlprefix\url{https://linkinghub.elsevier.com/retrieve/pii/0304399189900521}

\bibitem{Rodenburg1992}
J.~M. Rodenburg, R.~H.~T. Bates, \href{https://royalsocietypublishing.org/doi/10.1098/rsta.1992.0050}{{The theory of super-resolution electron microscopy via Wigner-distribution deconvolution}}, Philosophical Transactions of the Royal Society of London. Series A: Physical and Engineering Sciences 339~(1655) (1992) 521--553.
\newblock \href {https://doi.org/10.1098/rsta.1992.0050} {\path{doi:10.1098/rsta.1992.0050}}.
\newline\urlprefix\url{https://royalsocietypublishing.org/doi/10.1098/rsta.1992.0050}

\bibitem{Rodenburg1993}
J.~Rodenburg, B.~McCallum, P.~Nellist, \href{https://linkinghub.elsevier.com/retrieve/pii/0304399193901057}{{Experimental tests on double-resolution coherent imaging via STEM}}, Ultramicroscopy 48~(3) (1993) 304--314.
\newblock \href {https://doi.org/10.1016/0304-3991(93)90105-7} {\path{doi:10.1016/0304-3991(93)90105-7}}.
\newline\urlprefix\url{https://linkinghub.elsevier.com/retrieve/pii/0304399193901057}

\bibitem{Strauch2021}
A.~Strauch, D.~Weber, A.~Clausen, A.~Lesnichaia, A.~Bangun, B.~M{\"{a}}rz, F.~J. Lyu, Q.~Chen, A.~Rosenauer, R.~Dunin-Borkowski, K.~M{\"{u}}ller-Caspary, {Live Processing of Momentum-Resolved STEM Data for First Moment Imaging and Ptychography}, Microscopy and Microanalysis 27~(5) (2021) 1078--1092.
\newblock \href {http://arxiv.org/abs/2106.13457} {\path{arXiv:2106.13457}}, \href {https://doi.org/10.1017/S1431927621012423} {\path{doi:10.1017/S1431927621012423}}.

\bibitem{Bangun2023}
A.~Bangun, P.~F. Baumeister, A.~Clausen, D.~Weber, R.~E. Dunin-Borkowski, \href{https://doi.org/10.1093/micmic/ozad021}{{Wigner Distribution Deconvolution Adaptation for Live Ptychography Reconstruction}}, Microscopy and Microanalysis 29~(3) (2023) 994--1008.
\newblock \href {http://arxiv.org/abs/2212.01309} {\path{arXiv:2212.01309}}, \href {https://doi.org/10.1093/micmic/ozad021} {\path{doi:10.1093/micmic/ozad021}}.
\newline\urlprefix\url{https://doi.org/10.1093/micmic/ozad021}

\bibitem{Weber2024}
D.~Weber, S.~Ehrig, A.~Schropp, A.~Clausen, S.~Achilles, N.~Hoffmann, M.~Bussmann, R.~E. Dunin-Borkowski, C.~G. Schroer, \href{http://arxiv.org/abs/2308.10674 http://dx.doi.org/10.1093/mam/ozae004 https://academic.oup.com/mam/article/30/1/103/7611447}{{Live Iterative Ptychography}}, Microscopy and Microanalysis 30~(1) (2024) 103--117.
\newblock \href {http://arxiv.org/abs/2308.10674} {\path{arXiv:2308.10674}}, \href {https://doi.org/10.1093/mam/ozae004} {\path{doi:10.1093/mam/ozae004}}.
\newline\urlprefix\url{http://arxiv.org/abs/2308.10674 http://dx.doi.org/10.1093/mam/ozae004 https://academic.oup.com/mam/article/30/1/103/7611447}

\bibitem{Pennycook2015}
T.~J. Pennycook, A.~R. Lupini, H.~Yang, M.~F. Murfitt, L.~Jones, P.~D. Nellist, \href{http://www.sciencedirect.com/science/article/pii/S0304399114001934 https://linkinghub.elsevier.com/retrieve/pii/S0304399114001934}{{Efficient phase contrast imaging in STEM using a pixelated detector. Part 1: Experimental demonstration at atomic resolution}}, Ultramicroscopy 151~(0) (2015) 160--167.
\newblock \href {https://doi.org/10.1016/j.ultramic.2014.09.013} {\path{doi:10.1016/j.ultramic.2014.09.013}}.
\newline\urlprefix\url{http://www.sciencedirect.com/science/article/pii/S0304399114001934 https://linkinghub.elsevier.com/retrieve/pii/S0304399114001934}

\bibitem{Yang2015b}
H.~Yang, T.~J. Pennycook, P.~D. Nellist, \href{http://www.sciencedirect.com/science/article/pii/S0304399114002058}{{Efficient phase contrast imaging in STEM using a pixelated detector. Part II: Optimisation of imaging conditions}}, Ultramicroscopy 151~(0) (2015) 232--239.
\newblock \href {https://doi.org/10.1016/j.ultramic.2014.10.013} {\path{doi:10.1016/j.ultramic.2014.10.013}}.
\newline\urlprefix\url{http://www.sciencedirect.com/science/article/pii/S0304399114002058}

\bibitem{OLeary2020}
C.~M. O'Leary, C.~S. Allen, C.~Huang, J.~S. Kim, E.~Liberti, P.~D. Nellist, A.~I. Kirkland, \href{https://pubs.aip.org/apl/article/116/12/124101/570971/Phase-reconstruction-using-fast-binary-4D-STEM}{{Phase reconstruction using fast binary 4D STEM data}}, Applied Physics Letters 116~(12) (2020) 124101.
\newblock \href {https://doi.org/10.1063/1.5143213} {\path{doi:10.1063/1.5143213}}.
\newline\urlprefix\url{https://pubs.aip.org/apl/article/116/12/124101/570971/Phase-reconstruction-using-fast-binary-4D-STEM}

\bibitem{OLeary2021}
C.~M. O'Leary, G.~T. Martinez, E.~Liberti, M.~J. Humphry, A.~I. Kirkland, P.~D. Nellist, \href{https://doi.org/10.1016/j.ultramic.2020.113189 https://linkinghub.elsevier.com/retrieve/pii/S0304399120303314}{{Contrast transfer and noise considerations in focused-probe electron ptychography}}, Ultramicroscopy 221~(December 2020) (2021) 113189.
\newblock \href {https://doi.org/10.1016/j.ultramic.2020.113189} {\path{doi:10.1016/j.ultramic.2020.113189}}.
\newline\urlprefix\url{https://doi.org/10.1016/j.ultramic.2020.113189 https://linkinghub.elsevier.com/retrieve/pii/S0304399120303314}

\bibitem{Yang2016}
H.~Yang, R.~N. Rutte, L.~Jones, M.~Simson, R.~Sagawa, H.~Ryll, M.~Huth, T.~J. Pennycook, M.~L.~H. Green, H.~Soltau, Y.~Kondo, B.~G. Davis, P.~D. Nellist, \href{http://dx.doi.org/10.1038/ncomms12532}{{Simultaneous atomic-resolution electron ptychography and Z-contrast imaging of light and heavy elements in complex nanostructures}}, Nature Communications 7 (2016) 1--8.
\newblock \href {https://doi.org/10.1038/ncomms12532} {\path{doi:10.1038/ncomms12532}}.
\newline\urlprefix\url{http://dx.doi.org/10.1038/ncomms12532}

\bibitem{Yang2017}
H.~Yang, I.~MacLaren, L.~Jones, G.~T. Martinez, M.~Simson, M.~Huth, H.~Ryll, H.~Soltau, R.~Sagawa, Y.~Kondo, C.~Ophus, P.~Ercius, L.~Jin, A.~Kov{\'{a}}cs, P.~D. Nellist, {Electron ptychographic phase imaging of light elements in crystalline materials using Wigner distribution deconvolution}, Ultramicroscopy 180 (2017) 173--179.
\newblock \href {https://doi.org/10.1016/j.ultramic.2017.02.006} {\path{doi:10.1016/j.ultramic.2017.02.006}}.

\bibitem{Wang2017a}
P.~Wang, F.~Zhang, S.~Gao, M.~Zhang, A.~I. Kirkland, {Electron ptychographic diffractive imaging of boron atoms in LaB 6 crystals}, Scientific Reports 7~(1) (2017) 1--8.
\newblock \href {https://doi.org/10.1038/s41598-017-02778-x} {\path{doi:10.1038/s41598-017-02778-x}}.

\bibitem{Leidl2023}
M.~L. Leidl, C.~Sachse, K.~M{\"{u}}ller-Caspary, \href{https://scripts.iucr.org/cgi-bin/paper?S2052252523004505}{{Dynamical scattering in ice-embedded proteins in conventional and scanning transmission electron microscopy}}, IUCrJ 10~(4) (2023) 867--876.
\newblock \href {https://doi.org/10.1107/S2052252523004505} {\path{doi:10.1107/S2052252523004505}}.
\newline\urlprefix\url{https://scripts.iucr.org/cgi-bin/paper?S2052252523004505}

\bibitem{McCallum1992}
B.~McCallum, J.~Rodenburg, \href{https://linkinghub.elsevier.com/retrieve/pii/030439919290149E}{{Two-dimensional demonstration of Wigner phase-retrieval microscopy in the STEM configuration}}, Ultramicroscopy 45~(3-4) (1992) 371--380.
\newblock \href {https://doi.org/10.1016/0304-3991(92)90149-E} {\path{doi:10.1016/0304-3991(92)90149-E}}.
\newline\urlprefix\url{https://linkinghub.elsevier.com/retrieve/pii/030439919290149E}

\bibitem{Nellist1994}
P.~Nellist, J.~Rodenburg, \href{https://linkinghub.elsevier.com/retrieve/pii/0304399194900922}{{Beyond the conventional information limit: the relevant coherence function}}, Ultramicroscopy 54~(1) (1994) 61--74.
\newblock \href {https://doi.org/10.1016/0304-3991(94)90092-2} {\path{doi:10.1016/0304-3991(94)90092-2}}.
\newline\urlprefix\url{https://linkinghub.elsevier.com/retrieve/pii/0304399194900922}

\bibitem{Li2014}
P.~Li, T.~B. Edo, J.~M. Rodenburg, \href{http://dx.doi.org/10.1016/j.ultramic.2014.07.004}{{Ptychographic inversion via Wigner distribution deconvolution: Noise suppression and probe design}}, Ultramicroscopy 147 (2014) 106--113.
\newblock \href {https://doi.org/10.1016/j.ultramic.2014.07.004} {\path{doi:10.1016/j.ultramic.2014.07.004}}.
\newline\urlprefix\url{http://dx.doi.org/10.1016/j.ultramic.2014.07.004}

\bibitem{Muller2014}
K.~M{\"{u}}ller, F.~F. Krause, A.~B{\'{e}}ch{\'{e}}, M.~Schowalter, V.~Galioit, S.~L{\"{o}}ffler, J.~Verbeeck, J.~Zweck, P.~Schattschneider, A.~Rosenauer, \href{http://dx.doi.org/10.1038/ncomms6653 https://www.nature.com/articles/ncomms6653}{{Atomic electric fields revealed by a quantum mechanical approach to electron picodiffraction}}, Nature Communications 5~(1) (2014) 5653.
\newblock \href {https://doi.org/10.1038/ncomms6653} {\path{doi:10.1038/ncomms6653}}.
\newline\urlprefix\url{http://dx.doi.org/10.1038/ncomms6653 https://www.nature.com/articles/ncomms6653}

\bibitem{Lazic2016}
I.~Lazi{\'{c}}, E.~G.~T. Bosch, S.~Lazar, \href{http://www.sciencedirect.com/science/article/pii/S0304399115300449}{{Phase contrast \{STEM\} for thin samples: Integrated differential phase contrast}}, Ultramicroscopy 160 (2016) 265--280.
\newblock \href {https://doi.org/http://doi.org/10.1016/j.ultramic.2015.10.011} {\path{doi:http://doi.org/10.1016/j.ultramic.2015.10.011}}.
\newline\urlprefix\url{http://www.sciencedirect.com/science/article/pii/S0304399115300449}

\bibitem{Yang2016a}
H.~Yang, P.~Ercius, P.~D. Nellist, C.~Ophus, \href{http://dx.doi.org/10.1016/j.ultramic.2016.09.002}{{Enhanced phase contrast transfer using ptychography combined with a pre-specimen phase plate in a scanning transmission electron microscope}}, Ultramicroscopy 171 (2016) 117--125.
\newblock \href {https://doi.org/10.1016/j.ultramic.2016.09.002} {\path{doi:10.1016/j.ultramic.2016.09.002}}.
\newline\urlprefix\url{http://dx.doi.org/10.1016/j.ultramic.2016.09.002}

\bibitem{Hue2010}
F.~H{\"{u}}e, J.~M. Rodenburg, A.~M. Maiden, F.~Sweeney, P.~A. Midgley, {Wave-front phase retrieval in transmission electron microscopy via ptychography}, Physical Review B - Condensed Matter and Materials Physics 82~(12) (2010) 1--4.
\newblock \href {https://doi.org/10.1103/PhysRevB.82.121415} {\path{doi:10.1103/PhysRevB.82.121415}}.

\bibitem{Song2019}
J.~Song, C.~S. Allen, S.~Gao, C.~Huang, H.~Sawada, X.~Pan, J.~Warner, P.~Wang, A.~I. Kirkland, \href{http://www.nature.com/articles/s41598-019-40413-z}{{Atomic Resolution Defocused Electron Ptychography at Low Dose with a Fast, Direct Electron Detector}}, Scientific Reports 9~(1) (2019) 3919.
\newblock \href {https://doi.org/10.1038/s41598-019-40413-z} {\path{doi:10.1038/s41598-019-40413-z}}.
\newline\urlprefix\url{http://www.nature.com/articles/s41598-019-40413-z}

\bibitem{Vine2009a}
D.~J. Vine, G.~J. Williams, B.~Abbey, M.~A. Pfeifer, J.~N. Clark, M.~D. de~Jonge, I.~McNulty, A.~G. Peele, K.~A. Nugent, \href{https://link.aps.org/doi/10.1103/PhysRevA.80.063823}{{Ptychographic Fresnel coherent diffractive imaging}}, Physical Review A 80~(6) (2009) 063823.
\newblock \href {https://doi.org/10.1103/PhysRevA.80.063823} {\path{doi:10.1103/PhysRevA.80.063823}}.
\newline\urlprefix\url{https://link.aps.org/doi/10.1103/PhysRevA.80.063823}

\bibitem{Cowley1972}
J.~M. Cowley, I.~Sumio, S.~Iijima, \href{https://www.degruyter.com/document/doi/10.1515/zna-1972-0312/html https://www.degruyter.com/view/j/zna.1972.27.issue-3/zna-1972-0312/zna-1972-0312.xml}{{Electron Microscope Image Contrast for Thin Crystal}}, Zeitschrift f{\"{u}}r Naturforschung A 27~(3) (1972) 445--451.
\newblock \href {https://doi.org/10.1515/zna-1972-0312} {\path{doi:10.1515/zna-1972-0312}}.
\newline\urlprefix\url{https://www.degruyter.com/document/doi/10.1515/zna-1972-0312/html https://www.degruyter.com/view/j/zna.1972.27.issue-3/zna-1972-0312/zna-1972-0312.xml}

\bibitem{VanBenthem2005}
K.~van Benthem, A.~R. Lupini, M.~Kim, H.~S. Baik, S.~Doh, J.-H. Lee, M.~P. Oxley, S.~D. Findlay, L.~J. Allen, J.~T. Luck, S.~J. Pennycook, \href{https://pubs.aip.org/aip/apl/article/330557}{{Three-dimensional imaging of individual hafnium atoms inside a semiconductor device}}, Applied Physics Letters 87~(3) (2005) 034104.
\newblock \href {https://doi.org/10.1063/1.1991989} {\path{doi:10.1063/1.1991989}}.
\newline\urlprefix\url{https://pubs.aip.org/aip/apl/article/330557}

\bibitem{VanBenthem2006}
K.~van Benthem, A.~R. Lupini, M.~P. Oxley, S.~D. Findlay, L.~J. Allen, S.~J. Pennycook, \href{https://linkinghub.elsevier.com/retrieve/pii/S0304399106001069}{{Three-dimensional ADF imaging of individual atoms by through-focal series scanning transmission electron microscopy}}, Ultramicroscopy 106~(11-12) (2006) 1062--1068.
\newblock \href {https://doi.org/10.1016/j.ultramic.2006.04.020} {\path{doi:10.1016/j.ultramic.2006.04.020}}.
\newline\urlprefix\url{https://linkinghub.elsevier.com/retrieve/pii/S0304399106001069}

\bibitem{VanDyck1996}
D.~{Van Dyck}, M.~{Op de Beeck}, \href{https://linkinghub.elsevier.com/retrieve/pii/0304399196000083}{{A simple intuitive theory for electron diffraction}}, Ultramicroscopy 64~(1-4) (1996) 99--107.
\newblock \href {https://doi.org/10.1016/0304-3991(96)00008-3} {\path{doi:10.1016/0304-3991(96)00008-3}}.
\newline\urlprefix\url{https://linkinghub.elsevier.com/retrieve/pii/0304399196000083}

\bibitem{Humphreys1968}
C.~J. Humphreys, P.~B. Hirsch, {Absorption parameters in electron diffraction theory}, Philosophical Magazine 18~(151) (1968) 115--122.
\newblock \href {https://doi.org/10.1080/14786436808227313} {\path{doi:10.1080/14786436808227313}}.

\bibitem{Yoshioka1957}
H.~Yoshioka, \href{https://journals.jps.jp/doi/10.1143/JPSJ.12.618 http://jpsj.ipap.jp/link?JPSJ/12/618/}{{Effect of Inelastic Waves on Electron Diffraction}}, Journal of the Physical Society of Japan 12~(6) (1957) 618--628.
\newblock \href {https://doi.org/10.1143/JPSJ.12.618} {\path{doi:10.1143/JPSJ.12.618}}.
\newline\urlprefix\url{https://journals.jps.jp/doi/10.1143/JPSJ.12.618 http://jpsj.ipap.jp/link?JPSJ/12/618/}

\bibitem{Wang1998d}
Z.~L. Wang, W.~D. Mo, {the “ Optical Potential ” and Multiple Diffuse Scattering in Dynamical Electron Diffraction and Imaging}, Scanning Microscopy 12~(1) (1998) 91--107.

\bibitem{Mkhoyan2008}
K.~A. Mkhoyan, S.~E. MacCagnano-Zacher, M.~G. Thomas, J.~Silcox, \href{http://link.aps.org/doi/10.1103/PhysRevLett.100.025503 https://link.aps.org/doi/10.1103/PhysRevLett.100.025503}{{Critical role of inelastic interactions in quantitative electron microscopy}}, Physical Review Letters 100~(2) (2008) 1--4.
\newblock \href {https://doi.org/10.1103/PhysRevLett.100.025503} {\path{doi:10.1103/PhysRevLett.100.025503}}.
\newline\urlprefix\url{http://link.aps.org/doi/10.1103/PhysRevLett.100.025503 https://link.aps.org/doi/10.1103/PhysRevLett.100.025503}

\bibitem{Beyer2020}
A.~Beyer, F.~F. Krause, H.~L. Robert, S.~Firoozabadi, T.~Grieb, P.~K{\"{u}}kelhan, D.~Heimes, M.~Schowalter, K.~M{\"{u}}ller-Caspary, A.~Rosenauer, K.~Volz, \href{https://www.nature.com/articles/s41598-020-74434-w https://doi.org/10.1038/s41598-020-74434-w}{{Influence of plasmon excitations on atomic-resolution quantitative 4D scanning transmission electron microscopy}}, Scientific Reports 10~(1) (2020) 17890.
\newblock \href {https://doi.org/10.1038/s41598-020-74434-w} {\path{doi:10.1038/s41598-020-74434-w}}.
\newline\urlprefix\url{https://www.nature.com/articles/s41598-020-74434-w https://doi.org/10.1038/s41598-020-74434-w}

\bibitem{Barthel2020}
J.~Barthel, M.~Cattaneo, B.~G. Mendis, S.~D. Findlay, L.~J. Allen, {Angular dependence of fast-electron scattering from materials}, Physical Review B 101~(18) (2020) 1--9.
\newblock \href {https://doi.org/10.1103/PhysRevB.101.184109} {\path{doi:10.1103/PhysRevB.101.184109}}.

\bibitem{Robert2022}
H.~L. Robert, B.~Diederichs, K.~M{\"{u}}ller-Caspary, \href{https://aip.scitation.org/doi/10.1063/5.0129692}{{Contribution of multiple plasmon scattering in low-angle electron diffraction investigated by energy-filtered atomically resolved 4D-STEM}}, Applied Physics Letters 121~(21) (2022) 213502.
\newblock \href {https://doi.org/10.1063/5.0129692} {\path{doi:10.1063/5.0129692}}.
\newline\urlprefix\url{https://aip.scitation.org/doi/10.1063/5.0129692}

\bibitem{Hall1965}
C.~R. Hall, \href{http://www.tandfonline.com/doi/abs/10.1080/14786436508218919}{{The scattering of high energy electrons by the thermal vibrations of crystals}}, Philosophical Magazine 12~(118) (1965) 815--826.
\newblock \href {https://doi.org/10.1080/14786436508218919} {\path{doi:10.1080/14786436508218919}}.
\newline\urlprefix\url{http://www.tandfonline.com/doi/abs/10.1080/14786436508218919}

\bibitem{Hall1965a}
C.~R. Hall, P.~B. Hirsch, \href{http://rspa.royalsocietypublishing.org/content/286/1405/158.abstract}{{Effect of Thermal Diffuse Scattering on Propagation of High Energy Electrons Through Crystals}}, Proc. Roy. Soc. Lond. A 286~(1405) (1965) 158--177.
\newblock \href {https://doi.org/10.1098/rspa.1965.0136} {\path{doi:10.1098/rspa.1965.0136}}.
\newline\urlprefix\url{http://rspa.royalsocietypublishing.org/content/286/1405/158.abstract}

\bibitem{VanDyck2009}
D.~{Van Dyck}, {Is the frozen phonon model adequate to describe inelastic phonon scattering?}, Ultramicroscopy 109~(6) (2009) 677--682.
\newblock \href {https://doi.org/10.1016/j.ultramic.2009.01.001} {\path{doi:10.1016/j.ultramic.2009.01.001}}.

\bibitem{Fujiwara1961}
K.~Fujiwara, \href{https://journals.jps.jp/doi/10.1143/JPSJ.16.2226}{{Relativistic Dynamical Theory of Electron Diffraction}}, Journal of the Physical Society of Japan 16~(11) (1961) 2226--2238.
\newblock \href {https://doi.org/10.1143/JPSJ.16.2226} {\path{doi:10.1143/JPSJ.16.2226}}.
\newline\urlprefix\url{https://journals.jps.jp/doi/10.1143/JPSJ.16.2226}

\bibitem{Bunk2008}
O.~Bunk, M.~Dierolf, S.~Kynde, I.~Johnson, O.~Marti, F.~Pfeiffer, \href{https://linkinghub.elsevier.com/retrieve/pii/S0304399107001969}{{Influence of the overlap parameter on the convergence of the ptychographical iterative engine}}, Ultramicroscopy 108~(5) (2008) 481--487.
\newblock \href {https://doi.org/10.1016/j.ultramic.2007.08.003} {\path{doi:10.1016/j.ultramic.2007.08.003}}.
\newline\urlprefix\url{https://linkinghub.elsevier.com/retrieve/pii/S0304399107001969}

\bibitem{Cowley1969a}
J.~M. Cowley, \href{http://aip.scitation.org/doi/10.1063/1.1652901}{{IMAGE CONTRAST IN A TRANSMISSION SCANNING ELECTRON MICROSCOPE}}, Applied Physics Letters 15~(2) (1969) 58--59.
\newblock \href {https://doi.org/10.1063/1.1652901} {\path{doi:10.1063/1.1652901}}.
\newline\urlprefix\url{http://aip.scitation.org/doi/10.1063/1.1652901}

\bibitem{Krause2017}
F.~F. Krause, A.~Rosenauer, \href{http://www.sciencedirect.com/science/article/pii/S0968432816301421}{{Reciprocity relations in transmission electron microscopy: A rigorous derivation}}, Micron 92~(Supplement C) (2017) 1--5.
\newblock \href {https://doi.org/10.1016/j.micron.2016.09.007} {\path{doi:10.1016/j.micron.2016.09.007}}.
\newline\urlprefix\url{http://www.sciencedirect.com/science/article/pii/S0968432816301421}

\bibitem{Wigner1932}
E.~Wigner, {On the quantum correction for thermodynamic equilibrium}, Physical Review 40~(5) (1932) 749--759.
\newblock \href {https://doi.org/10.1103/PhysRev.40.749} {\path{doi:10.1103/PhysRev.40.749}}.

\bibitem{Bates1986}
R.~T. Bates, M.~J. McDonnell, {Image restoration and reconstruction}, Oxford University Press, Inc., 1986.

\bibitem{Mollenstedt1955}
G.~M{\"{o}}llenstedt, H.~D{\"{u}}ker, \href{http://dx.doi.org/10.1007/BF00621530}{{Fresnelscher Interferenzversuch mit einem Biprisma f{\"{u}}r Elektronenwellen}}, Naturwissenschaften 42 (1955) 41.
\newblock \href {https://doi.org/10.1007/BF00621530} {\path{doi:10.1007/BF00621530}}.
\newline\urlprefix\url{http://dx.doi.org/10.1007/BF00621530}

\bibitem{Tonomura1978}
A.~Tonomura, T.~Matsuda, T.~Komoda, \href{https://dx.doi.org/10.1143/JJAP.17.1137 https://iopscience.iop.org/article/10.1143/JJAP.17.1137}{{Two Beam Interference with Field Emission Electron Beam}}, Japanese Journal of Applied Physics 17~(6) (1978) 1137--1138.
\newblock \href {https://doi.org/10.1143/JJAP.17.1137} {\path{doi:10.1143/JJAP.17.1137}}.
\newline\urlprefix\url{https://dx.doi.org/10.1143/JJAP.17.1137 https://iopscience.iop.org/article/10.1143/JJAP.17.1137}

\bibitem{Winkler2020}
F.~Winkler, J.~Barthel, R.~E. Dunin-Borkowski, K.~M{\"{u}}ller-Caspary, \href{https://doi.org/10.1016/j.ultramic.2019.112926 https://linkinghub.elsevier.com/retrieve/pii/S0304399119303092}{{Direct measurement of electrostatic potentials at the atomic scale: A conceptual comparison between electron holography and scanning transmission electron microscopy}}, Ultramicroscopy 210~(September 2019) (2020) 112926.
\newblock \href {https://doi.org/10.1016/j.ultramic.2019.112926} {\path{doi:10.1016/j.ultramic.2019.112926}}.
\newline\urlprefix\url{https://doi.org/10.1016/j.ultramic.2019.112926 https://linkinghub.elsevier.com/retrieve/pii/S0304399119303092}

\bibitem{Sayre1952}
D.~Sayre, \href{https://doi.org/10.1107/S0365110X52002276}{{Some implications of a theorem due to Shannon}}, Acta Crystallographica 5~(6) (1952) 843.
\newblock \href {https://doi.org/10.1107/S0365110X52002276} {\path{doi:10.1107/S0365110X52002276}}.
\newline\urlprefix\url{https://doi.org/10.1107/S0365110X52002276}

\bibitem{Nellist1995}
P.~D. Nellist, B.~C. McCallum, J.~M. Rodenburg, \href{https://www.nature.com/articles/374630a0}{{Resolution beyond the 'information limit' in transmission electron microscopy}}, Nature 374~(6523) (1995) 630--632.
\newblock \href {https://doi.org/10.1038/374630a0} {\path{doi:10.1038/374630a0}}.
\newline\urlprefix\url{https://www.nature.com/articles/374630a0}

\bibitem{Maiden2011}
A.~M. Maiden, M.~J. Humphry, F.~Zhang, J.~M. Rodenburg, {Superresolution imaging via ptychography}, Journal of the Optical Society of America A 28~(4) (2011) 604.
\newblock \href {https://doi.org/10.1364/josaa.28.000604} {\path{doi:10.1364/josaa.28.000604}}.

\bibitem{Humphry2012}
M.~Humphry, B.~Kraus, A.~Hurst, A.~Maiden, J.~Rodenburg, \href{http://dx.doi.org/10.1038/ncomms1733 https://www.nature.com/articles/ncomms1733}{{Ptychographic electron microscopy using high-angle dark-field scattering for sub-nanometre resolution imaging}}, Nature Communications 3~(1) (2012) 730.
\newblock \href {https://doi.org/10.1038/ncomms1733} {\path{doi:10.1038/ncomms1733}}.
\newline\urlprefix\url{http://dx.doi.org/10.1038/ncomms1733 https://www.nature.com/articles/ncomms1733}

\bibitem{Jiang2018}
Y.~Jiang, Z.~Chen, Y.~Han, P.~Deb, H.~Gao, S.~Xie, P.~Purohit, M.~W. Tate, J.~Park, S.~M. Gruner, V.~Elser, D.~A. Muller, \href{http://www.nature.com/articles/s41586-018-0298-5}{{Electron ptychography of 2D materials to deep sub-{\aa}ngstr{\"{o}}m resolution}} (jul 2018).
\newblock \href {https://doi.org/10.1038/s41586-018-0298-5} {\path{doi:10.1038/s41586-018-0298-5}}.
\newline\urlprefix\url{http://www.nature.com/articles/s41586-018-0298-5}

\bibitem{Li2025a}
S.~Li, N.~Gauquelin, H.~L. {Lalandec Robert}, A.~Annys, C.~Gao, C.~Hofer, T.~J. Pennycook, J.~Verbeeck, \href{http://arxiv.org/abs/2505.09555}{{Improving the low-dose performance of aberration correction in single sideband ptychography}} (may 2025).
\newblock \href {http://arxiv.org/abs/2505.09555} {\path{arXiv:2505.09555}}.
\newline\urlprefix\url{http://arxiv.org/abs/2505.09555}

\bibitem{Seki2018}
T.~Seki, Y.~Ikuhara, N.~Shibata, \href{https://doi.org/10.1016/j.ultramic.2018.06.014 http://www.sciencedirect.com/science/article/pii/S0304399118300603 https://linkinghub.elsevier.com/retrieve/pii/S0304399118300603}{{Theoretical framework of statistical noise in scanning transmission electron microscopy}}, Ultramicroscopy 193~(June) (2018) 118--125.
\newblock \href {https://doi.org/10.1016/j.ultramic.2018.06.014} {\path{doi:10.1016/j.ultramic.2018.06.014}}.
\newline\urlprefix\url{https://doi.org/10.1016/j.ultramic.2018.06.014 http://www.sciencedirect.com/science/article/pii/S0304399118300603 https://linkinghub.elsevier.com/retrieve/pii/S0304399118300603}

\bibitem{Luczka1991}
J.~Luczka, M.~Niemiec, \href{https://iopscience.iop.org/article/10.1088/0305-4470/24/17/010}{{A master equation for quantum systems driven by Poisson white noise}}, Journal of Physics A: Mathematical and General 24~(17) (1991) L1021--L1024.
\newblock \href {https://doi.org/10.1088/0305-4470/24/17/010} {\path{doi:10.1088/0305-4470/24/17/010}}.
\newline\urlprefix\url{https://iopscience.iop.org/article/10.1088/0305-4470/24/17/010}

\bibitem{Clark2023}
L.~Clark, G.~T. Martinez, C.~M. O'Leary, H.~Yang, Z.~Ding, T.~C. Petersen, S.~D. Findlay, P.~D. Nellist, \href{https://academic.oup.com/mam/article/29/1/384/6948181}{{The Effect of Dynamical Scattering on Single-plane Phase Retrieval in Electron Ptychography}}, Microscopy and Microanalysis 29~(1) (2023) 384--394.
\newblock \href {https://doi.org/10.1093/micmic/ozac022} {\path{doi:10.1093/micmic/ozac022}}.
\newline\urlprefix\url{https://academic.oup.com/mam/article/29/1/384/6948181}

\bibitem{Hofer2023}
C.~Hofer, T.~J. Pennycook, \href{https://doi.org/10.1016/j.ultramic.2023.113829}{{Reliable phase quantification in focused probe electron ptychography of thin materials}}, Ultramicroscopy 254~(August) (2023) 113829.
\newblock \href {http://arxiv.org/abs/2307.14171} {\path{arXiv:2307.14171}}, \href {https://doi.org/10.1016/j.ultramic.2023.113829} {\path{doi:10.1016/j.ultramic.2023.113829}}.
\newline\urlprefix\url{https://doi.org/10.1016/j.ultramic.2023.113829}

\bibitem{Verbeeck2018}
J.~Verbeeck, A.~B{\'{e}}ch{\'{e}}, K.~M{\"{u}}ller-Caspary, G.~Guzzinati, M.~A. Luong, M.~{Den Hertog}, \href{https://linkinghub.elsevier.com/retrieve/pii/S0304399117305041}{{Demonstration of a 2 × 2 programmable phase plate for electrons}}, Ultramicroscopy 190 (2018) 58--65.
\newblock \href {https://doi.org/10.1016/j.ultramic.2018.03.017} {\path{doi:10.1016/j.ultramic.2018.03.017}}.
\newline\urlprefix\url{https://linkinghub.elsevier.com/retrieve/pii/S0304399117305041}

\bibitem{VegaIbanez2023}
F.~{Vega Ib{\'{a}}{\~{n}}ez}, A.~B{\'{e}}ch{\'{e}}, J.~Verbeeck, \href{https://academic.oup.com/mam/article/29/1/341/6987564}{{Can a Programmable Phase Plate Serve as an Aberration Corrector in the Transmission Electron Microscope (TEM)?}}, Microscopy and Microanalysis 29~(1) (2023) 341--351.
\newblock \href {http://arxiv.org/abs/2205.07697} {\path{arXiv:2205.07697}}, \href {https://doi.org/10.1017/S1431927622012260} {\path{doi:10.1017/S1431927622012260}}.
\newline\urlprefix\url{https://academic.oup.com/mam/article/29/1/341/6987564}

\bibitem{Yu2023}
C.-P. Yu, F.~{Vega Iba{\~{n}}ez}, A.~B{\'{e}}ch{\'{e}}, J.~Verbeeck, \href{http://arxiv.org/abs/2308.16304 https://scipost.org/10.21468/SciPostPhys.15.6.223}{{Quantum wavefront shaping with a 48-element programmable phase plate for electrons}}, SciPost Physics 15~(6) (2023) 223.
\newblock \href {http://arxiv.org/abs/2308.16304} {\path{arXiv:2308.16304}}, \href {https://doi.org/10.21468/SciPostPhys.15.6.223} {\path{doi:10.21468/SciPostPhys.15.6.223}}.
\newline\urlprefix\url{http://arxiv.org/abs/2308.16304 https://scipost.org/10.21468/SciPostPhys.15.6.223}

\bibitem{Shibata2010}
N.~Shibata, Y.~Kohno, S.~D. Findlay, H.~Sawada, Y.~Kondo, Y.~Ikuhara, \href{http://jmicro.oxfordjournals.org/content/59/6/473.abstract}{{New area detector for atomic-resolution scanning transmission electron microscopy}}, J. Electron Microsc. 59~(6) (2010) 473--479.
\newblock \href {https://doi.org/10.1093/jmicro/dfq014} {\path{doi:10.1093/jmicro/dfq014}}.
\newline\urlprefix\url{http://jmicro.oxfordjournals.org/content/59/6/473.abstract}

\bibitem{Lohr2012}
M.~Lohr, R.~Schregle, M.~Jetter, C.~W{\"{a}}chter, T.~Wunderer, F.~Scholz, J.~Zweck, \href{https://linkinghub.elsevier.com/retrieve/pii/S0304399112000629}{{Differential phase contrast 2.0—Opening new “fields” for an established technique}}, Ultramicroscopy 117 (2012) 7--14.
\newblock \href {https://doi.org/10.1016/j.ultramic.2012.03.020} {\path{doi:10.1016/j.ultramic.2012.03.020}}.
\newline\urlprefix\url{https://linkinghub.elsevier.com/retrieve/pii/S0304399112000629}

\bibitem{Dekkers1974}
N.~H. Dekkers, H.~de~Lang, \href{http://xrm.phys.northwestern.edu/research/pdf_papers/1974/dekkers_optik_1974.pdf}{{Differential Phase Contrast in a STEM}}, Optik 41 (1974) 452--456.
\newline\urlprefix\url{http://xrm.phys.northwestern.edu/research/pdf_papers/1974/dekkers_optik_1974.pdf}

\bibitem{Rose1977}
H.~Rose, \href{https://linkinghub.elsevier.com/retrieve/pii/S0304399176915382 https://doi.org/10.1016/S0304-3991(76)91538-2}{{Nonstandard imaging methods in electron microscopy}}, Ultramicroscopy 2~(1) (1977) 251--267.
\newblock \href {https://doi.org/10.1016/S0304-3991(76)91538-2} {\path{doi:10.1016/S0304-3991(76)91538-2}}.
\newline\urlprefix\url{https://linkinghub.elsevier.com/retrieve/pii/S0304399176915382 https://doi.org/10.1016/S0304-3991(76)91538-2}

\bibitem{Seki2017}
T.~Seki, G.~Sanchez-Santolino, R.~Ishikawa, S.~D. Findlay, Y.~Ikuhara, N.~Shibata, G.~S{\'{a}}nchez-Santolino, R.~Ishikawa, S.~D. Findlay, Y.~Ikuhara, N.~Shibata, \href{http://dx.doi.org/10.1016/j.ultramic.2017.07.013 http://www.sciencedirect.com/science/article/pii/S0304399117302711}{{Quantitative electric field mapping in thin specimens using a segmented detector: Revisiting the transfer function for differential phase contrast}}, Ultramicroscopy 182 (2017) 258--263.
\newblock \href {https://doi.org/10.1016/j.ultramic.2017.07.013} {\path{doi:10.1016/j.ultramic.2017.07.013}}.
\newline\urlprefix\url{http://dx.doi.org/10.1016/j.ultramic.2017.07.013 http://www.sciencedirect.com/science/article/pii/S0304399117302711}

\bibitem{Schwarzhuber2018}
F.~Schwarzhuber, P.~Melzl, S.~P{\"{o}}llath, J.~Zweck, \href{https://www.sciencedirect.com/science/article/pii/S0304399118300615}{{Introducing a non-pixelated and fast centre of mass detector for differential phase contrast microscopy}}, Ultramicroscopy 192 (2018) 21--28.
\newblock \href {https://doi.org/10.1016/j.ultramic.2018.05.003} {\path{doi:10.1016/j.ultramic.2018.05.003}}.
\newline\urlprefix\url{https://www.sciencedirect.com/science/article/pii/S0304399118300615}

\bibitem{Yucelen2018}
E.~Y{\"{u}}celen, I.~Lazic, E.~G.~T. Bosch, I.~Lazi{\'{c}}, E.~G.~T. Bosch, \href{https://doi.org/10.1038/s41598-018-20377-2}{{Phase contrast scanning transmission electron microscopy imaging of light and heavy atoms at the limit of contrast and resolution}}, Scientific Reports 8~(1) (2018) 1--10.
\newblock \href {https://doi.org/10.1038/s41598-018-20377-2} {\path{doi:10.1038/s41598-018-20377-2}}.
\newline\urlprefix\url{https://doi.org/10.1038/s41598-018-20377-2}

\bibitem{Close2015}
R.~Close, Z.~Chen, N.~Shibata, S.~D. Findlay, \href{http://dx.doi.org/10.1016/j.ultramic.2015.09.002 http://www.sciencedirect.com/science/article/pii/S0304399115300255}{{Towards quantitative, atomic-resolution reconstruction of the electrostatic potential via differential phase contrast using electrons}}, Ultramicroscopy 159 (2015) 124--137.
\newblock \href {https://doi.org/10.1016/j.ultramic.2015.09.002} {\path{doi:10.1016/j.ultramic.2015.09.002}}.
\newline\urlprefix\url{http://dx.doi.org/10.1016/j.ultramic.2015.09.002 http://www.sciencedirect.com/science/article/pii/S0304399115300255}

\bibitem{Addiego2020}
C.~Addiego, W.~Gao, X.~Pan, \href{https://doi.org/10.1016/j.ultramic.2019.112850}{{Thickness and defocus dependence of inter-atomic electric fields measured by scanning diffraction}}, Ultramicroscopy 208~(September 2019) (2020) 112850.
\newblock \href {https://doi.org/10.1016/j.ultramic.2019.112850} {\path{doi:10.1016/j.ultramic.2019.112850}}.
\newline\urlprefix\url{https://doi.org/10.1016/j.ultramic.2019.112850}

\bibitem{Burger2020}
J.~B{\"{u}}rger, T.~Riedl, J.~K. Lindner, \href{https://doi.org/10.1016/j.ultramic.2020.113118}{{Influence of lens aberrations, specimen thickness and tilt on differential phase contrast STEM images}}, Ultramicroscopy 219~(February) (2020) 113118.
\newblock \href {https://doi.org/10.1016/j.ultramic.2020.113118} {\path{doi:10.1016/j.ultramic.2020.113118}}.
\newline\urlprefix\url{https://doi.org/10.1016/j.ultramic.2020.113118}

\bibitem{Robert2021}
H.~Robert, I.~Lobato, F.~Lyu, Q.~Chen, S.~{Van Aert}, D.~{Van Dyck}, K.~M{\"{u}}ller-Caspary, \href{https://doi.org/10.1016/j.ultramic.2021.113425 https://linkinghub.elsevier.com/retrieve/pii/S030439912100200X}{{Dynamical diffraction of high-energy electrons investigated by focal series momentum-resolved scanning transmission electron microscopy at atomic resolution}}, Ultramicroscopy 233~(October 2021) (2022) 113425.
\newblock \href {https://doi.org/10.1016/j.ultramic.2021.113425} {\path{doi:10.1016/j.ultramic.2021.113425}}.
\newline\urlprefix\url{https://doi.org/10.1016/j.ultramic.2021.113425 https://linkinghub.elsevier.com/retrieve/pii/S030439912100200X}

\bibitem{Liang2023}
Z.~Liang, D.~Song, B.~Ge, \href{https://doi.org/10.1016/j.ultramic.2023.113686}{{Optimizing experimental parameters of integrated differential phase contrast (iDPC) for atomic resolution imaging}}, Ultramicroscopy 246~(June 2022) (2023) 113686.
\newblock \href {https://doi.org/10.1016/j.ultramic.2023.113686} {\path{doi:10.1016/j.ultramic.2023.113686}}.
\newline\urlprefix\url{https://doi.org/10.1016/j.ultramic.2023.113686}

\bibitem{Black1957}
G.~Black, E.~H. Linfoot, \href{https://royalsocietypublishing.org/doi/10.1098/rspa.1957.0059}{{Spherical aberration and the information content of optical images}}, Proceedings of the Royal Society of London. Series A. Mathematical and Physical Sciences 239~(1219) (1957) 522--540.
\newblock \href {https://doi.org/10.1098/rspa.1957.0059} {\path{doi:10.1098/rspa.1957.0059}}.
\newline\urlprefix\url{https://royalsocietypublishing.org/doi/10.1098/rspa.1957.0059}

\bibitem{Gao2022}
C.~Gao, C.~Hofer, D.~Jannis, A.~B{\'{e}}ch{\'{e}}, J.~Verbeeck, T.~J. Pennycook, \href{https://pubs.aip.org/aip/apl/article/2834118}{{Overcoming contrast reversals in focused probe ptychography of thick materials: An optimal pipeline for efficiently determining local atomic structure in materials science}}, Applied Physics Letters 121~(8) (2022) 081906.
\newblock \href {https://doi.org/10.1063/5.0101895} {\path{doi:10.1063/5.0101895}}.
\newline\urlprefix\url{https://pubs.aip.org/aip/apl/article/2834118}

\bibitem{Shibata2015}
N.~Shibata, S.~D. Findlay, H.~Sasaki, T.~Matsumoto, H.~Sawada, Y.~Kohno, S.~Otomo, R.~Minato, Y.~Ikuhara, {Imaging of built-in electric field at a p-n junction by scanning transmission electron microscopy}, Sci. Rep. 5 (2015) 10040.

\bibitem{Chapman1984}
J.~N. Chapman, \href{http://stacks.iop.org/0022-3727/17/i=4/a=003}{{The investigation of magnetic domain structures in thin foils by electron microscopy}}, J. Phys. D: Appl. Phys. 17~(4) (1984) 623.
\newline\urlprefix\url{http://stacks.iop.org/0022-3727/17/i=4/a=003}

\bibitem{Mahr2022}
C.~Mahr, T.~Grieb, F.~F. Krause, M.~Schowalter, A.~Rosenauer, \href{https://doi.org/10.1016/j.ultramic.2022.113503 https://linkinghub.elsevier.com/retrieve/pii/S0304399122000389}{{Towards the interpretation of a shift of the central beam in nano-beam electron diffraction as a change in mean inner potential}}, Ultramicroscopy 236~(March) (2022) 113503.
\newblock \href {https://doi.org/10.1016/j.ultramic.2022.113503} {\path{doi:10.1016/j.ultramic.2022.113503}}.
\newline\urlprefix\url{https://doi.org/10.1016/j.ultramic.2022.113503 https://linkinghub.elsevier.com/retrieve/pii/S0304399122000389}

\bibitem{Wu2017}
M.~Wu, E.~Spiecker, \href{http://dx.doi.org/10.1016/j.ultramic.2017.03.029 https://linkinghub.elsevier.com/retrieve/pii/S0304399117301341}{{Correlative micro-diffraction and differential phase contrast study of mean inner potential and subtle beam-specimen interaction}}, Ultramicroscopy 176 (2017) 233--245.
\newblock \href {https://doi.org/10.1016/j.ultramic.2017.03.029} {\path{doi:10.1016/j.ultramic.2017.03.029}}.
\newline\urlprefix\url{http://dx.doi.org/10.1016/j.ultramic.2017.03.029 https://linkinghub.elsevier.com/retrieve/pii/S0304399117301341}

\bibitem{McVitie2018}
S.~McVitie, S.~Hughes, K.~Fallon, S.~McFadzean, D.~McGrouther, M.~Krajnak, W.~Legrand, D.~Maccariello, S.~Collin, K.~Garcia, N.~Reyren, V.~Cros, A.~Fert, K.~Zeissler, C.~H. Marrows, \href{http://dx.doi.org/10.1038/s41598-018-23799-0 https://www.nature.com/articles/s41598-018-23799-0}{{A transmission electron microscope study of N{\'{e}}el skyrmion magnetic textures in multilayer thin film systems with large interfacial chiral interaction}}, Scientific Reports 8~(1) (2018) 5703.
\newblock \href {http://arxiv.org/abs/1711.05552} {\path{arXiv:1711.05552}}, \href {https://doi.org/10.1038/s41598-018-23799-0} {\path{doi:10.1038/s41598-018-23799-0}}.
\newline\urlprefix\url{http://dx.doi.org/10.1038/s41598-018-23799-0 https://www.nature.com/articles/s41598-018-23799-0}

\bibitem{Dushimineza2023}
J.~F. Dushimineza, J.~Jo, R.~E. Dunin-Borkowski, K.~M{\"{u}}ller-Caspary, \href{https://doi.org/10.1016/j.ultramic.2023.113808}{{Quantitative electric field mapping between electrically biased needles by scanning transmission electron microscopy and electron holography}}, Ultramicroscopy 253~(July) (2023) 113808.
\newblock \href {https://doi.org/10.1016/j.ultramic.2023.113808} {\path{doi:10.1016/j.ultramic.2023.113808}}.
\newline\urlprefix\url{https://doi.org/10.1016/j.ultramic.2023.113808}

\bibitem{Paszke2019}
A.~Paszke, S.~Gross, F.~Massa, A.~Lerer, J.~Bradbury, G.~Chanan, T.~Killeen, Z.~Lin, N.~Gimelshein, L.~Antiga, A.~Desmaison, A.~Kopf, E.~Yang, Z.~DeVito, M.~Raison, A.~Tejani, S.~Chilamkurthy, B.~Steiner, L.~Fang, J.~Bai, S.~Chintala, \href{http://papers.neurips.cc/paper/9015-pytorch-an-imperative-style-high-performance-deep-learning-library.pdf}{{PyTorch: An Imperative Style, High-Performance Deep Learning Library}}, in: Advances in Neural Information Processing Systems 32, Curran Associates, Inc., 2019, pp. 8024--8035.
\newblock \href {https://doi.org/10.5555/3454287.3455008} {\path{doi:10.5555/3454287.3455008}}.
\newline\urlprefix\url{http://papers.neurips.cc/paper/9015-pytorch-an-imperative-style-high-performance-deep-learning-library.pdf}

\bibitem{Self1983}
P.~G. Self, M.~A. O'Keefe, P.~R. Buseck, A.~E.~C. Spargo, \href{http://www.sciencedirect.com/science/article/pii/0304399183900530}{{Practical computation of amplitudes and phases in electron diffraction}}, Ultramicroscopy 11~(1) (1983) 35--52.
\newblock \href {https://doi.org/10.1016/0304-3991(83)90053-0} {\path{doi:10.1016/0304-3991(83)90053-0}}.
\newline\urlprefix\url{http://www.sciencedirect.com/science/article/pii/0304399183900530}

\bibitem{Kirkland2020}
E.~J. Kirkland, \href{http://link.springer.com/10.1007/978-3-030-33260-0}{{Advanced Computing in Electron Microscopy}}, Springer International Publishing, Cham, 2020.
\newblock \href {https://doi.org/10.1007/978-3-030-33260-0} {\path{doi:10.1007/978-3-030-33260-0}}.
\newline\urlprefix\url{http://link.springer.com/10.1007/978-3-030-33260-0}

\bibitem{Cowley1979a}
J.~M. Cowley, {Coherent interference in convergent-beam electron diffraction and shadow imaging}, Ultramicroscopy 4 (1979) 435--450.

\bibitem{Capitani2006}
G.~C. Capitani, P.~Oleynikov, S.~Hovm{\"{o}}ller, M.~Mellini, {A practical method to detect and correct for lens distortion in the TEM}, Ultramicroscopy 106~(2) (2006) 66--74.
\newblock \href {https://doi.org/10.1016/j.ultramic.2005.06.003} {\path{doi:10.1016/j.ultramic.2005.06.003}}.

\bibitem{Cowley1957}
J.~M. Cowley, A.~F. Moodie, \href{http://dx.doi.org/10.1107/S0365110X57002194}{{The scattering of electrons by atoms and crystals. I. A new theoretical approach}}, Acta Crystallographica 10~(10) (1957) 609--619.
\newblock \href {https://doi.org/10.1107/s0365110x57002194} {\path{doi:10.1107/s0365110x57002194}}.
\newline\urlprefix\url{http://dx.doi.org/10.1107/S0365110X57002194}

\bibitem{Goodman1974}
P.~Goodman, A.~F. Moodie, {Numerical evaluations of N‐beam wave functions in electron scattering by the multi‐slice method}, Acta Crystallographica Section A 30~(2) (1974) 280--290.
\newblock \href {https://doi.org/10.1107/S056773947400057X} {\path{doi:10.1107/S056773947400057X}}.

\bibitem{Ishizuka1977}
K.~Ishizuka, N.~Uyeda, \href{http://dx.doi.org/10.1107/S0567739477001879}{{A new theoretical and practical approach to the multislice method}}, Acta Crystallogr., Sect. A 33~(5) (1977) 740--749.
\newblock \href {https://doi.org/10.1107/S0567739477001879} {\path{doi:10.1107/S0567739477001879}}.
\newline\urlprefix\url{http://dx.doi.org/10.1107/S0567739477001879}

\bibitem{Lobato2014}
I.~Lobato, D.~{Van Dyck}, \href{http://scripts.iucr.org/cgi-bin/paper?S205327331401643X}{{An accurate parameterization for scattering factors, electron densities and electrostatic potentials for neutral atoms that obey all physical constraints}}, Acta Crystallographica Section A Foundations and Advances 70~(6) (2014) 636--649.
\newblock \href {https://doi.org/10.1107/S205327331401643X} {\path{doi:10.1107/S205327331401643X}}.
\newline\urlprefix\url{http://scripts.iucr.org/cgi-bin/paper?S205327331401643X}

\bibitem{Wang1998a}
Z.~L. Wang, \href{http://dx.doi.org/10.1107/S0108767398001457}{{The 'Frozen-Lattice' Approach for Incoherent Phonon Excitation in Electron Scattering. How Accurate Is It?}}, Acta Crystallographica Section A: Foundations of Crystallography 54~(4) (1998) 460--467.
\newblock \href {https://doi.org/10.1107/S0108767398001457} {\path{doi:10.1107/S0108767398001457}}.
\newline\urlprefix\url{http://dx.doi.org/10.1107/S0108767398001457}

\bibitem{Muller2001}
D.~A. Muller, B.~Edwards, E.~J. Kirkland, J.~Silcox, {Simulation of thermal diffuse scattering including a detailed phonon dispersion curve}, Ultramicroscopy 86~(86) (2001) 371--380.

\bibitem{Loane1991}
R.~F. Loane, P.~Xu, J.~Silcox, {Thermal vibrations in convergent‐beam electron diffraction}, Acta Crystallographica Section A 47~(3) (1991) 267--278.
\newblock \href {https://doi.org/10.1107/S0108767391000375} {\path{doi:10.1107/S0108767391000375}}.

\bibitem{Denisov2023}
N.~Denisov, D.~Jannis, A.~Orekhov, K.~M{\"{u}}ller-Caspary, J.~Verbeeck, \href{https://doi.org/10.1016/j.ultramic.2023.113777}{{Characterization of a Timepix detector for use in SEM acceleration voltage range}}, Ultramicroscopy 253~(April) (2023) 113777.
\newblock \href {https://doi.org/10.1016/j.ultramic.2023.113777} {\path{doi:10.1016/j.ultramic.2023.113777}}.
\newline\urlprefix\url{https://doi.org/10.1016/j.ultramic.2023.113777}

\bibitem{LeBeau2008}
J.~M. LeBeau, S.~Stemmer, {Experimental quantification of annular dark-field images in scanning transmission electron microscopy}, Ultramicroscopy 108 (2008) 1653--1658.
\newblock \href {https://doi.org/10.1016/j.ultramic.2008.07.001} {\path{doi:10.1016/j.ultramic.2008.07.001}}.

\bibitem{Rosenauer2009}
A.~Rosenauer, K.~Gries, K.~M{\"{u}}ller, A.~Pretorius, M.~Schowalter, A.~Avramescu, K.~Engl, S.~Lutgen, \href{http://www.sciencedirect.com/science/article/B6TW1-4W91PY1-2/2/377d9008104d5889aa5e83786389aaaa}{{Measurement of specimen thickness and composition in Alx Ga1 - x N / GaN using high-angle annular dark field images}}, Ultramicroscopy 109~(9) (2009) 1171--1182.
\newblock \href {https://doi.org/10.1016/j.ultramic.2009.05.003} {\path{doi:10.1016/j.ultramic.2009.05.003}}.
\newline\urlprefix\url{http://www.sciencedirect.com/science/article/B6TW1-4W91PY1-2/2/377d9008104d5889aa5e83786389aaaa}

\bibitem{DAlfonso2016}
A.~J. D'Alfonso, L.~J. Allen, H.~Sawada, A.~I. Kirkland, \href{http://dx.doi.org/10.1063/1.4941269}{{Dose-dependent high-resolution electron ptychography}}, Journal of Applied Physics 119~(5) (2016) 0--5.
\newblock \href {https://doi.org/10.1063/1.4941269} {\path{doi:10.1063/1.4941269}}.
\newline\urlprefix\url{http://dx.doi.org/10.1063/1.4941269}

\bibitem{Cederquist1987}
J.~N. Cederquist, C.~C. Wackerman, \href{https://opg.optica.org/abstract.cfm?URI=josaa-4-9-1788}{{Phase-retrieval error: a lower bound}}, Journal of the Optical Society of America A 4~(9) (1987) 1788.
\newblock \href {https://doi.org/10.1364/JOSAA.4.001788} {\path{doi:10.1364/JOSAA.4.001788}}.
\newline\urlprefix\url{https://opg.optica.org/abstract.cfm?URI=josaa-4-9-1788}

\bibitem{Wei2020}
X.~Wei, H.~P. Urbach, W.~M. Coene, {Cram{\'{e}}r-Rao lower bound and maximum-likelihood estimation in ptychography with Poisson noise}, Physical Review A 102~(4) (2020).
\newblock \href {https://doi.org/10.1103/PhysRevA.102.043516} {\path{doi:10.1103/PhysRevA.102.043516}}.

\bibitem{Bouchet2021}
D.~Bouchet, J.~Dong, D.~Maestre, T.~Juffmann, \href{https://doi.org/10.1103/PhysRevApplied.15.024047 https://link.aps.org/doi/10.1103/PhysRevApplied.15.024047}{{Fundamental Bounds on the Precision of Classical Phase Microscopes}}, Physical Review Applied 15~(2) (2021) 024047.
\newblock \href {http://arxiv.org/abs/2011.04799} {\path{arXiv:2011.04799}}, \href {https://doi.org/10.1103/PhysRevApplied.15.024047} {\path{doi:10.1103/PhysRevApplied.15.024047}}.
\newline\urlprefix\url{https://doi.org/10.1103/PhysRevApplied.15.024047 https://link.aps.org/doi/10.1103/PhysRevApplied.15.024047}

\bibitem{Koppell2021}
S.~Koppell, M.~Kasevich, \href{https://opg.optica.org/abstract.cfm?URI=optica-8-4-493}{{Information transfer as a framework for optimized phase imaging}}, Optica 8~(4) (2021) 493.
\newblock \href {http://arxiv.org/abs/2010.09786} {\path{arXiv:2010.09786}}, \href {https://doi.org/10.1364/OPTICA.412129} {\path{doi:10.1364/OPTICA.412129}}.
\newline\urlprefix\url{https://opg.optica.org/abstract.cfm?URI=optica-8-4-493}

\bibitem{Dwyer2024}
C.~Dwyer, D.~M. Paganin, \href{http://arxiv.org/abs/2309.04701 https://link.aps.org/doi/10.1103/PhysRevB.110.024110}{{Quantum and classical Fisher information in four-dimensional scanning transmission electron microscopy}}, Physical Review B 110~(2) (2024) 024110.
\newblock \href {http://arxiv.org/abs/2309.04701} {\path{arXiv:2309.04701}}, \href {https://doi.org/10.1103/PhysRevB.110.024110} {\path{doi:10.1103/PhysRevB.110.024110}}.
\newline\urlprefix\url{http://arxiv.org/abs/2309.04701 https://link.aps.org/doi/10.1103/PhysRevB.110.024110}

\bibitem{VegaIbanez2025}
F.~{Vega Ib{\'{a}}{\~{n}}ez}, J.~Verbeeck, \href{http://arxiv.org/abs/2408.10590 https://academic.oup.com/mam/advance-article/doi/10.1093/mam/ozae125/7953286}{{Retrieval of Phase Information from Low-Dose Electron Microscopy Experiments: Are We at the Limit Yet?}}, Microscopy and Microanalysis (2025) 1--14\href {http://arxiv.org/abs/2408.10590} {\path{arXiv:2408.10590}}, \href {https://doi.org/10.1093/mam/ozae125} {\path{doi:10.1093/mam/ozae125}}.
\newline\urlprefix\url{http://arxiv.org/abs/2408.10590 https://academic.oup.com/mam/advance-article/doi/10.1093/mam/ozae125/7953286}

\bibitem{Rao1945}
C.~R. Rao, \href{http://link.springer.com/10.1007/978-1-4612-0919-5_16}{{Information and the Accuracy Attainable in the Estimation of Statistical Parameters}}, in: Bull. Calcutta Math. Soc., Vol.~37, 1992, pp. 235--247.
\newblock \href {https://doi.org/10.1007/978-1-4612-0919-5_16} {\path{doi:10.1007/978-1-4612-0919-5_16}}.
\newline\urlprefix\url{http://link.springer.com/10.1007/978-1-4612-0919-5_16}

\bibitem{Egerton2013}
R.~F. Egerton, \href{http://dx.doi.org/10.1016/j.ultramic.2012.07.006}{{Control of radiation damage in the TEM}}, Ultramicroscopy 127 (2013) 100--108.
\newblock \href {https://doi.org/10.1016/j.ultramic.2012.07.006} {\path{doi:10.1016/j.ultramic.2012.07.006}}.
\newline\urlprefix\url{http://dx.doi.org/10.1016/j.ultramic.2012.07.006}

\bibitem{VanHeel1982}
M.~G. van Heel, W.~Keegstra, W.~Schutter, E.~F.~J. van Bruggen, {Life Chemistry Reports}, Life Chemistry Reports 5, suppl. (1982) 69--73.

\bibitem{Saxton1982}
W.~O. Saxton, W.~Baumeister, \href{https://onlinelibrary.wiley.com/doi/10.1111/j.1365-2818.1982.tb00405.x}{{The correlation averaging of a regularly arranged bacterial cell envelope protein}}, Journal of Microscopy 127~(2) (1982) 127--138.
\newblock \href {https://doi.org/10.1111/j.1365-2818.1982.tb00405.x} {\path{doi:10.1111/j.1365-2818.1982.tb00405.x}}.
\newline\urlprefix\url{https://onlinelibrary.wiley.com/doi/10.1111/j.1365-2818.1982.tb00405.x}

\bibitem{Edo2013}
T.~B. Edo, D.~J. Batey, A.~M. Maiden, C.~Rau, U.~Wagner, Z.~D. Pe{\v{s}}i{\'{c}}, T.~A. Waigh, J.~M. Rodenburg, \href{https://link.aps.org/doi/10.1103/PhysRevA.87.053850}{{Sampling in x-ray ptychography}}, Physical Review A 87~(5) (2013) 053850.
\newblock \href {https://doi.org/10.1103/PhysRevA.87.053850} {\path{doi:10.1103/PhysRevA.87.053850}}.
\newline\urlprefix\url{https://link.aps.org/doi/10.1103/PhysRevA.87.053850}

\bibitem{Kabius2009}
B.~Kabius, P.~Hartel, M.~Haider, H.~M{\"{u}}ller, S.~Uhlemann, U.~Loebau, J.~Zach, H.~Rose, {First application of Cc-corrected imaging for high-resolution and energy-filtered TEM}, Journal of Electron Microscopy 58~(3) (2009) 147--155.
\newblock \href {https://doi.org/10.1093/jmicro/dfp021} {\path{doi:10.1093/jmicro/dfp021}}.

\bibitem{Sawada2015a}
H.~Sawada, T.~Sasaki, F.~Hosokawa, K.~Suenaga, \href{https://link.aps.org/doi/10.1103/PhysRevLett.114.166102}{{Atomic-Resolution STEM Imaging of Graphene at Low Voltage of 30 kV with Resolution Enhancement by Using Large Convergence Angle}}, Physical Review Letters 114~(16) (2015) 166102.
\newblock \href {https://doi.org/10.1103/PhysRevLett.114.166102} {\path{doi:10.1103/PhysRevLett.114.166102}}.
\newline\urlprefix\url{https://link.aps.org/doi/10.1103/PhysRevLett.114.166102}

\bibitem{Ishikawa2015}
R.~Ishikawa, A.~R. Lupini, Y.~Hinuma, S.~J. Pennycook, \href{http://dx.doi.org/10.1016/j.ultramic.2014.11.009}{{Large-angle illumination STEM: Toward three-dimensional atom-by-atom imaging}}, Ultramicroscopy 151~(1) (2015) 122--129.
\newblock \href {https://doi.org/10.1016/j.ultramic.2014.11.009} {\path{doi:10.1016/j.ultramic.2014.11.009}}.
\newline\urlprefix\url{http://dx.doi.org/10.1016/j.ultramic.2014.11.009}

\bibitem{Brown2019}
H.~G. Brown, R.~Ishikawa, G.~S´anchez-Santolino, N.~Shibata, Y.~Ikuhara, L.~J. Allen, S.~D. Findlay, \href{https://linkinghub.elsevier.com/retrieve/pii/S030439911830336X https://doi.org/10.1016/j.ultramic.2018.12.010}{{Large angle illumination enabling accurate structure reconstruction from thick samples in scanning transmission electron microscopy}}, Ultramicroscopy 197~(October 2018) (2019) 112--121.
\newblock \href {https://doi.org/10.1016/j.ultramic.2018.12.010} {\path{doi:10.1016/j.ultramic.2018.12.010}}.
\newline\urlprefix\url{https://linkinghub.elsevier.com/retrieve/pii/S030439911830336X https://doi.org/10.1016/j.ultramic.2018.12.010}

\bibitem{Ma2024}
Y.~Ma, J.~Shi, R.~Guzman, A.~Li, W.~Zhou, \href{https://doi.org/10.1093/micmic/ozae027 https://academic.oup.com/mam/article/30/2/226/7641388}{{Aberration Correction for Large-Angle Illumination Scanning Transmission Electron Microscopy by Using Iterative Electron Ptychography Algorithms}}, Microscopy and Microanalysis 30~(2) (2024) 226--235.
\newblock \href {https://doi.org/10.1093/mam/ozae027} {\path{doi:10.1093/mam/ozae027}}.
\newline\urlprefix\url{https://doi.org/10.1093/micmic/ozae027 https://academic.oup.com/mam/article/30/2/226/7641388}

\bibitem{Susi2019a}
T.~Susi, J.~C. Meyer, J.~Kotakoski, \href{http://dx.doi.org/10.1038/s42254-019-0058-y https://www.nature.com/articles/s42254-019-0058-y}{{Quantifying transmission electron microscopy irradiation effects using two-dimensional materials}}, Nature Reviews Physics 1~(6) (2019) 397--405.
\newblock \href {https://doi.org/10.1038/s42254-019-0058-y} {\path{doi:10.1038/s42254-019-0058-y}}.
\newline\urlprefix\url{http://dx.doi.org/10.1038/s42254-019-0058-y https://www.nature.com/articles/s42254-019-0058-y}

\bibitem{Muller2023}
J.~M{\"{u}}ller, M.~Heyl, T.~Schultz, K.~Elsner, M.~Schloz, S.~R{\"{u}}hl, H.~Seiler, N.~Koch, E.~J. List-Kratochvil, C.~T. Koch, {Probing crystallinity and grain structure of 2D materials and 2D‐like van der Waals heterostructures by low‐voltage electron diffraction}, Physica Status Solidi (a) (2023).
\newblock \href {https://doi.org/10.1002/pssa.202300148} {\path{doi:10.1002/pssa.202300148}}.

\bibitem{Cheng2015}
Y.~Cheng, N.~Grigorieff, P.~A. Penczek, T.~Walz, \href{http://dx.doi.org/10.1016/j.cell.2015.03.050 https://linkinghub.elsevier.com/retrieve/pii/S0092867415003700}{{A Primer to Single-Particle Cryo-Electron Microscopy}}, Cell 161~(3) (2015) 438--449.
\newblock \href {https://doi.org/10.1016/j.cell.2015.03.050} {\path{doi:10.1016/j.cell.2015.03.050}}.
\newline\urlprefix\url{http://dx.doi.org/10.1016/j.cell.2015.03.050 https://linkinghub.elsevier.com/retrieve/pii/S0092867415003700}

\bibitem{Nakane2020}
T.~Nakane, A.~Kotecha, A.~Sente, G.~McMullan, S.~Masiulis, P.~M. G.~E. Brown, I.~T. Grigoras, L.~Malinauskaite, T.~Malinauskas, J.~Miehling, T.~Ucha{\'{n}}ski, L.~Yu, D.~Karia, E.~V. Pechnikova, E.~de~Jong, J.~Keizer, M.~Bischoff, J.~McCormack, P.~Tiemeijer, S.~W. Hardwick, D.~Y. Chirgadze, G.~Murshudov, A.~R. Aricescu, S.~H.~W. Scheres, \href{http://dx.doi.org/10.1038/s41586-020-2829-0 https://www.nature.com/articles/s41586-020-2829-0}{{Single-particle cryo-EM at atomic resolution}}, Nature 587~(7832) (2020) 152--156.
\newblock \href {https://doi.org/10.1038/s41586-020-2829-0} {\path{doi:10.1038/s41586-020-2829-0}}.
\newline\urlprefix\url{http://dx.doi.org/10.1038/s41586-020-2829-0 https://www.nature.com/articles/s41586-020-2829-0}

\bibitem{Bethe1930}
H.~Bethe, \href{http://dx.doi.org/10.1002/andp.19303970303}{{Zur Theorie des Durchgangs schneller Korpuskularstrahlen durch Materie}}, Ann. Phys. 5 (1930) 325.
\newblock \href {https://doi.org/10.1002/andp.19303970303} {\path{doi:10.1002/andp.19303970303}}.
\newline\urlprefix\url{http://dx.doi.org/10.1002/andp.19303970303}

\bibitem{Mott1930}
N.~F. Mott, \href{http://rspa.royalsocietypublishing.org/content/127/806/658.short}{{The Scattering of Electrons by Atoms}}, Proc. Roy. Soc. Lond. A 127~(806) (1930) 658--665.
\newblock \href {https://doi.org/10.1098/rspa.1930.0082} {\path{doi:10.1098/rspa.1930.0082}}.
\newline\urlprefix\url{http://rspa.royalsocietypublishing.org/content/127/806/658.short}

\bibitem{Dubochet1988}
J.~Dubochet, M.~Adrian, J.-J. Chang, J.-C. Homo, J.~Lepault, A.~W. McDowall, P.~Schultz, \href{https://www.cambridge.org/core/product/identifier/S0033583500004297/type/journal_article}{{Cryo-electron microscopy of vitrified specimens}}, Quarterly Reviews of Biophysics 21~(2) (1988) 129--228.
\newblock \href {https://doi.org/10.1017/S0033583500004297} {\path{doi:10.1017/S0033583500004297}}.
\newline\urlprefix\url{https://www.cambridge.org/core/product/identifier/S0033583500004297/type/journal_article}

\bibitem{Henderson1995}
R.~Henderson, \href{https://www.cambridge.org/core/product/identifier/S003358350000305X/type/journal_article}{{The potential and limitations of neutrons, electrons and X-rays for atomic resolution microscopy of unstained biological molecules}}, Quarterly Reviews of Biophysics 28~(2) (1995) 171--193.
\newblock \href {https://doi.org/10.1017/S003358350000305X} {\path{doi:10.1017/S003358350000305X}}.
\newline\urlprefix\url{https://www.cambridge.org/core/product/identifier/S003358350000305X/type/journal_article}

\bibitem{Gao2024}
C.~Gao, C.~Hofer, T.~Pennycook, \href{http://arxiv.org/abs/2306.08587 https://linkinghub.elsevier.com/retrieve/pii/S0304399123001961}{{On central focusing for contrast optimization in direct electron ptychography of thick samples}}, Ultramicroscopy 256~(November 2023) (2024) 113879.
\newblock \href {http://arxiv.org/abs/2306.08587} {\path{arXiv:2306.08587}}, \href {https://doi.org/10.1016/j.ultramic.2023.113879} {\path{doi:10.1016/j.ultramic.2023.113879}}.
\newline\urlprefix\url{http://arxiv.org/abs/2306.08587 https://linkinghub.elsevier.com/retrieve/pii/S0304399123001961}

\bibitem{Baxter2009}
W.~T. Baxter, R.~A. Grassucci, H.~Gao, J.~Frank, \href{http://dx.doi.org/10.1016/j.jsb.2009.02.012 https://linkinghub.elsevier.com/retrieve/pii/S1047847709000641}{{Determination of signal-to-noise ratios and spectral SNRs in cryo-EM low-dose imaging of molecules}}, Journal of Structural Biology 166~(2) (2009) 126--132.
\newblock \href {https://doi.org/10.1016/j.jsb.2009.02.012} {\path{doi:10.1016/j.jsb.2009.02.012}}.
\newline\urlprefix\url{http://dx.doi.org/10.1016/j.jsb.2009.02.012 https://linkinghub.elsevier.com/retrieve/pii/S1047847709000641}

\bibitem{Vulovic2013}
M.~Vulovi{\'{c}}, R.~B. Ravelli, L.~J. van Vliet, A.~J. Koster, I.~Lazi{\'{c}}, U.~L{\"{u}}cken, H.~Rullg{\aa}rd, O.~{\"{O}}ktem, B.~Rieger, \href{https://linkinghub.elsevier.com/retrieve/pii/S1047847713001226}{{Image formation modeling in cryo-electron microscopy}}, Journal of Structural Biology 183~(1) (2013) 19--32.
\newblock \href {https://doi.org/10.1016/j.jsb.2013.05.008} {\path{doi:10.1016/j.jsb.2013.05.008}}.
\newline\urlprefix\url{https://linkinghub.elsevier.com/retrieve/pii/S1047847713001226}

\bibitem{Parkhurst2024}
J.~M. Parkhurst, A.~Cavalleri, M.~Dumoux, M.~Basham, D.~Clare, C.~A. Siebert, G.~Evans, J.~H. Naismith, A.~Kirkland, J.~W. Essex, \href{https://doi.org/10.1016/j.ultramic.2023.113882 https://linkinghub.elsevier.com/retrieve/pii/S0304399123001997}{{Computational models of amorphous ice for accurate simulation of cryo-EM images of biological samples}}, Ultramicroscopy 256~(November 2023) (2024) 113882.
\newblock \href {https://doi.org/10.1016/j.ultramic.2023.113882} {\path{doi:10.1016/j.ultramic.2023.113882}}.
\newline\urlprefix\url{https://doi.org/10.1016/j.ultramic.2023.113882 https://linkinghub.elsevier.com/retrieve/pii/S0304399123001997}

\bibitem{Plamann1998}
T.~Plamann, J.~M. Rodenburg, \href{https://scripts.iucr.org/cgi-bin/paper?S0108767397010507}{{Electron Ptychography. II. Theory of Three-Dimensional Propagation Effects}}, Acta Crystallographica Section A Foundations of Crystallography 54~(1) (1998) 61--73.
\newblock \href {https://doi.org/10.1107/S0108767397010507} {\path{doi:10.1107/S0108767397010507}}.
\newline\urlprefix\url{https://scripts.iucr.org/cgi-bin/paper?S0108767397010507}

\bibitem{Maiden2012}
A.~M. Maiden, M.~J. Humphry, J.~M. Rodenburg, \href{https://opg.optica.org/abstract.cfm?URI=josaa-29-8-1606}{{Ptychographic transmission microscopy in three dimensions using a multi-slice approach}}, Journal of the Optical Society of America A 29~(8) (2012) 1606.
\newblock \href {https://doi.org/10.1364/JOSAA.29.001606} {\path{doi:10.1364/JOSAA.29.001606}}.
\newline\urlprefix\url{https://opg.optica.org/abstract.cfm?URI=josaa-29-8-1606}

\bibitem{Tsai2016}
E.~H.~R. Tsai, I.~Usov, A.~Diaz, A.~Menzel, M.~Guizar-Sicairos, \href{https://opg.optica.org/abstract.cfm?URI=oe-24-25-29089}{{X-ray ptychography with extended depth of field}}, Optics Express 24~(25) (2016) 29089.
\newblock \href {https://doi.org/10.1364/OE.24.029089} {\path{doi:10.1364/OE.24.029089}}.
\newline\urlprefix\url{https://opg.optica.org/abstract.cfm?URI=oe-24-25-29089}

\bibitem{Gao2017}
S.~Gao, P.~Wang, F.~Zhang, G.~T. Martinez, P.~D. Nellist, X.~Pan, A.~I. Kirkland, \href{https://doi.org/10.1038/s41467-017-00150-1 http://dx.doi.org/10.1038/s41467-017-00150-1}{{Electron ptychographic microscopy for three-dimensional imaging}}, Nature Communications 8~(1) (2017) 163.
\newblock \href {https://doi.org/10.1038/s41467-017-00150-1} {\path{doi:10.1038/s41467-017-00150-1}}.
\newline\urlprefix\url{https://doi.org/10.1038/s41467-017-00150-1 http://dx.doi.org/10.1038/s41467-017-00150-1}

\bibitem{Chen2021}
Z.~Chen, Y.~Jiang, Y.-T.~T. Shao, M.~E. Holtz, M.~Odstr{\v{c}}il, M.~Guizar-Sicairos, I.~Hanke, S.~Ganschow, D.~G. Schlom, D.~A. Muller, \href{https://www.sciencemag.org/lookup/doi/10.1126/science.abg2533}{{Electron ptychography achieves atomic-resolution limits set by lattice vibrations}}, Science 372~(6544) (2021) 826--831.
\newblock \href {http://arxiv.org/abs/2101.00465} {\path{arXiv:2101.00465}}, \href {https://doi.org/10.1126/science.abg2533} {\path{doi:10.1126/science.abg2533}}.
\newline\urlprefix\url{https://www.sciencemag.org/lookup/doi/10.1126/science.abg2533}

\bibitem{Susi2025}
T.~Susi, \href{http://arxiv.org/abs/2502.09938 https://onlinelibrary.wiley.com/doi/10.1111/jmi.13409}{{Quantifying phase magnitudes of open‐source focused‐probe 4D‐STEM ptychography reconstructions}}, Journal of Microscopy~(February) (2025) 1--16.
\newblock \href {http://arxiv.org/abs/2502.09938} {\path{arXiv:2502.09938}}, \href {https://doi.org/10.1111/jmi.13409} {\path{doi:10.1111/jmi.13409}}.
\newline\urlprefix\url{http://arxiv.org/abs/2502.09938 https://onlinelibrary.wiley.com/doi/10.1111/jmi.13409}

\bibitem{Ballabriga2018}
R.~Ballabriga, M.~Campbell, X.~Llopart, \href{http://dx.doi.org/10.1016/j.nima.2017.07.029}{{Asic developments for radiation imaging applications: The medipix and timepix family}}, Nuclear Instruments and Methods in Physics Research, Section A: Accelerators, Spectrometers, Detectors and Associated Equipment 878~(May 2017) (2018) 10--23.
\newblock \href {https://doi.org/10.1016/j.nima.2017.07.029} {\path{doi:10.1016/j.nima.2017.07.029}}.
\newline\urlprefix\url{http://dx.doi.org/10.1016/j.nima.2017.07.029}

\bibitem{Pennicard2011}
D.~Pennicard, R.~Ballabriga, X.~Llopart, M.~Campbell, H.~Graafsma, \href{https://linkinghub.elsevier.com/retrieve/pii/S0168900211002105 http://dx.doi.org/10.1016/j.nima.2011.01.124}{{Simulations of charge summing and threshold dispersion effects in Medipix3}}, Nuclear Instruments and Methods in Physics Research Section A: Accelerators, Spectrometers, Detectors and Associated Equipment 636~(1) (2011) 74--81.
\newblock \href {https://doi.org/10.1016/j.nima.2011.01.124} {\path{doi:10.1016/j.nima.2011.01.124}}.
\newline\urlprefix\url{https://linkinghub.elsevier.com/retrieve/pii/S0168900211002105 http://dx.doi.org/10.1016/j.nima.2011.01.124}

\bibitem{VanSchayck2020}
J.~P. van Schayck, E.~van Genderen, E.~Maddox, L.~Roussel, H.~Boulanger, E.~Fr{\"{o}}jdh, J.-P. Abrahams, P.~J. Peters, R.~B. Ravelli, \href{https://doi.org/10.1016/j.ultramic.2020.113091 https://linkinghub.elsevier.com/retrieve/pii/S0304399120302424}{{Sub-pixel electron detection using a convolutional neural network}}, Ultramicroscopy 218~(February) (2020) 113091.
\newblock \href {https://doi.org/10.1016/j.ultramic.2020.113091} {\path{doi:10.1016/j.ultramic.2020.113091}}.
\newline\urlprefix\url{https://doi.org/10.1016/j.ultramic.2020.113091 https://linkinghub.elsevier.com/retrieve/pii/S0304399120302424}

\bibitem{VanSchayck2023}
J.~P. van Schayck, Y.~Zhang, K.~Knoops, P.~J. Peters, R.~B.~G. Ravelli, {Integration of an Event-driven Timepix3 Hybrid Pixel Detector into a Cryo-EM Workflow}, Microscopy and Microanalysis 29~(1) (2023) 352--363.
\newblock \href {https://doi.org/10.1093/micmic/ozac009} {\path{doi:10.1093/micmic/ozac009}}.

\end{thebibliography}

\end{document}